\documentclass{IEEEtran}
\usepackage{cite}
\usepackage{amsmath,amssymb,amsfonts}
\usepackage{balance}
\usepackage{algorithmic}
\usepackage{graphicx}
\usepackage{textcomp}
\def\BibTeX{{\rm B\kern-.05em{\sc i\kern-.025em b}\kern-.08em
    T\kern-.1667em\lower.7ex\hbox{E}\kern-.125emX}}
\usepackage{cite}
\usepackage{float}
\usepackage{amsmath}
\usepackage[utf8]{inputenc}
\usepackage[final]{pdfpages}
\usepackage{pdfpages}
\usepackage{lipsum}
\usepackage{textcase}
\usepackage{url}
\usepackage{amsmath,esint} 
\usepackage{epstopdf}
\usepackage{array}
\usepackage{color}
\usepackage{subcaption} 
\usepackage{lettrine}
\usepackage{bm}

\DeclareRobustCommand{\uvec}[1]{{%
  \ifcsname uvec#1\endcsname
     \csname uvec#1\endcsname
   \else
    \bm{\hat{\mathbf{#1}}}%
   \fi
}}

\newcolumntype{P}[1]{>{\centering\arraybackslash}p{#1}}
\newcolumntype{M}[1]{>{\centering\arraybackslash}m{#1}}

\begin{document}

\title{Coverage Enhancement for NLOS mmWave Links Using Passive Reflectors}%: Measurements and Ray Tracing Simulations}

\author{Wahab Khawaja, Ozgur Ozdemir, \IEEEmembership{Member, IEEE}, Yavuz Yapici, \IEEEmembership{Member, IEEE}, Fatih Erden,  Martins Ezuma, and Ismail Guvenc, \IEEEmembership{Member, IEEE}
\thanks{Manuscript submitted April 14, 2019. This  work  was  supported  in  part  by  NASA  under  the  Federal Award ID  NNX17AJ94A, and in part by DOCOMO Innovations. Wahab Khawaja is supported by a Fulbright scholarship. } 
\thanks{
The authors are with the Department of Electrical and Computer Engineering, North Carolina State University, Raleigh, NC 27606 USA (e-mails:\{wkhawaj, oozdemi, yyapici, ferden, mcezuma, iguvenc\}@ncsu.edu). }\vspace{-2mm}
}

% \author{
% \IEEEauthorblockN{Wahab Khawaja, Ozgur Ozdemir, Yavuz Yapici, Fatih Erden, Martins Ezuma and Ismail Guvenc
% }
% %\IEEEauthorblockA{Department of Electrical and Computer Engineering, North Carolina State University, Raleigh, NC}
% }  

\maketitle

\begin{abstract}
The future 5G networks are expected to use millimeter wave~(mmWave) frequency bands to take advantage of large unused spectrum. However, due to the high path loss at mmWave frequencies, coverage of mmWave signals can get severely reduced, especially for non-line-of-sight~(NLOS) scenarios as mmWave signals are severely attenuated when going through obstructions. In this work, we study the use of passive metallic reflectors of different shapes/sizes to improve 28~GHz mmWave signal coverage for both indoor and outdoor NLOS scenarios. We quantify the gains that can be achieved in the link quality with metallic reflectors using measurements, analytical expressions, and ray tracing simulations. In particular, we provide an analytical model for the end-to-end received power in an NLOS scenario using reflectors of different shapes and sizes. %It is observed that 
For a given size of the flat metallic sheet reflector approaching to the size of incident plane waves, we show that the reflected received power for the NLOS link is same as line-of-sight~(LOS) free space received power of the same link distance. Extensive results are provided to study impact of environmental features and reflector characteristics on NLOS link quality.
%Software defined radio based mmWave transceiver platforms operating at $28$~GHz are used for measurements. Subsequently, ray tracing~(RT) simulations are carried out in a similar indoor environment using Remcom Wireless InSite software. 
%We provide extensive results for comparing measurements and RT simulations in NLOS propagation scenarios and quantify coverage gains due to using passive reflectors. 

\begin{IEEEkeywords}
Coverage, indoor, mmWave, non-line-of-sight~(NLOS), outdoor, PXI, ray tracing~(RT), reflector.
\end{IEEEkeywords}
\end{abstract}
\vspace{-1mm}
\IEEEpeerreviewmaketitle

\section{Introduction}
% Importance of mmWaves (motivation), basic problem and solutions
\lettrine[lines=2]{T}{HE} use of smart communication devices and the higher data rate applications supported by them have seen a surge in the recent decade. These applications require higher communication bandwidths, whereas the available sub-$6$~GHz spectrum is reaching its limits due to spectrum congestion. With the opening of millimeter wave~(mmWave) spectrum by FCC~\cite{FCC_28G}, various research efforts are underway to use mmWave spectrum for future $5$G communications. However, a major bottleneck for propagation at mmWave frequencies is the high free space attenuation,  especially for the non-line-of-sight~(NLOS) paths. This makes radio frequency planning very difficult for long distance communications.

% \blfootnote{This  work  has  been  supported  in  part  by  NASA  under  the  Federal Award ID number NNX17AJ94A and by DOCOMO Innovations Inc., Palo Alto, USA. The authors are currently affiliated with Electrical and Computer Engineering Department of North Carolina State University. Email: \{wkhawaj, oozdemi, yyapici, ferden, mcezuma iguvenc\}@ncsu.edu}

Various solutions to this problem have been proposed in the literature, including, high transmit power, high sensitivity receivers, deployment of multiple access points or repeaters, and beam-forming. However, there are limitations to each of these solutions. Increasing the transmit power beyond a certain level becomes impractical due to regulations, whereas the receiver sensitivity, is constrained by the sophisticated and expensive equipment required. Similarly, using a large number of access points may not be feasible economically. The beam-forming requires expensive, complex and power hungry devices, and it may still suffer from NLOS propagation. 

\begin{figure*}[!t]
		
	\begin{subfigure}{0.365\textwidth}
	\centering
    \includegraphics[width=\textwidth]{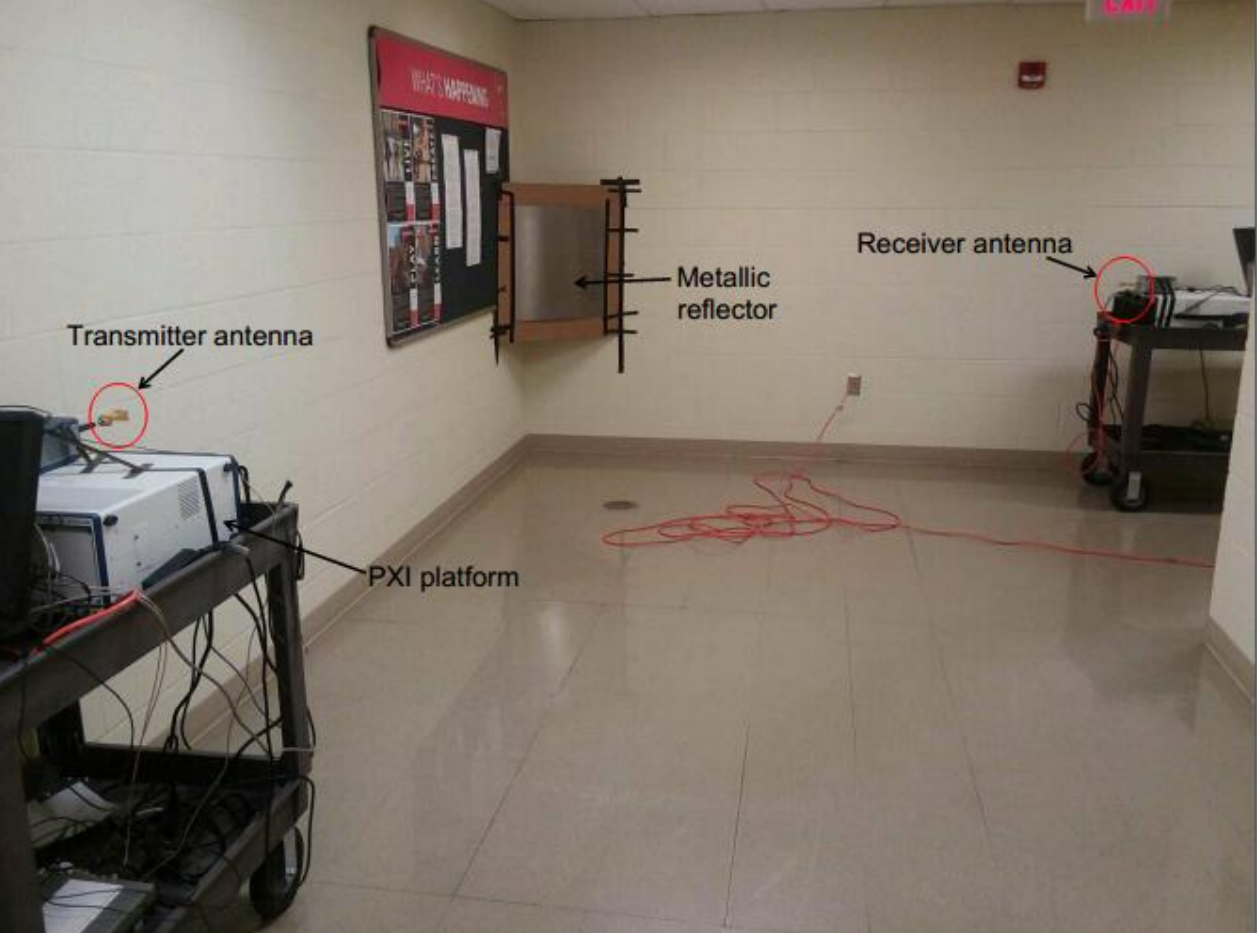}
	 \caption{}
     \end{subfigure}
     \begin{subfigure}{0.245\textwidth}
	\centering
    \includegraphics[width=\textwidth]{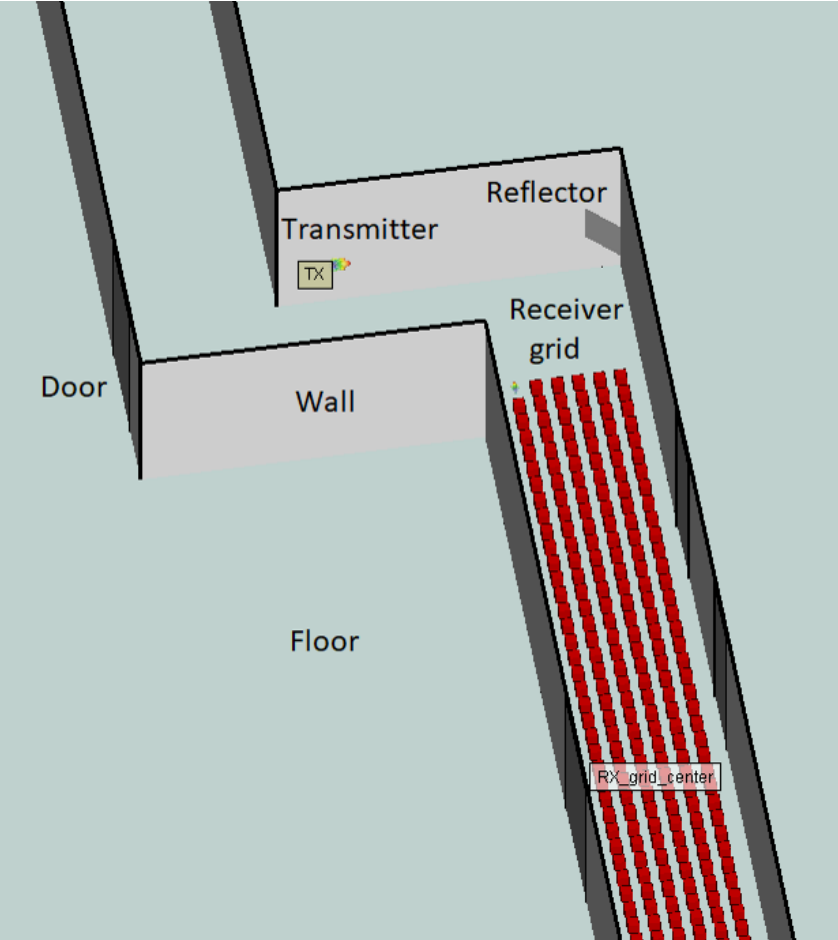}
	 \caption{}
     \end{subfigure}
     \begin{subfigure}{0.39\textwidth}
	\centering
    \includegraphics[width=\textwidth]{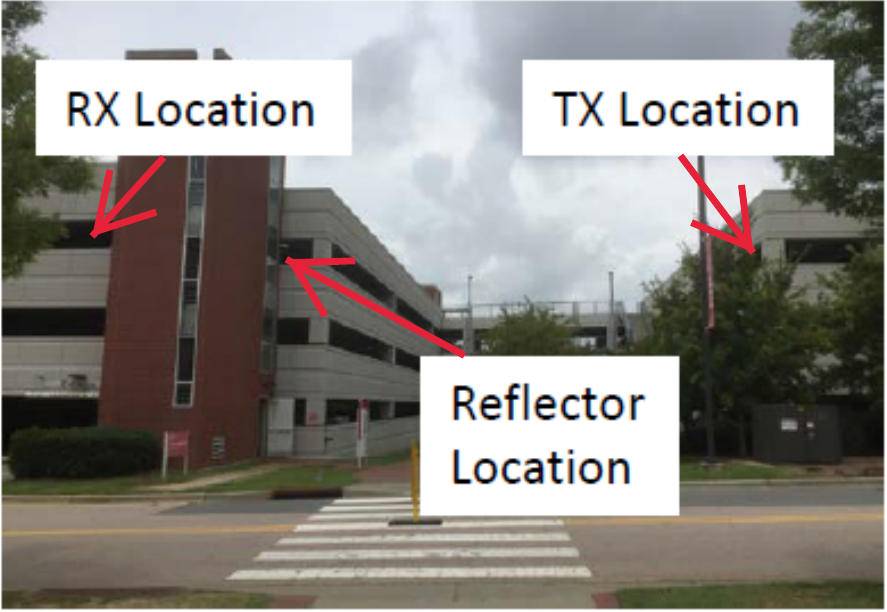}
	 \caption{}
     \end{subfigure}
      \vspace{-1mm}
    \caption{(a) Measurement setup in the basement corridor of Engineering Building~II at North Carolina State University for flat square sheet aluminum reflector $0.61\times0.61$~m$^2$, oriented at an azimuth angle of $45^{\circ}$ with respect to the boresight of the transmit antenna, (b) simulation environment of the measurement scenario in Wireless InSite, and (c) measurement setup for the outdoor scenario.}\label{Fig:scenario}\vspace{-5mm}
\end{figure*}

A convenient %feasible and economical 
solution for NLOS mmWave signal coverage is by introducing metallic passive reflectors. This stems from the fact that electromagnetic waves behave similarly to light~\cite{Light_EM}. The reflection properties of electromagnetic waves are better at higher frequencies due to smaller skin depth~\cite{reflection} and lower material penetration. Similarly, the diffraction around the edges of reflectors is smaller at mmWave frequencies. These reflectors can act similar to a communication repeater but can operate without electricity and negligible maintenance. They have longer life spans, and small initial investment cost when compared with repeaters consisting of active elements. They may even be part of everyday objects, such as street signs, lamp posts, advertisement boards, that can additionally improve mmWave signal coverage. 

Passive metallic reflectors have been studied and employed in the past for long distance satellite communications~\cite{NASA_refl,Literature4,Literature5}. However, these studies are limited to point-to-point links, whereas, for cellular networks, we may require wide coverage. There are also limited studies available for downlink communications using passive reflectors~\cite{Microwave_refl,Literature6}. This is due to the fact that most of the downlink civilian communications operate at sub-$6$~GHz, where the communication radius is in the kilometer range and few communication repeaters are required. Due to the large wavelength, the electromagnetic waves can easily penetrate through most of the building structures without high attenuation, resulting in mostly NLOS communications for the downlink. On the other hand, mmWave signals observe higher free space path loss and higher penetration loss due to smaller wavelengths. As a result, the communication radius generally shrinks to few hundred of meters. This requires a large number of communication repeaters for the downlink and commonly used active repeaters may not be feasible. 

The studies available to date in the literature on using passive reflectors for mmWave coverage enhancement are limited. In~\cite{lit_60GHz_indoor}, indoor coverage analysis at $60$~GHz was carried out due to reflections using simulations. It was observed that at $60$~GHz, the coverage in the NLOS was dependent solely on the reflections. A parabolic passive reflector is used for outdoor coverage enhancement at mmWave frequencies in~\cite{Literature3}, that reflects incoming signal power from the base station to users in the building shadowed zones. Numerical results indicate better coverage in the shadowed zones using reflectors. In \cite{Literature1}, a parabolic reflector is used behind a patch antenna operating at $60$~GHz of a hand held device. A gain of $19$~dB - $25$~dB is reported after the introduction of a parabolic reflector that helps to counter the finger shadowing while operating the device. In~\cite{Literature2}, reflecting properties of different building materials both in the indoor and outdoor environments were calculated using channel measurements at $60$~GHz.

% Contribution of our work
To the best of our knowledge, there are no empirical studies available in the literature on the use of metallic reflectors for downlink coverage enhancement at $28$~GHz except our previous works~\cite{indoor_wahab, outdoor_wahab}. This work is a major extension of our earlier studies~\cite{indoor_wahab, outdoor_wahab}, where key our contributions and findings can be summarized as follows.
%, as additional contributions, we now  provide a comprehensive analytical modeling of the end-to-end received power in the NLOS region using reflectors. Moreover, several new measurements are provided in indoor and outdoor environments for different reflector sizes. The new analytical results for indoor are compared with the measurement and ray tracing simulation results at $28$~GHz, and additional ray tracing results are provided to study the impact of central frequency. The contributions of this work can be summarized as follows:
%\begin{enumerate}
%\item 

$1)$ We have performed measurements for indoor and outdoor NLOS mmWave propagation scenarios at $28$~GHz using National Instruments PXI platform shown in Fig.~\ref{Fig:scenario}. Different sizes and shapes of metallic reflectors are used for enhancing coverage in the NLOS region. The received power was measured over an NLOS grid in an indoor corridor. Ray tracing simulations were also carried out in a similar environment at $28$~GHz and other center frequencies. For the outdoor, the received power was measured at a given NLOS point at different azimuth and elevation angles, %. The received power was measured 
in the presence and absence of flat metallic sheet reflectors.

%\item 
$2)$ An analytical model for end-to-end reflected received power is developed for NLOS propagation. %The model is applicable for different shapes and sizes of the reflectors. 
The model is obtained by considering reflectors as secondary sources of transmission towards the receiver, and it is applicable for different shapes and sizes of the reflectors. The received power due to first order reflections from the flat metallic sheet reflector of a given size approaches to line-of-sight~(LOS) received power. However, for a non-metallic reflector, the received power is significantly smaller compared to a metallic reflector. The reduction in the received power for non-metallic reflectors is mainly due to absorption. 

%\noindent 
$3)$ The received power due to first order reflections 
%from the reflector 
is dependent on the size of the incident plane waves~ given by the pointing vector. Once the size of the reflector is equivalent to the size of the incident plane waves, the received power is independent of the size of the reflector. Similarly, we used a secondary reflector of size comparable to the primary reflector in order to further steer the power in another direction.

%\noindent 
$4)$ For flat reflectors, the reflected received power is mainly dependent on the orientation of the transmitter and receiver antenna's boresight with the surface normal of the reflector. In addition, the received power is also dependent on the orientation of the reflected plane waves from a reflector of given size towards the receiver antenna. Different sizes of flat reflectors provide maximum received power at different orientations for given incident plane waves. However, for outward curved reflectors e.g. cylinder and sphere, the orientation of the reflector towards the receiver is less significant compared to flat reflectors. This is because curved reflectors diverge the incoming energy in different directions.
%\end{enumerate}

%The measurement results are compared with the simulation results obtained using Wireless InSite RT software by creating a similar indoor environment and incorporating diffuse scattering phenomenon in the simulations. \textcolor{red}{Similarly, the analytical results are compared with the measurement and simulation results.}
\vspace{-1mm}
\section{Reflection Characteristics and Assumptions} 
In this section, we will discuss the reflection characteristics of metallic reflectors, size of incident plane waves obtained through measurements, and the effective area of different shaped reflectors. 
\vspace{-1mm}
\subsection{Factors Affecting the Reflection Characteristics}\label{Section:Shapes}
Any solid object in the path of the radio waves can act as a reflector. The reflection characteristics for radio waves are dependent on the following main factors: 1) radiation pattern of the transmit antenna, 2) size and material of the reflector, 3) shape of the reflector, 4) orientation of the reflector~(for flat reflectors). Moreover, the diffraction around the edges of the reflector also affects the reflection characteristics. However, the energy due to diffraction around the edges of the flat reflectors is significantly small. The diffraction loss is greater than $20$~dB~(from single edge diffraction~\cite{diffraction}) for all the flat reflectors. For curved reflectors the diffracted energy is even smaller than the flat reflectors. 

% Transmit antenna pattern
Reflectors can be used effectively for steering directional transmissions to desired NLOS regions. The directional transmission from a wave-guide antenna e.g. horn can be approximated to plane waves in space at a given distance from the source~\cite{balanis}. These plane waves have a given area~(provided by the pointing vector) and shape. The shape and size of plane waves are dependent on the radiation pattern of the antenna, %. The shape and size of plane waves 
and they can be approximated from the half-power beamwidths of the antenna in the azimuth and elevation planes.

% Size and material of reflector
The size of the reflector is an important factor that affects the amount of reflected power. Considering planar wave transmission, the size of the reflector should be at least equal to the size of the incident plane waves. This ensures maximum reflected power. However, increasing the size of the reflector beyond the size of the plane waves will not contribute to the increase in the reflected power. Moreover, the material of the reflector also affects the reflected power. If the material is a perfect conductor with polished surface, we have maximum reflected power. In such a case, the skin depth is zero. However, if the material is not a perfect conductor, then we have additional power losses due to absorption.

% Shape of metallic reflectors
%Shapes of reflectors can vary dependent on the application: e.g., well known primary reflectors used at the backside of the antenna are usually designed to focus the energy beam in a given direction and are usually parabolic in shape. Whereas, for the secondary reflectors, shapes may vary. In the past, large flat metallic sheet reflectors are used for point to point communications among the base stations. However, for down link communications, different shapes of the reflectors can be considered dependent on the coverage requirements. In the rest of the paper, we will focus on the downlink mmWave communication links employing secondary reflectors.   

The shape of a reflector determines the reflection pattern. Radar cross section~(RCS) can be used to distinguish the reflection pattern of different shaped reflectors. Flat reflectors have larger RCS compared to the curved shaped reflectors. Additionally, the RCS pattern is highly directive for flat reflectors. For curved shaped reflectors including cylinder and sphere, the reflection characteristics are mainly dependent on the curve angle. For a given curve angle, the incident plane waves can either converge or diverge in different directions. Other complex shape reflectors such as saw-tooth reflectors can be employed for obtaining different scattering patterns. %\textcolor{red}{Analytical expression for received power due to first order reflections from flat, cylinder and sphere reflectors are provided in Section~\ref{Section:Power_distribution_modeling}.}

%\subsection{Size of Reflectors} \label{Section:Sizes}
%The ideal size of the reflector is an important design consideration and it is related to the radiation pattern of the transmitted energy, distance, and location of the reflector from the transmitter for a given coverage area. Considering a directional transmit source with respective half-power radiation beamwidths at the elevation and azimuth planes, the transmitted directional beam can be represented as occupying a certain part of a sphere. The radius of sphere and surface area of the transmitted radiation beam on the sphere is dependent on the distance from the source. For simplicity, we can approximate the spherical surface of the transmitted radiation beam on the sphere as a rectangle\cite{Antenna_approx}. It can be observed that as the radius doubles, the energy from the source spreads at four times the surface area, resulting in one fourth of intensity as compared to the source. Therefore, in order to reflect maximum incident energy at a given distance from the source, the size of the flat reflector should be comparable to the surface area extended by the beam at that distance. On the other hand, from a communication operator's point of view, the size of reflector should be as small as possible considering construction, deployment costs and regulations~(e.g. strong winds, earthquakes). 

\begin{figure}[!t]
\centering
    \includegraphics[width=0.9\columnwidth]{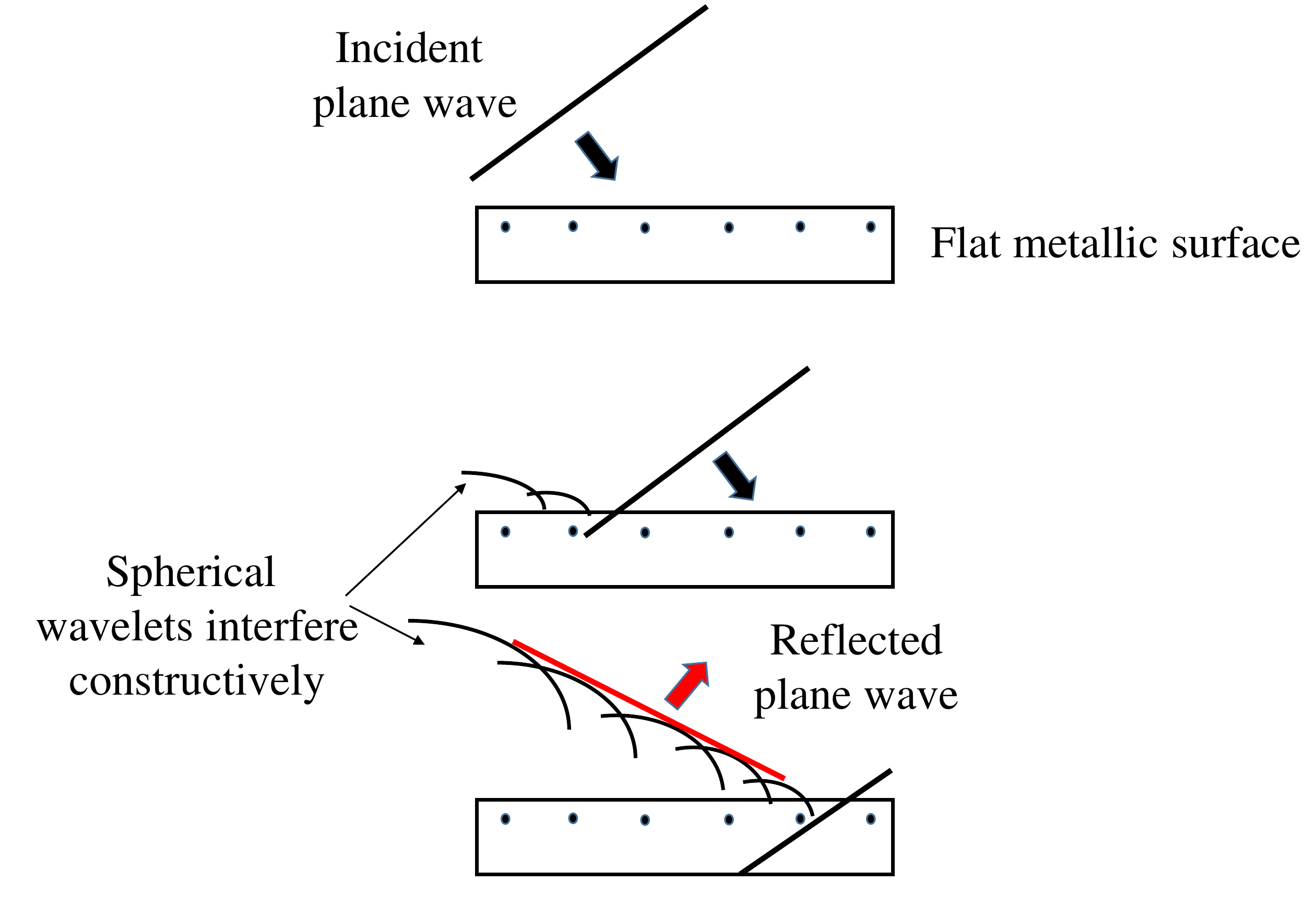}
     \vspace{-1mm}
    \caption{Reflection of an incident plane wave from the flat metallic surface.} \label{Fig:wavefronts} 
    \vspace{-2mm}
\end{figure}

% Orientation of reflector
As discussed earlier, the orientation of flat reflectors are important for obtaining maximum reflected power. The boresight of transmitter and receiver antennas should be aligned towards the center of the reflector. This results in maximum reflected power. However, it was observed during measurements that the region of maximum reflection is different for different sizes of reflectors. This is mainly due to the plane wave nature of the transmission and its interaction with the surface of the reflector, shown in Fig.~\ref{Fig:wavefronts}. This is contrary to the case if we consider transmitted wave as a point light source. On the other hand, for the curved shaped reflectors, change in geometrical orientation is generally not required. For curved reflectors, reflected power is dependent on the inherent curve angle in the azimuth and elevation planes. 

\vspace{-1mm}
\subsection{Effective Plane Waves Area through Measurements}  \label{Section:Plane_wave_theory}
The radiation from all practical antennas is in the form of wavefronts. Furthermore, we can assign rays~(that are perpendicular to the wavefronts) to different regions of the wavefronts. These rays provide antenna gain at specific regions in space during propagation. Therefore, theoretically, the minimum area of a reflector required to steer the maximum energy from a source antenna is equal to the area spanned by the ray with maximum gain~($17$~dBi in our case). However, for small sized reflectors, it becomes extremely difficult to align the incoming ray~(with maximum gain) with the maximum gain region~($17$~dBi) of the receiving antenna. Therefore, we require a given size of the reflector for steering the energy towards the receiver conveniently. 

\begin{figure}[!t]
	\centering           
	\includegraphics[width=0.85\columnwidth]{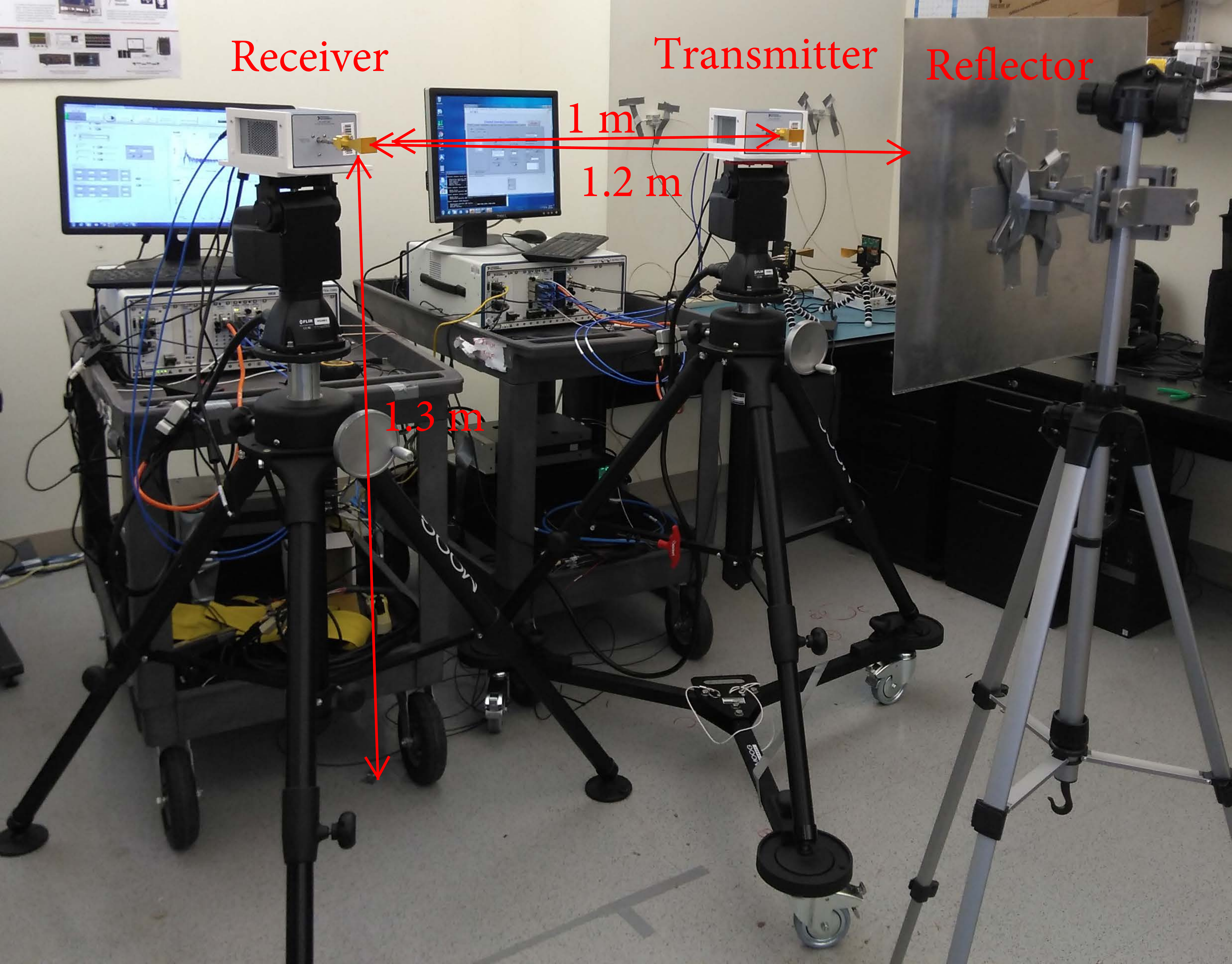}
	\caption{Verification setup using flat square metallic sheet reflectors of different sizes at $1.2$~m from the transmitter and receiver antennas. }\vspace{-4mm} \label{Fig:Verification_setup}
\end{figure}

In our measurements, we used horn antennas. The wavefront transmission from the horn antenna at a given distance in the far field can be approximated to plane waves. These plane waves have a power density given by the pointing vector $S$~\cite{planewaves}. This pointing vector $S$ spans a given area in space called \emph{effective plane wave area} represented as $A_{\rm pw}$. In the rest of the paper, we refer to the effective plane wave area as simply plane wave area. The area, $A_{\rm pw}$ is comparable to the area spanned by the half power antenna radiation beamwidths in the azimuth and elevation planes on a sphere at far field distance. The area, $A_{\rm pw}$ can provide us an estimate of the minimum area of a reflector required to steer the maximum energy conveniently from the transmitter to the receiver. In order to find $A_{\rm pw}$, we conducted a simple experiment in the lab. The setup is shown in Fig.~\ref{Fig:Verification_setup}. We used PXI transmitter, receiver setup, and metallic square sheet reflectors of sizes in the range from $0.12\times0.12$~m$^2$ to $0.30\times0.30$~m$^2$.  

\begin{figure}[!t]
\begin{center}
	\includegraphics[width=0.9\columnwidth]{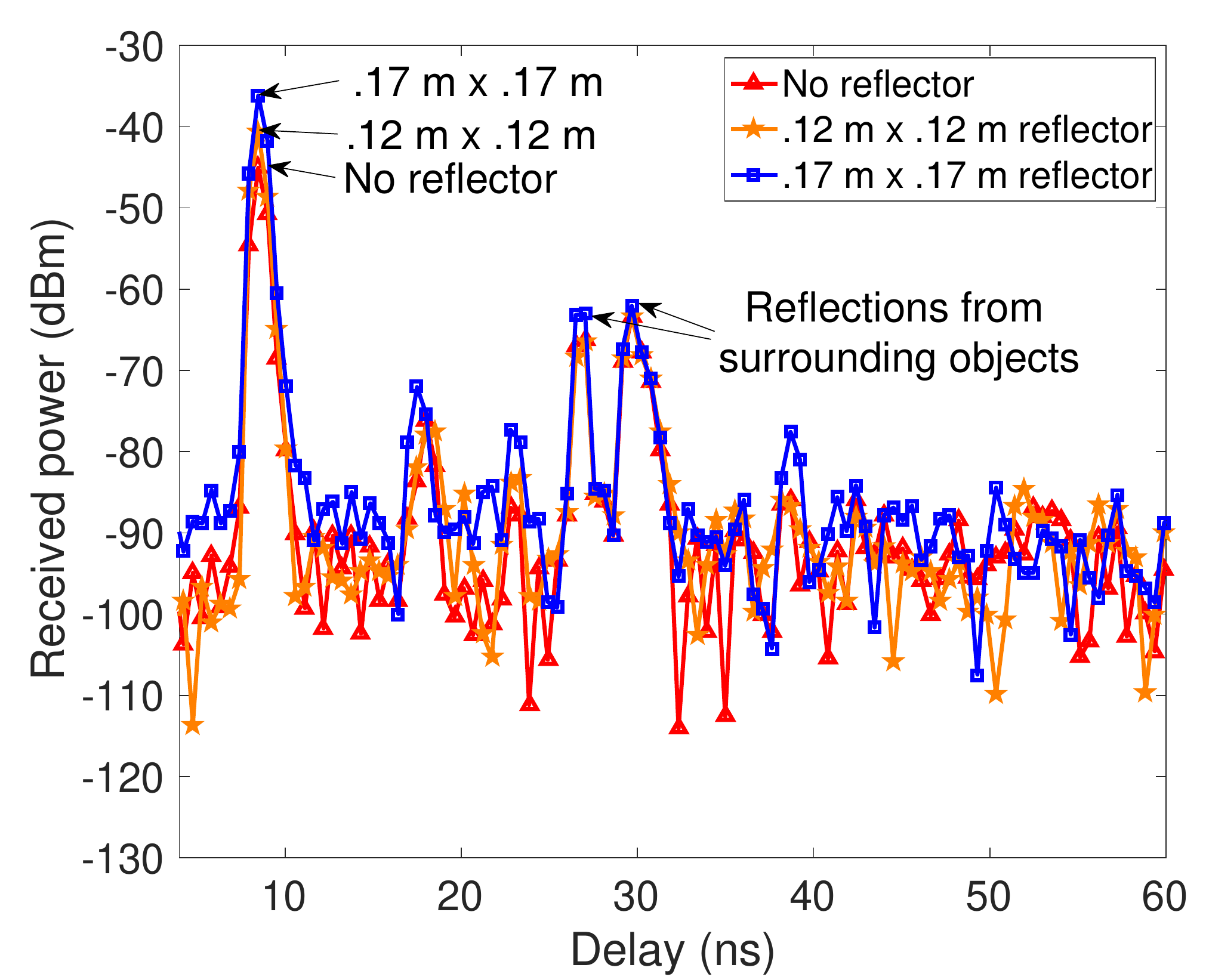}
	\caption{PDP showing received power~(from individual MPCs) for no reflector and flat square metallic sheet reflectors of sizes $0.12\times0.12$~m$^2$ and $0.17\times0.17$~m$^2$. This PDP corresponds to a distance of $2.4$~m~(two-way) from the transmitter to the receiver via the reflector. The time of flight for the reflector position is $8$~ns.} \vspace{-5mm} \label{Fig:Reflector_sizes_PDP}
\end{center}
\end{figure}

The transmit power was kept at $0$~dBm. The transmitter and the receiver antennas were placed at a distance of $1$~m from each other. The reflector was placed on a tripod such that it had equal distance from the transmitter and receiver antennas. The reflector was placed at $1.2$~m, and $3.6$~m distance from the transmitter/receiver antennas. The height of the transmitter and the receiver antennas were kept at $1.3$~m from the ground. The transmitter and the receiver antennas pointed to the center of the sheet reflectors. The tripod carrying reflector was rotated around its center in order to observe the maximum reflected power. This was performed in order to capture the maximum flux.~The time resolution of our PXI setup is $0.65$~ns per sample. Therefore, we can distinguish reflections from any two objects at a physical distance of $19.5$~cm. These reflections are observed as multipath components~(MPCs) in the power delay profile~(PDP). The PDP when the reflector is placed at $1.2$~m from the transmitter/receiver antennas is shown in Fig.~\ref{Fig:Reflector_sizes_PDP}. The reflections from the reflector were observed at $8$~ns~($2.4$~m; two-way distance) for different sizes of reflectors. %In addition, to reflection from the reflector at $8$~ns, there are reflections from the nearby objects at different time delays noticeably from wall, and metallic desk. }

\begin{figure}[!t]
\begin{center}          
	\includegraphics[width=0.9\columnwidth]{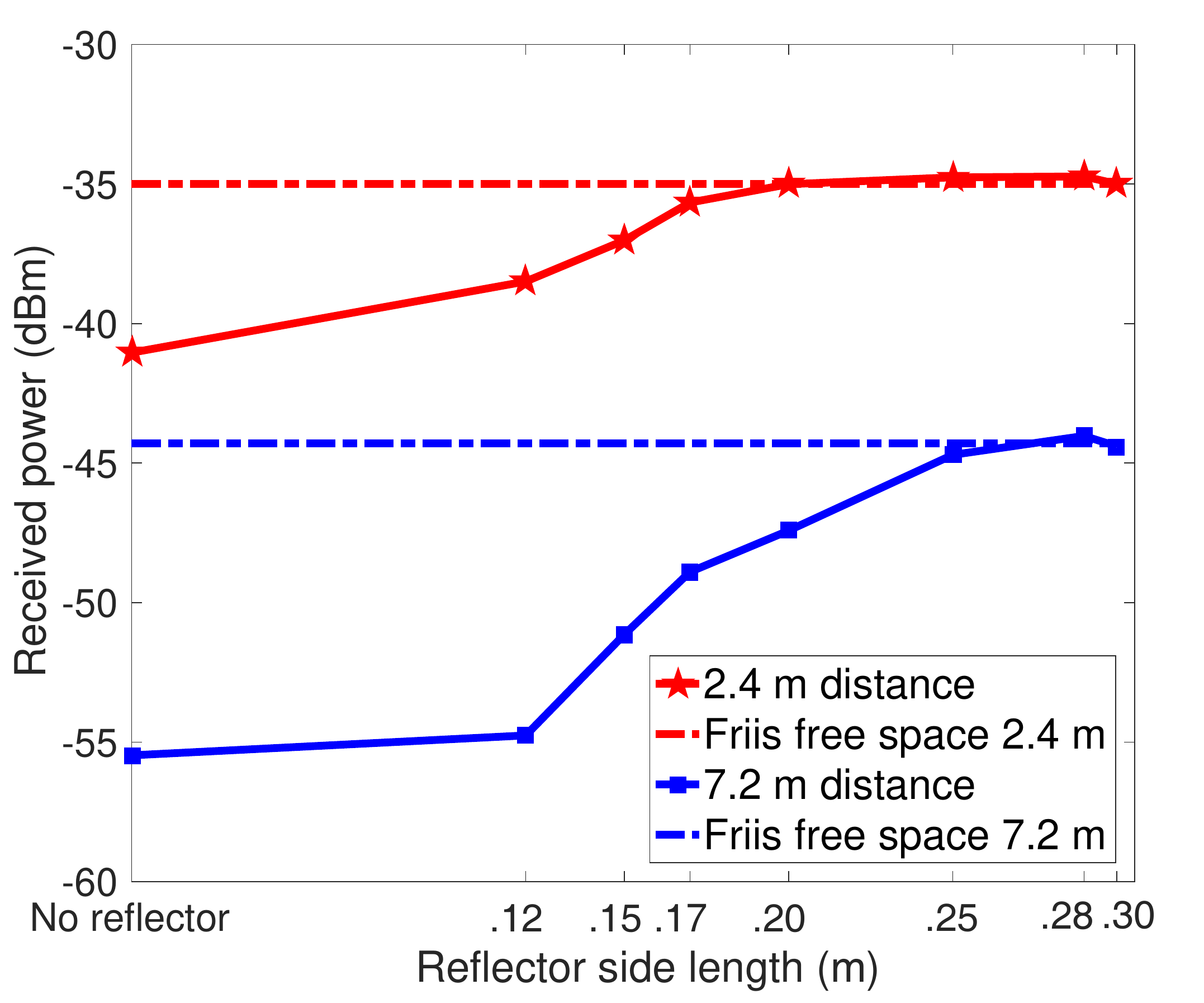}
	\caption{Maximum received power due to flat square metallic reflectors of different sizes at two distances of the transmitter to the receiver via the reflector. Friis free space received power is also plotted at given distances for comparison. } \vspace{-4mm} \label{Fig:Reflector_sizes}
\end{center}
\end{figure}

The maximum reflected received powers from different sized metallic sheet reflectors are shown in Fig.~\ref{Fig:Reflector_sizes}. These reflected powers are obtained from the PDP shown in Fig.~\ref{Fig:Reflector_sizes_PDP}. The first received power in Fig.~\ref{Fig:Reflector_sizes} for no reflector is due to reflections from the tripod body. Increasing the size of the reflector results in an increase of the received power. However, it becomes constant after a given reflector size. This received power approaches to the free space LOS power~(given by Friis equation) at the given distance. Moreover, this reflector size corresponds to approximately the size of the plane waves at that distance from the transmitter antenna. However, after $7.2$~m~(two-way distance), there is no further increase in received power beyond $0.25\times0.25$~m$^2$. Therefore, this reflector area can be approximated to the area of the transmitted plane waves given by $A_{\rm pw}$.  

\begin{figure*}[!t]
		
	\begin{subfigure}{0.31\textwidth}
	\centering
    \includegraphics[width=\textwidth]{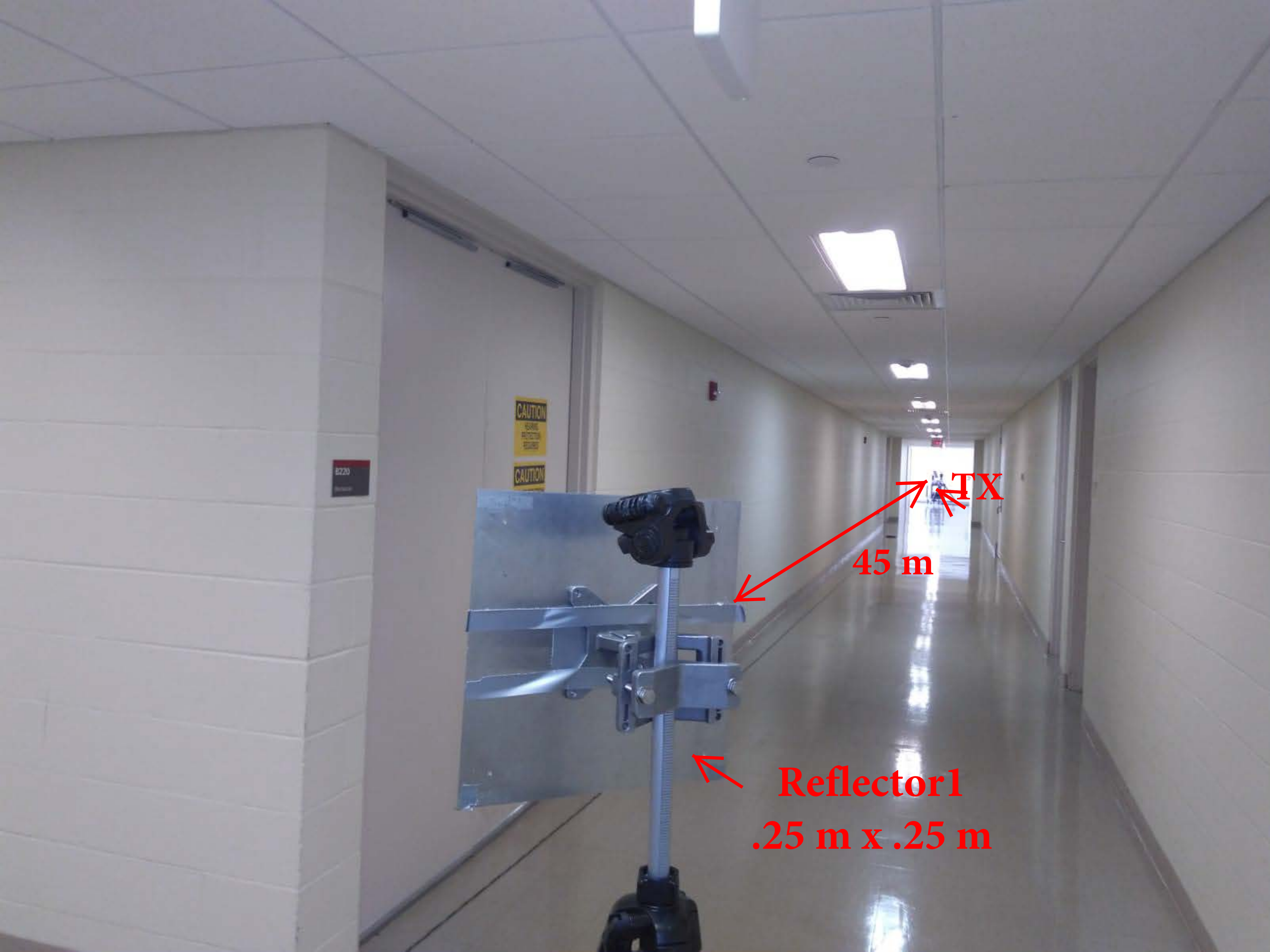}
	 \caption{}
     \end{subfigure}
     \begin{subfigure}{0.31\textwidth}
	\centering
    \includegraphics[width=\textwidth]{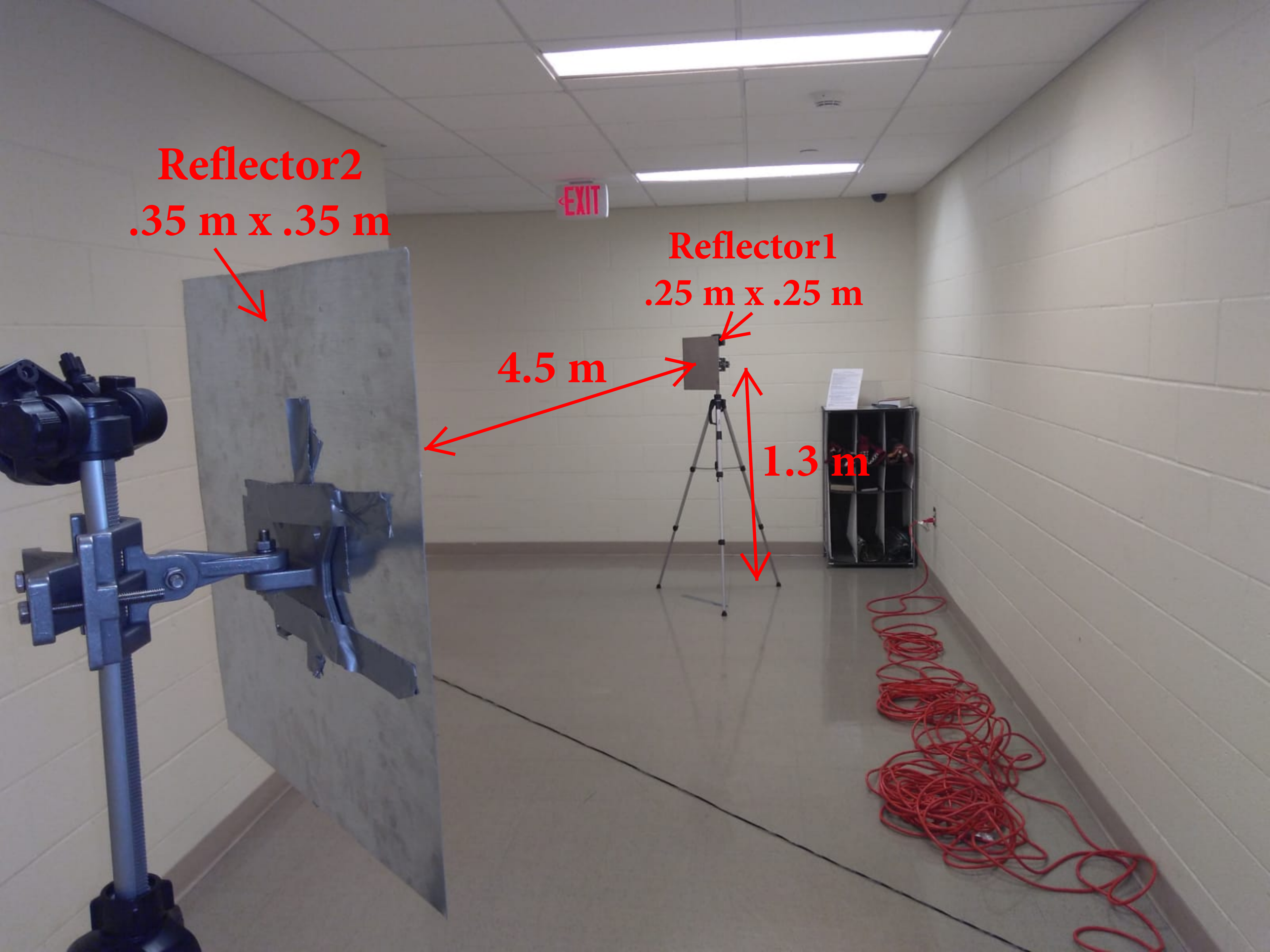}
	 \caption{}
     \end{subfigure}
     \begin{subfigure}{0.31\textwidth}
	\centering
    \includegraphics[width=\textwidth]{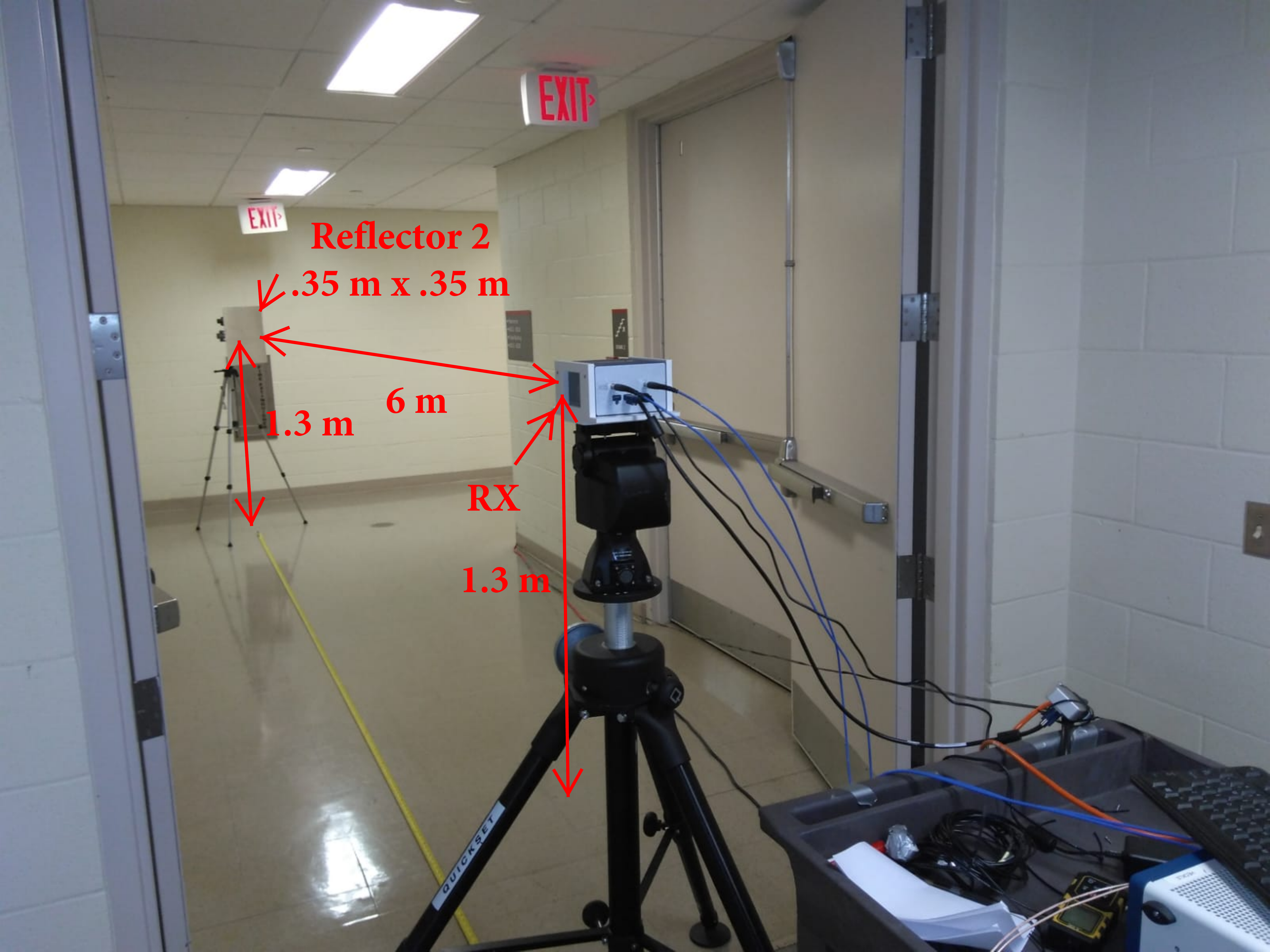}
	 \caption{}
     \end{subfigure}
      \vspace{-1mm}
    \caption{(a) A $0.25\times0.25$~m$^2$ flat square metallic sheet reflector placed at $45^\circ$~(azimuth) with respect to the boresight of the transmit antenna. The distance from the transmitter to the reflector is $45$~m. (b) A secondary reflector of size $0.35\times0.35$~m$^2$ placed at the azimuth plane as the first reflector, oriented at $45^\circ$ with respect to the receiver antenna's boresight. The two reflectors are apart by a distance of $4.5$~m. (c) A receiver placed at $6$~m distance from the secondary reflector.}\label{Fig:plane_basement}\vspace{-5mm}
\end{figure*}

In order to further verify the approximate size of the plane waves at large distances, we extended the same experiment to the corridor of a basement shown in Fig.~\ref{Fig:plane_basement}. The size of the first reflector was $0.25\times0.25$~m$^2$. The received power from this reflector was first measured at $49.5$~m distance from the transmitter. This was later the position of the second reflector, shown in Fig.~\ref{Fig:plane_basement}(b). The received power at this position was the same as the Friis free space power. The received power was also measured at another position, shown in Fig.~\ref{Fig:plane_basement}(c) at $55.5$~m distance from the transmitter. Here, the second reflector of size $0.35\times0.35$~m$^2$ was introduced at $49.5$~m distance from the transmitter. At this position also, the received power was the same as the Friis free space power. However, the size of the second reflector required was larger than the first reflector in order to get to the Friis free space power. This is mainly due to dispersion and possibly an enlargement of the reflected plane waves from the first reflector.  
\vspace{-1mm}
\vspace{-1mm}
\subsection{Effective Area of Different Shaped Reflectors}\label{Section:Effective_area}
\begin{figure*}
	\begin{subfigure}{0.3\textwidth}
	\centering
	\includegraphics[width=\textwidth]{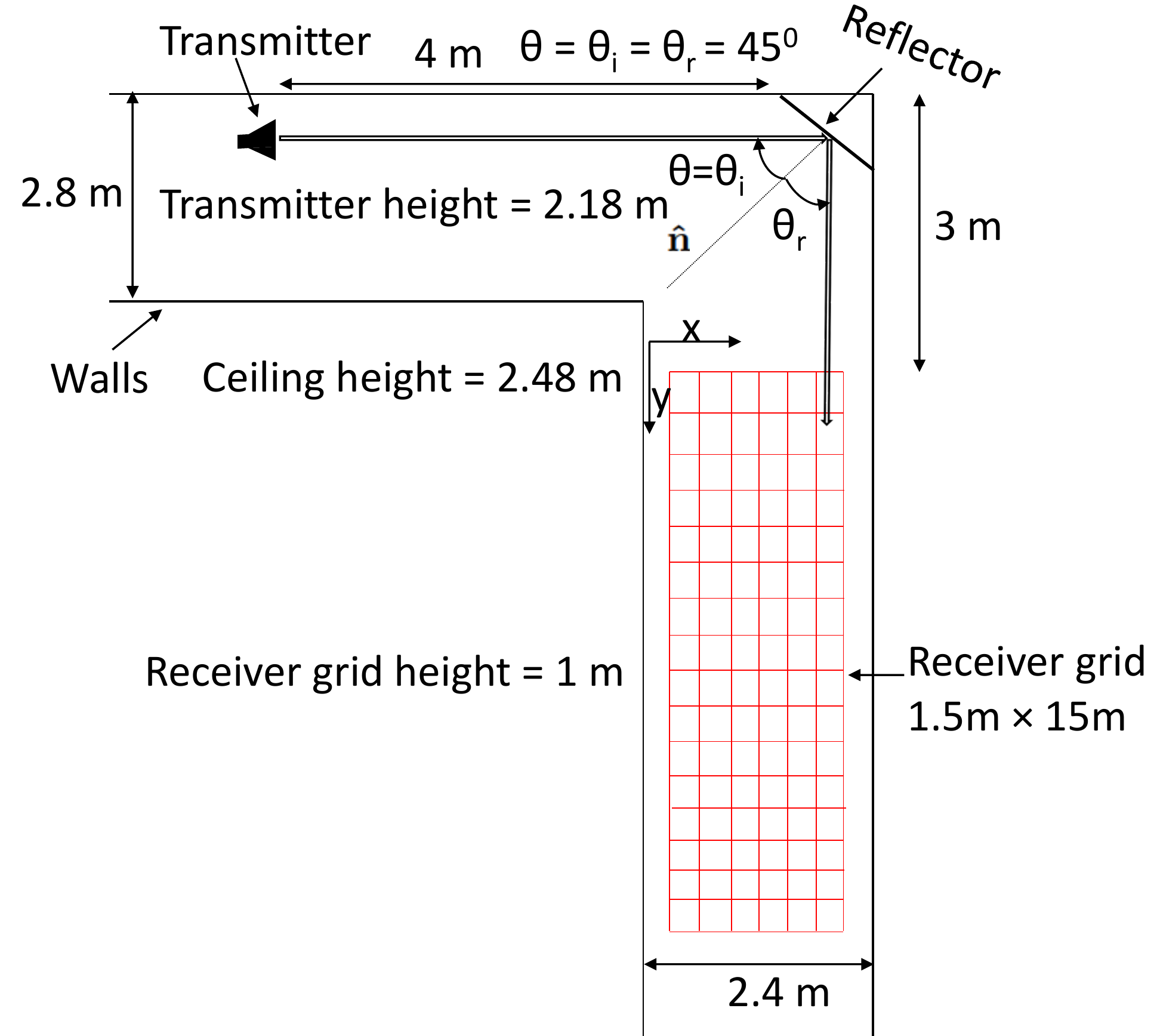}
	\caption{}
    \end{subfigure}	
	\begin{subfigure}{0.31\textwidth}
	\centering
	\includegraphics[width=\textwidth]{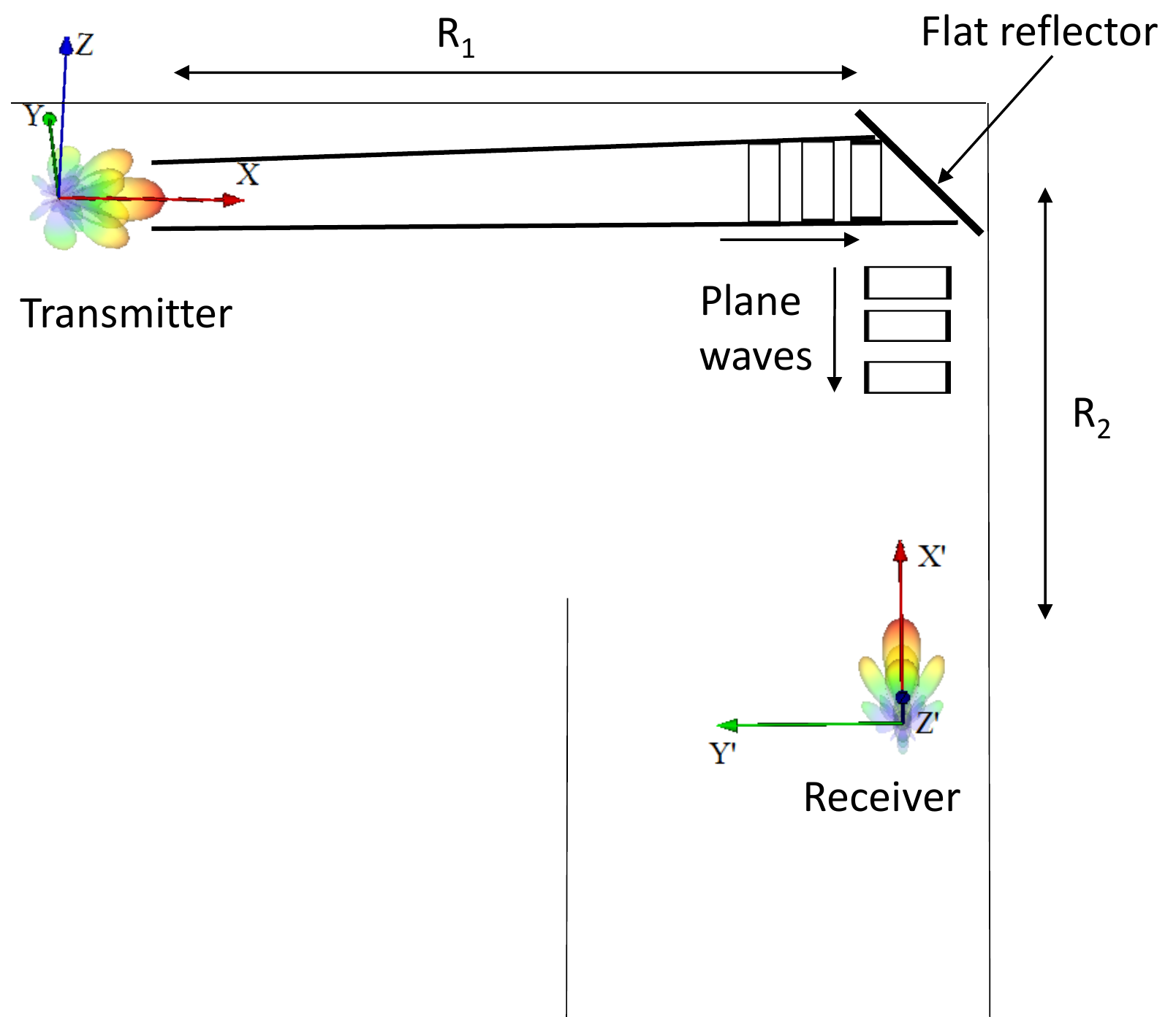}
	\caption{}
    \end{subfigure}			
	\begin{subfigure}{0.31\textwidth}
	\centering
    \includegraphics[width=\textwidth]{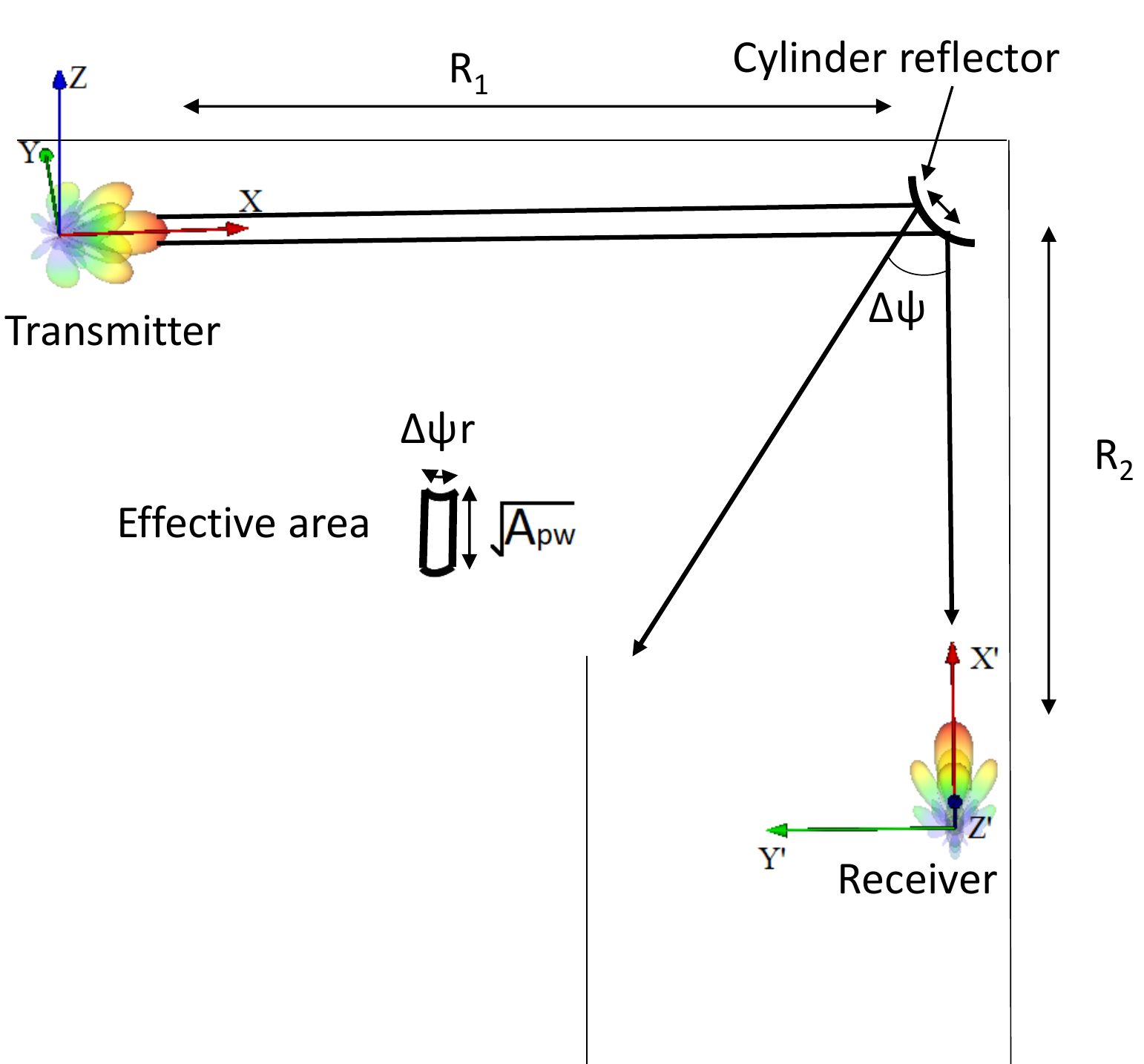}
	 \caption{}
     \end{subfigure}
     \begin{subfigure}{0.45\textwidth}
	\centering
    \includegraphics[width=\textwidth]{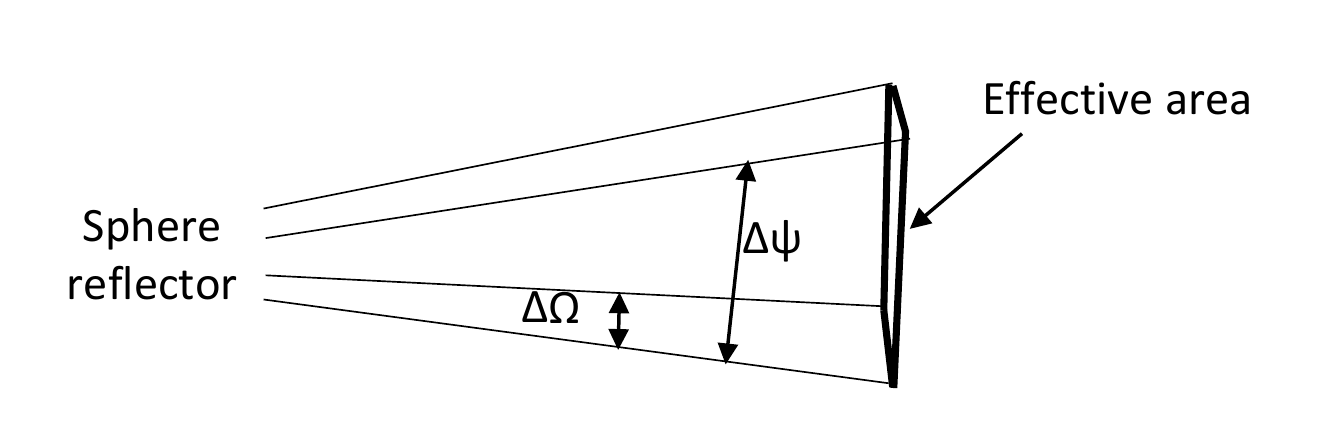}
	 \caption{}
     \end{subfigure}
     \begin{subfigure}{0.45\textwidth}
	\centering
    \includegraphics[width=\textwidth]{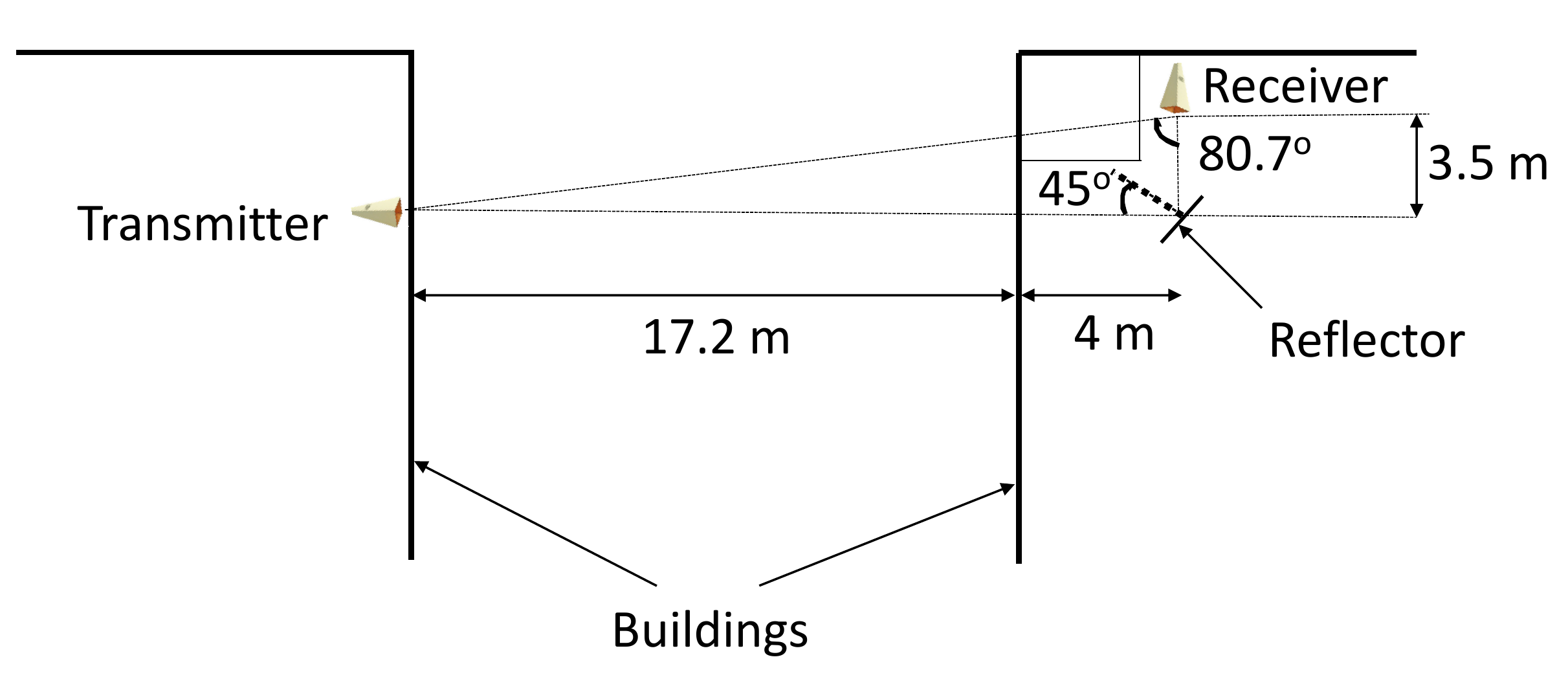}
	 \caption{}
     \end{subfigure}
       \caption{(a) Geometrical model in the azimuth plane for the indoor scenario with a reflecting surface deployed at the corner of a corridor, (b) top view (azimuth plane) for the transmitted and reflected plane waves from a flat reflector, (c) top view for transmission and reflection on the receiver grid corresponding to the effective area of the cylinder reflector, (d) side view of secondary transmission from the sphere reflector corresponding to its effective area, (e) top view of the layout of the outdoor measurement scenario.}\label{Fig:Core_area} \vspace{-4mm}
\end{figure*}

The effective reflector area is the area that captures and redirects the incoming plane waves towards the receiver. The effective area is, therefore, similar to the RCS of the reflector. For flat reflectors this effective area is $A_{\rm refl} = wh$, where $w$ and $h$ are the width and height of the reflector, respectively. For the cylindrical reflector, the effective area for our setup~(cylinder placed vertically) is a fraction of $2\pi rh$, where $r$ and $h$ represent the radius and height of the cylinder, respectively. This effective area corresponds to the width of the receiver grid shown in Fig.~\ref{Fig:Core_area}(c). This effective area is calculated based on the angle $\Delta \Psi$~(in radian) from the geometrical setup. The height of the cylinder is scaled to the height of the incident plane waves. The overall effective area is $A_{\rm refl} = \Delta \Psi r\sqrt{A_{\rm pw}}$. 

The reflected energy from the sphere is scaled in both azimuth and elevation planes. Therefore, the effective area is obtained based on the solid angle from the reflector towards the receiver grid. The solid angle is approximated as the product of the angular widths in the azimuth and elevation planes denoted as $\Delta \Psi$, $\Delta \Omega$, respectively, shown in Fig.~\ref{Fig:Core_area}(d). The width in the azimuth plane corresponds to the width of the receiver grid. The width in the elevation plane corresponds to a small region around the receiver antenna. Therefore, the effective area for our setup is $A_{\rm refl} = \Delta \Psi \Delta \Omega R_{\rm m}^2$, where $R_{\rm m}$ is the minimum distance of the receiver antenna from reflector.

\vspace{-1mm}
\section{Modeling End-to-End NLOS Received Power} % with Reflectors}  
\label{Section:Power_distribution_modeling}
In this section, we will provide an analytical model for NLOS end-to-end received power using reflectors of different sizes and shapes. 
\vspace{-1.5mm}
\subsection{Total Received Power} \label{Section:Total_received_power}
In this subsection, we provide the total received power modeling for NLOS end-to-end propagation in the presence of reflectors. The NLOS area is such that there is a direct path between the transmitter and the receiver through the reflector. Fig.~\ref{Fig:Core_area} shows typical indoor and outdoor NLOS propagation scenarios with a reflector placed at specific positions. In our modeling, we consider the reflector as a secondary transmission source. Therefore, a major contribution of the received power comes from first order reflections. In addition to first order reflections, second-order reflections are also significant for large-sized primary reflectors. Moreover, there are also other sources in the surroundings that contribute to the total received power. Taking all these into account, the total NLOS received power $P$ can be represented as:
\vspace{-1mm}
\begin{align}
P =  P_{\rm refl}^{(1)} + P_{\rm refl}^{(2)} + P_{\rm olos} + P_{\rm s},  \vspace{-1mm} \label{Eq:Sum_power} 
\end{align}
where $P_{\rm refl}^{(1)}$ and $P_{\rm refl}^{(2)}$ are the received powers due to first and  second-order reflections from the reflectors, respectively, $P_{\rm olos}$ is the power from the obstructed LOS (OLOS) path and $P_{\rm s}$ is the received power from other surrounding objects. We have considered the contribution of only two dominant reflectors to the received power, however, it can be extended to any number of reflectors. Moreover, the received power from $P_{\rm olos}$ is independent of the reflector/s and not necessarily exist for every NLOS scenario, whereas, $P_{\rm s}$ is generally always present and can change with the position or size of the reflector. 

\subsection{First Order Reflected Power Modeling} \label{Section:First_order_modeling}
The received power due to first order reflections from the reflector contributes mainly to the total received power. In addition, for flat reflectors, the orientation of the reflector with respect to transmitter and receiver antennas is important. However, for curved reflectors, the orientation is less significant. The received power due to first order reflections from the reflector is calculated based on the transmit power density. The transmitted power density at the reflector is denoted as $p_{\rm refl}(R_1)$, at distance $R_1$ from transmitter~\cite{balanis}, and given as:
\begin{equation}
   p_{\rm refl}(R_1) = \frac{P_{\rm tx}G_{\rm tx}(\theta_{\rm tx},\phi_{\rm tx})}{4\pi R_1^2}, 
\end{equation}
where $P_{\rm tx}$ and $G_{\rm tx}(\theta_{\rm tx},\phi_{\rm tx})$ are the transmitted isotropic power and gain~(directivity) of the transmit antenna at respective azimuth and elevation angles of $\theta_{\rm tx}$ and $\phi_{\rm tx}$. This power density is received by the reflector with a given effective area $A_{\rm refl}$~(see Section~\ref{Section:Effective_area}). The effective area $A_{\rm refl}$, that captures and redirects the incoming power density towards the receiver is also known as the RCS, denoted by $\sigma$. In other words, the reflector acts as a secondary source of transmission towards the receiver. This secondary transmission can be explained by Huygens' principle~\cite{huygens} shown in Fig.~\ref{Fig:wavefronts}. 

The captured and redirected power density from the reflector is denoted as $p_{\rm refl}(R_1,R_2)$, where $R_2$ is the distance of the reflector to another reflector or the receiver. The redirected power density $p_{\rm refl}(R_1,R_2)$ is given by:
\begin{equation}
   p_{\rm refl}(R_1,R_2) = \frac{P_{\rm tx}G_{\rm tx}(\theta_{\rm tx},\phi_{\rm tx})}{(4\pi R_1R_2)^2}\sigma\Gamma, 
\end{equation}
where $\Gamma$ is the reflection efficiency of the first reflector material. If the reflector is a conductor with a polished surface, then, $\Gamma=1$. However, for dielectrics or lossy reflector materials, $\Gamma<1$. The plane waves after reflection from the reflector travels towards the receiver shown in Fig.~\ref{Fig:Core_area}(b). If $A_{\rm rx}$ is the receiver antenna's aperture, then the received power captured at the receiver antenna, $P_{\rm r}^{(1)}(R_1,R_2)$ is given by:
%\textcolor{blue}{Ismail: Align equal signs, not sure how many times I commented on this before! Include and sign right before two equation signs before. Check all paper. This paragraph is also almost one page long if you check it, divide into multiple paragraphs!}
\begin{align}
   P_{\rm refl}^{(1)}(R_1,R_2) = &~p_{\rm refl}(R_1,R_2) A_{\rm rx}, \\
P_{\rm refl}^{(1)}(R_1,R_2) = &~p_{\rm refl}(R_1,R_2)G_{\rm rx}(\theta_{\rm rx},\phi_{\rm rx})\frac{\lambda^2}{4\pi},  \label{Eq:RX_2}
\end{align}
where $G_{\rm rx}(\theta_{\rm rx},\phi_{\rm rx})$ is the gain of the receiver antenna at respective azimuth and elevation angles. The reflected received power is also dependent on the relative orientation of the transmitter, receiver antennas and the reflector~(\emph{for flat reflectors}). There are two orientations that need to be considered for reflected power from the transmitter to the receiver. The first is the physical orientation between the transmitter, receiver antenna's boresight and the reflector's surface normal. 

In order to get maximum received power, the angles between the transmitter and receiver antenna's boresight with the reflector's surface normal should be the same. Let $\theta_{\rm i}$ and $\theta_{\rm r}$ represent the angles formed between the unit vectors of the boresight of the transmitter and receiver antennas, with the unit normal of the reflector, respectively, in the azimuth plane. This is shown in Fig.~\ref{Fig:Core_area}(a). Similarly, $\phi_{\rm i}$, $\phi_{\rm r}$ represent the corresponding angles in the elevation plane. In our setup~(both indoor and outdoor) the transmitter antenna and reflector are fixed, therefore, $\theta_{\rm i}$ and $\phi_{\rm i}$ are constant. Let us represent, $\Delta \theta = |\theta_{\rm opt} - \theta_{\rm r}|$, where $\theta_{\rm opt}$ is the optimum reflection angle corresponding to the transmit antenna's boresight and surface normal of the reflector. For example, for our indoor setup in Fig.~\ref{Fig:Core_area}(a), $\theta_{\rm opt} = 45^\circ$. The change in $\Delta \theta$ due to shifting from optimum angle results in an exponential decrease in the received power given by $\alpha_{\rm f}^{\Delta \theta}$, where $\alpha_{\rm f}<1$. The constant $\alpha_{\rm f}$ incorporates the decrease due to shift from the maximum reflected power region and decrease due to receiver antenna's gain compared to the boresight. Similarly, the decrease in the elevation plane can be represented by $\alpha_{\rm f}^{\Delta \phi}$.

The second orientation is between the reflected plane waves from the reflector to the receiver. This orientation is dependent on the surface characteristics of a given reflector generating the reflected plane waves. For example, the surface normal of the reflected plane waves from a small sized reflector may not be in the same direction as generated by a large-sized reflector, even though both are flat shaped reflectors. This is mainly due to the surface currents generated by the incident plane waves over the surface of the reflector. 

 %Additionally, due to angular rotation of the reflector shown in Fig.~\ref{Fig:Core_area}(a) in the azimuth plane, we have a constant factor given by $\cos\theta_{\rm i}$ for flat reflectors. This arises from the dot product between the pointing vector $\Vec{S}$ and the surface normal of the reflector. This factor is constant as the boresight of the transmit antenna and reflector surface normal remains fixed.

The second orientation takes into account the angle between the unit normal vector of the reflected plane waves and the unit normal vector of the boresight of the receiver antenna. Let $\uvec{n}_{\rm rp}(r_{\rm rp},\theta_{\rm rp},\phi_{\rm rp}), \uvec{n}_{\rm rx}(r_{\rm rx},\theta_{\rm rx},\phi_{\rm rx})$ represents the unit normal vectors of the reflected plane wave surface and the unit vector of the boresight of the receiver antenna, respectively. In order to express the reduction in the received power due to orientation mismatch between the reflected plane waves and the receiver antenna's boresight we define a term, $\beta^{1-|\uvec{n}_{\rm rp}\cdot\uvec{n}_{\rm rx}|^2}$, where $\beta<1$. Now, the received power from (\ref{Eq:RX_2}) is given as:  
\begin{align}
P_{\rm refl}^{(1)}(R_1,R_2) &= \frac{p_{\rm refl}(R_1,R_2)G_{\rm rx}(\theta_{\rm rx},\phi_{\rm rx})\lambda{^2}}{4\pi} \nonumber \\
&\times \alpha_{\rm f}^{\Delta \theta} \alpha_{\rm f}^{\Delta \phi}\beta^{1-|\uvec{n}_{\rm rp}.\uvec{n}_{\rm rx}|^2} \Gamma, \vspace{-1mm} \label{Eq:RX_3}
\end{align}
%|n_{\rm {c,refl}}.n_{\rm RX}|^2 \eta_{\rm T}\eta_{\rm R}|\rho_{\rm TX}\cdot\rho_{\rm sc}|^2, 
if the orientations are perfect i.e. $\theta_{\rm i} = \theta_{\rm r} = \theta_{\rm opt}$, $\phi_{\rm i}= \phi_{\rm r} = \phi_{\rm opt}$, $\uvec{n}_{\rm rp}.\uvec{n}_{\rm rx}=1$ and the reflector is perfect polished conductor with $\Gamma=1$. Moreover, if the RCS of the reflector is equal or larger than the area of the incident plane waves, denoted by $\sigma'$, then, the reflected received power in (\ref{Eq:RX_3}) approaches to Friis free space received power. In other words, all the transmitted power is captured by the reflector and redirected towards the receiver at a given angle. Therefore, we can equate (\ref{Eq:RX_3}) and Friis free space equation~\cite{Friis} equal to each other as follows:
\begin{align}
&\frac{P_{\rm tx}G_{\rm tx}(\theta_{\rm tx},\phi_{\rm tx})G_{\rm rx}(\theta_{\rm rx},\phi_{\rm rx})\lambda^2 \sigma'}{4\pi(4\pi R_1R_2)^2} \nonumber =\\ & \frac{P_{\rm tx}G_{\rm tx}(\theta_{\rm tx},\phi_{\rm tx})G_{\rm rx}(\theta_{\rm rx},\phi_{\rm rx})\lambda^2}{4\pi (R_1 + R_2)^2}, \vspace{-1mm} \label{Eq:Power_Eq1}
\end{align}
where the area of the incident plane waves is given by
\vspace{-1mm}
\begin{equation}
\sigma' = \frac{4\pi(R_1R_2)^2}{(R_1+R_2)^2}. \vspace{-1mm} \label{Eq:Power_Eq2}
\end{equation}

Let $A_{\rm refl}$ represent the effective area of the reflector. Additionally, considering the polarization mismatch losses between the transmitted and reflected plane waves from the reflector, we can write (\ref{Eq:RX_3}) and (\ref{Eq:Power_Eq2}) as follows:  
\begin{align}
    P_{\rm R}^{(1)}(R_1,R_2) &= %\nonumber \\ 
    %&
    \frac{P_{\rm tx}G_{\rm tx}(\theta_{\rm tx},\phi_{\rm tx})G_{\rm rx}(\theta_{\rm rx},\phi_{\rm rx})\lambda^2 \sigma'}{4\pi(4\pi R_1R_2)^2} \nonumber \\
    &\times\alpha_{\rm f}^{\Delta \theta} \alpha_{\rm f}^{\Delta \phi} 
    \beta^{1-|\uvec{n}_{\rm rp}\cdot\uvec{n}_{\rm rx}|^2} \nonumber \\
    &\times \frac{\rm{min}\big( A_{\rm pw}, A_{\rm refl}\big)}{A_{\rm pw}}\Gamma|\rho_{\rm TX}\cdot\rho_{\rm refl}|^2, \vspace{-1mm} \label{Eq:PowerEq3}
\end{align}
where $\rho_{\rm TX}$ and $\rho_{\rm refl}$ represents the polarization unit vectors at the transmitter and at the reflector~(after reflection), respectively, and $|\rho_{\rm TX}\cdot\rho_{\rm refl}|^2$ represents the polarization mismatch loss. From (\ref{Eq:PowerEq3}), if there are no orientation losses, and size of the reflector is equal or larger than the size of the plane waves, the received power approaches to Friis free space received power. However, the rate of power decay~(generally given by path loss exponent) can change depending on the environment.

For cylinder and sphere reflectors, the orientation losses are considered to be negligible. Because the cylinder or sphere provides spreading of the incoming energy equally in respective directions. For cylinder and sphere reflectors, the decrease in the received power on the grid is mainly due to two factors. One is due to the decrease in the gain of the receiver antenna as we move away from the boresight. The second is due to the distance on the grid. Similar to flat reflectors, the decrease in the received power due to shift from the receiver antenna's boresight is represented by $\alpha_{\rm c}^{\Delta \theta}$ and $\alpha_{\rm c}^{\Delta \phi}$ in the azimuth and elevation planes. However, $\alpha_{\rm c}>\alpha_{\rm f}$. This is because we do not have additional decrease due to shift from a maximum reflected power region as observed for flat reflectors. In addition, the reflected power from cylinder and sphere is distributed over the indoor receiver grid based on their effective areas given in~Section~\ref{Section:Effective_area}. Therefore, the received power for cylinder and sphere reflectors is given as:
\begin{align}
    P_{\rm R}^{(1)}(R_1,R_2) &= \frac{P_{\rm tx}G_{\rm tx}(\theta_{\rm tx},\phi_{\rm tx})G_{\rm rx}(\theta_{\rm rx},\phi_{\rm rx})}{4\pi(4\pi R_1R_2)^2}\nonumber\\
    &\times\lambda^2 \sigma' \alpha_{\rm c}^{\Delta \theta} \alpha_{\rm c}^{\Delta \phi}\frac{\rm{min}\big( A_{\rm pw}, A_{\rm refl}\big)}{A_{\rm pw}}\Gamma. \vspace{-2mm} \label{Eq:PowerEq4}
\end{align}
\vspace{-2mm}
\subsection{Second Order Reflected and OLOS Power Modeling}
The received power due to second-order reflections from an additional reflector can be obtained in a similar way as first order reflections. If $R_1$ represents the distance from the transmitter to the first reflector and $R_2$ represent the distance from the first reflector to the second reflector, and $R_3$ represents the distance from the second reflector to the receiver antenna, the received power is given as: 
\vspace{-1mm}
\begin{align}
    P_{\rm R}^{(2)}(R_1,R_2,R_3) &= \frac{P_{\rm tx}G_{\rm tx}(\theta_{\rm tx},\phi_{\rm tx})G_{\rm rx}(\theta_{\rm rx},\phi_{\rm rx})\lambda^2 \sigma''}{(4\pi)^4 (R_1R_2R_3)^2} \nonumber \\
    &\times \alpha_{\rm f}^{\Delta \theta^{(1)}} \alpha_{\rm f}^{\Delta \phi^{(1)}} \alpha_{\rm f}^{\Delta \theta^{(2)}}  \alpha_{\rm f}^{\Delta \phi^{(2)}}\nonumber\\ &\times\beta^{1-\big|\uvec{n}_{\rm rp}\cdot\uvec{n}_{\rm refl}\big|^2} \beta^{1-\big|\uvec{n}_{\rm refl}\cdot\uvec{n}_{\rm rx}\big|^2}\nonumber \\
    &\times \frac{\rm{min}\big( A_{\rm pw}^{(1)}, A_{\rm refl}^{(1)}\big)}{A_{\rm pw}^{(1)}}  \frac{\rm{min}\big( A_{\rm pw}^{(2)}, A_{\rm refl}^{(2)}\big)}{A_{\rm pw}^{(2)}} \nonumber \\
    &\times \Gamma_1\Gamma_2 |\rho_{\rm TX}\cdot\rho_{\rm refl}|^2 |\rho_{\rm refl}\cdot\rho_{\rm refl}^{(2)}|^2, \label{Eq:PowerEq5}
\end{align}
where we have 
\begin{equation}
\sigma'' = \frac{(4\pi R_1R_2R_3)^2}{(R_1 + R_2 + R_3)^2},\vspace{-1mm}
\end{equation} 
while $\Delta \theta^{(1)}$ and $\Delta \theta^{(2)}$ represent the absolute difference between the optimum reflection angle~(the expected reflection angle) and the current receiver angle~(due to its position) in the azimuth plane for first and second reflectors, respectively. Similarly, $\Delta \phi^{(1)}$, and $\Delta \phi^{(2)}$ represent the absolute difference of angles in the elevation plane for first and second reflectors, respectively. Moreover, $\uvec{n}_{\rm refl}$ is the unit vector for the surface normal of the plane waves reflected from the second reflector, $A_{\rm pw}^{(1)}$ and $A_{\rm pw}^{(2)}$ are the areas of the plane waves incident on the first and second reflectors, respectively, $A_{\rm refl}^{(1)}$ and $A_{\rm refl}^{(2)}$ represent the effective areas of the first and second reflectors, respectively. The reflection efficiency of the first and second reflectors are represented by $\Gamma_1$ and $\Gamma_2$, respectively, whereas, $\rho_{\rm refl}^{(2)}$ represents the polarization vector at the second reflector~(after reflection).
\begin{figure}[!t]
\centering
	\begin{subfigure}{.49\textwidth}
	\centering
	\includegraphics[width=\textwidth]{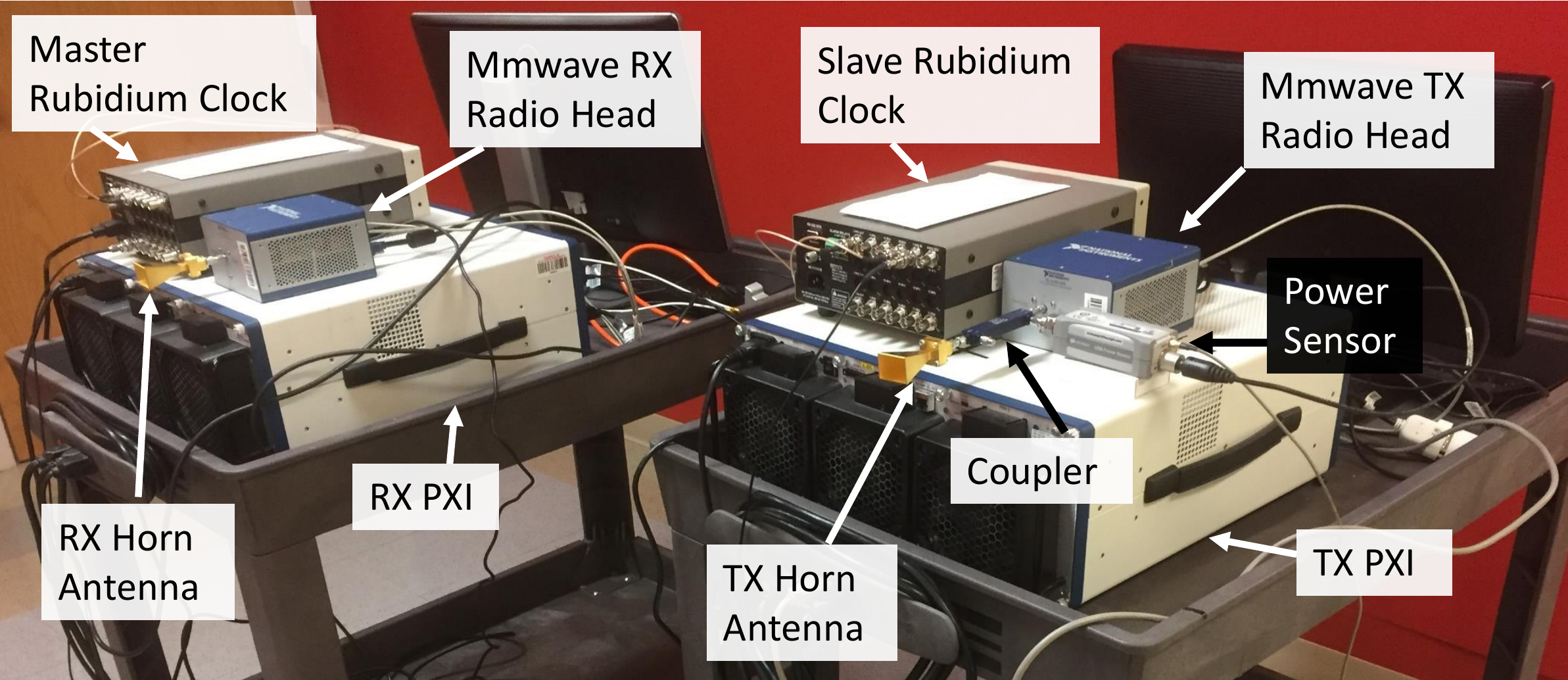}
	\caption{\smallskip}
    \end{subfigure}
    
	\begin{subfigure}{0.2\textwidth}
	\centering
    \includegraphics[width=\textwidth]{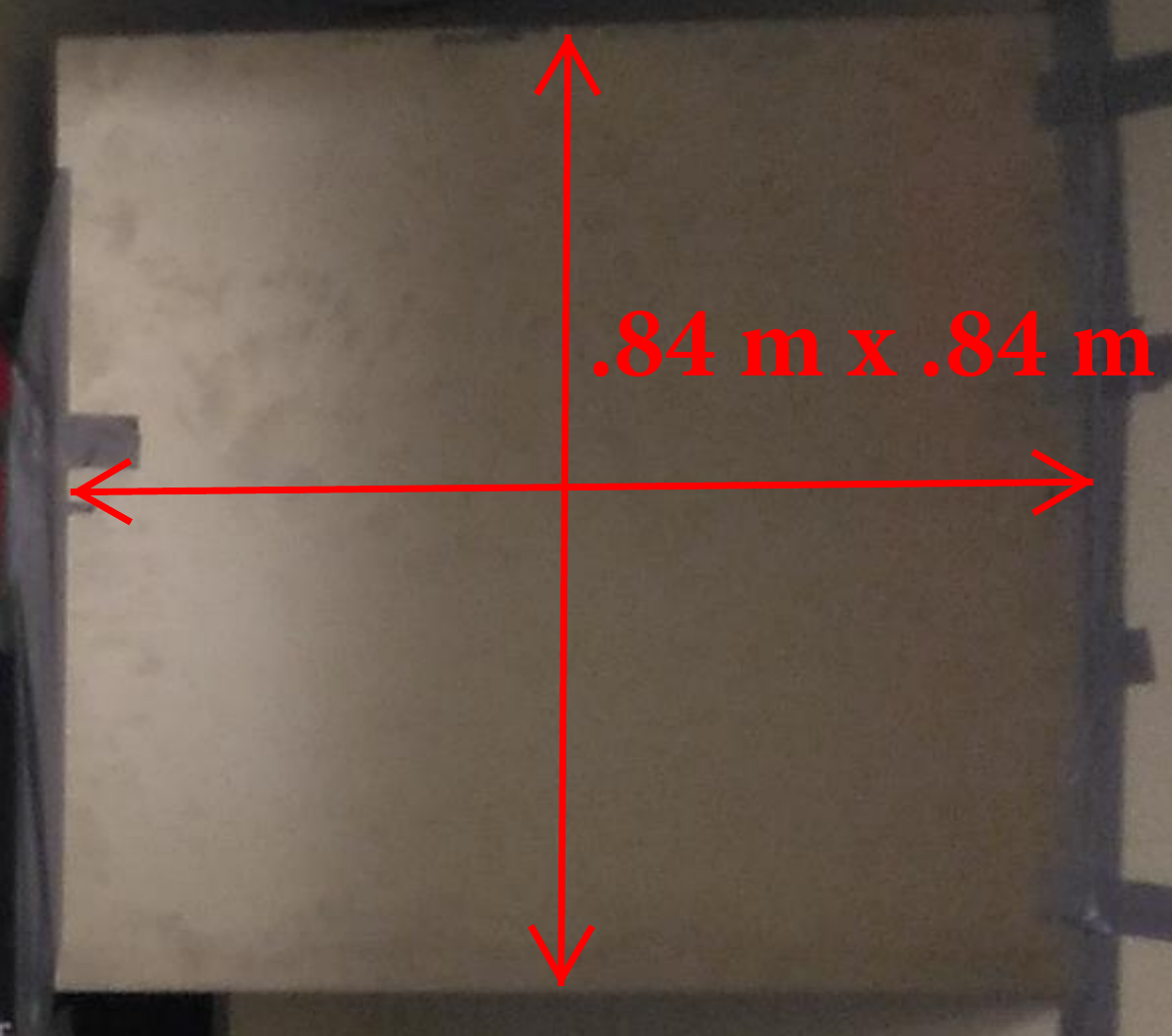}
	 \caption{} 
     \end{subfigure}
     \begin{subfigure}{0.08\textwidth}
	\centering
    \includegraphics[width=\textwidth]{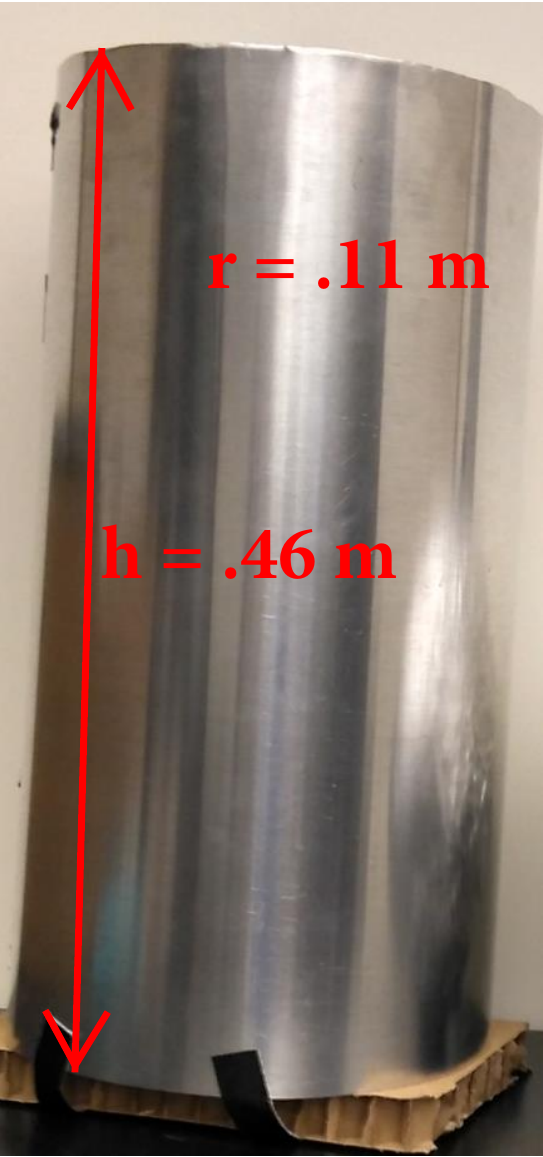}
	 \caption{}
     \end{subfigure}
     \begin{subfigure}{0.18\textwidth}
	\centering
    \includegraphics[width=\textwidth]{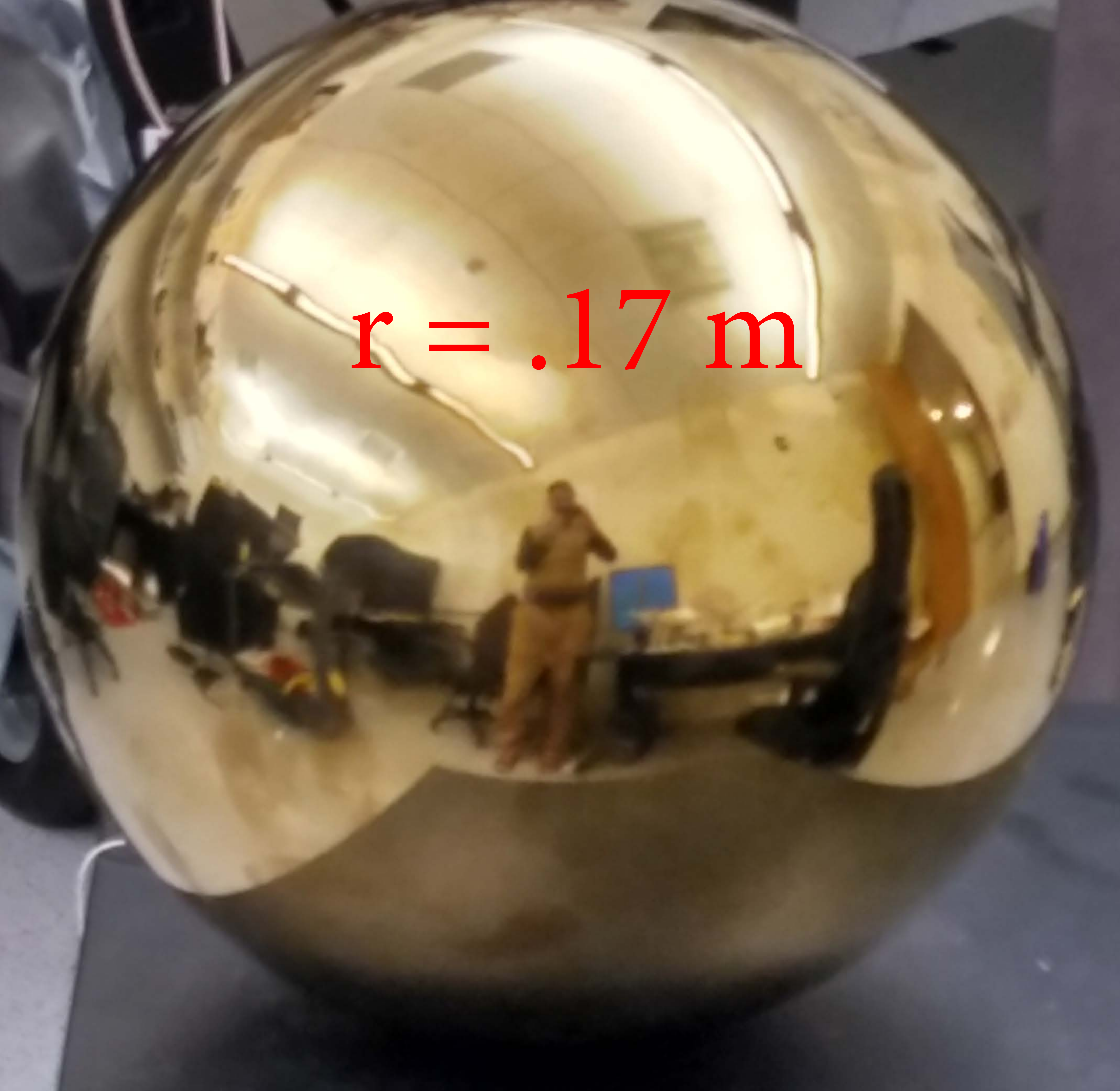}
	 \caption{}
     \end{subfigure}
     \vspace{-1mm}
    \caption{(a) Channel sounder setup, (b) $0.84\times0.84$~m$^2$ flat reflector, (c) cylinder reflector, (d) sphere reflector. } \vspace{-4mm} \label{Fig:setup_reflectors} 
\end{figure}

Apart from the received power due to dominant reflectors, we can have received power from the OLOS also if one exists. Let us consider that a significant OLOS path is present due to obstruction with the propagation loss coefficient, $\eta<1$. The coefficient $\eta$ incorporates the obstruction losses. Then, the OLOS received power at a distance of $R$ between the transmitter and the receiver is given as:\vspace{-1mm}
\begin{align}
    P_{\rm{olos}}(R) = \frac{P_{\rm tx}G_{\rm tx}(\theta_{\rm tx}^{\rm{}},\phi_{\rm tx}^{\rm{}})G_{\rm rx}(\theta_{\rm rx}^{\rm{}},\phi_{\rm rx}^{\rm{}})\lambda^2 \eta^{\rm{}}}{(4\pi R)^2}.  \label{Eq:OLOS}
\end{align}

The remaining constituent of the total received power $P_{\rm s}$ incorporates mostly received power from weaker or higher order reflections from surrounding objects. It can also incorporate the received power due to diffraction in NLOS scenarios depending on the operating frequency and propagation geometry. 

\vspace{-1mm}
\section{Propagation Measurements and Ray Tracing Simulations Setup}
Propagation measurements were performed in typical NLOS scenarios for indoor and outdoor. For indoor, reflectors of different shapes and sizes were placed at the edge of the corridor. The received power was measured over a receiver gird in the corridor. Ray tracing simulations were also carried out for this propagation setup. For outdoor, different sizes of flat metallic sheet reflectors were used for an NLOS area due to obstruction. The received power was measured at a single receiver position at different azimuth and elevation angles.
\vspace{-1mm}
\subsection{Indoor Measurement Setup}
Indoor measurements were carried out in the basement corridor of Engineering Building~II at North Carolina State University. The measurement and simulation environment for indoor and outdoor channel measurements are shown in Fig.~\ref{Fig:scenario}. The receiver is moved at different positions in the $(x,y)$ plane of the corridor to form a receiver grid shown in Fig.~\ref{Fig:Core_area}(a). The size of the $(x,y)$ receiver grid is $(1.5~\rm{m},15~\rm{m})$ such that each measurement block is $0.3~$m$\times0.3$~m. A similar geometry is generated using  the Remcom Wireless InSite RT software to compare with the measurement outcomes and will be explained in Section~\ref{Section:Ray_tracing_setup}.

The measurements were performed using NI mmWave transceiver system at $28$~GHz~\cite{NImmwave} as shown in Fig.~\ref{Fig:setup_reflectors}(a). The system consists of two PXI platforms: one transmitter and one receiver. There are two rubidium~(Rb) clocks used at the transmitter and the receiver sides that provide common $10$~MHz clock and pulse per second (PPS) signal. The output from the PXI intermediate frequency (IF) module is connected to the mmWave transmitter radio head that converts the IF to $28$~GHz. Similarly, at the receiver side, the mmWave radio head down converts $28$~GHz RF signal to IF.

The digital to analog converter at the transmitter and the analog to digital converter at the receiver has a sampling rate of $3.072$~GS/s. The channel sounder supports $1$~GHz and $2$~GHz modes of operation. The measurements for this paper are performed using the $2$~GHz mode where the sounding signal duration is $1.33$~$\mu$s, which is the maximum measurable excess delay. This mode provides a $0.65$~ns delay resolution in the delay domain, corresponding to $19.5$~cm distance resolution. The analog to digital converter has around $60$~dB dynamic range and this system can measure path loss up to $185$~dB. The transmit power for the experiment is set to $0$~dBm. A power sensor measures the power at the output of the mmWave transmitter front end using an RF coupler. The power sensor lets us convert measurements in dB units into dBm units.

In order to get accurate channel measurements, we need to characterize the non-flat frequency response of the measurement hardware itself, and subsequently do a calibration to compensate for the impulse response due to the hardware. For calibration purposes, a cable with fixed attenuators connects the transmitter to the receiver. Assuming the cable and the attenuators have a flat response, the channel response of the hardware is measured. During actual measurements, the hardware response is equalized assuming hardware response does not vary over time. After this equalization, we obtain the response of the actual over the air channel.

The antennas that are used at the transmitter and the receiver are linearly polarized pyramidal horn antennas \cite{Horn_antenna_sage}, having a gain of $17$~dBi and half-power beam-widths of $26$~ and $24$ degrees in the E and H planes, respectively. To improve the coverage area in NLOS receiver region in the corridors, we use aluminum flat sheet reflectors with different sizes, a cylinder, and, a sphere as shown in Fig.~\ref{Fig:setup_reflectors}(b), Fig.~\ref{Fig:setup_reflectors}(c), and Fig.~\ref{Fig:setup_reflectors}(d). These reflectors are placed at the corner of the walls facing the corridor as shown in~Fig.~\ref{Fig:scenario}(a). The aluminum sheet used is 5086-H32 having a thickness of $1.6$~mm. Three flat sheets with side lengths of $0.30$~m, $0.61$~m, and $0.84$~m, respectively, are used in the measurements. A metallic cylinder of radius $0.11$~m and height $0.46$~m is used, whereas a mirror ball covered with an aluminum sheet having a diameter of $0.34$~m is used. The surface areas of $0.61\times0.61$~m$^2$, flat reflector, cylinder, sphere have similar cross-sectional area. 

In order to place different flat reflectors on the same plane, a cardboard of size $0.84\times0.84$~m$^2$ is used as a reference as shown in Fig.~\ref{Fig:scenario}(a). The center of the cardboard is aligned to the center of the bore-sight axis of the antenna. Different sized reflectors are placed such that their centers are aligned to the center of the cardboard. Similarly, the bore-sight axis of the antenna is aligned to the center of the cylinder and sphere. There is no orientation of the reflectors in the vertical plane.
\vspace{-1mm}
\begin{figure}[!t]
    \centering
     \begin{subfigure}{0.18\textwidth}
	\centering
     \includegraphics[width=\textwidth]{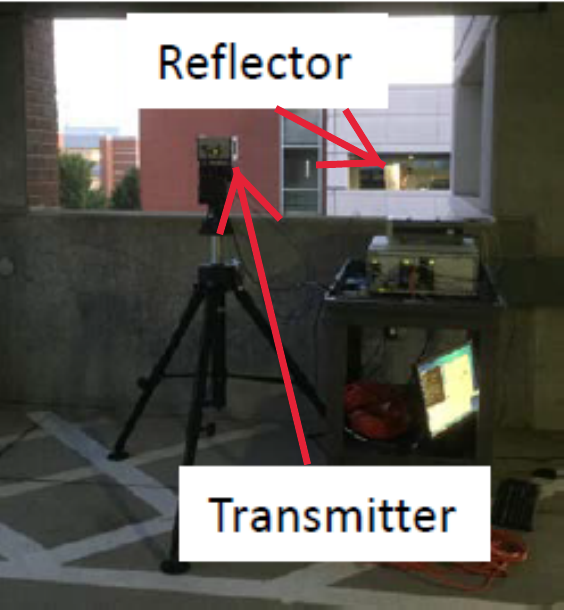}
	 \caption{}
     \end{subfigure}
      \begin{subfigure}{0.21\textwidth}
	\centering
    \includegraphics[width=\textwidth, height = 3.68cm]{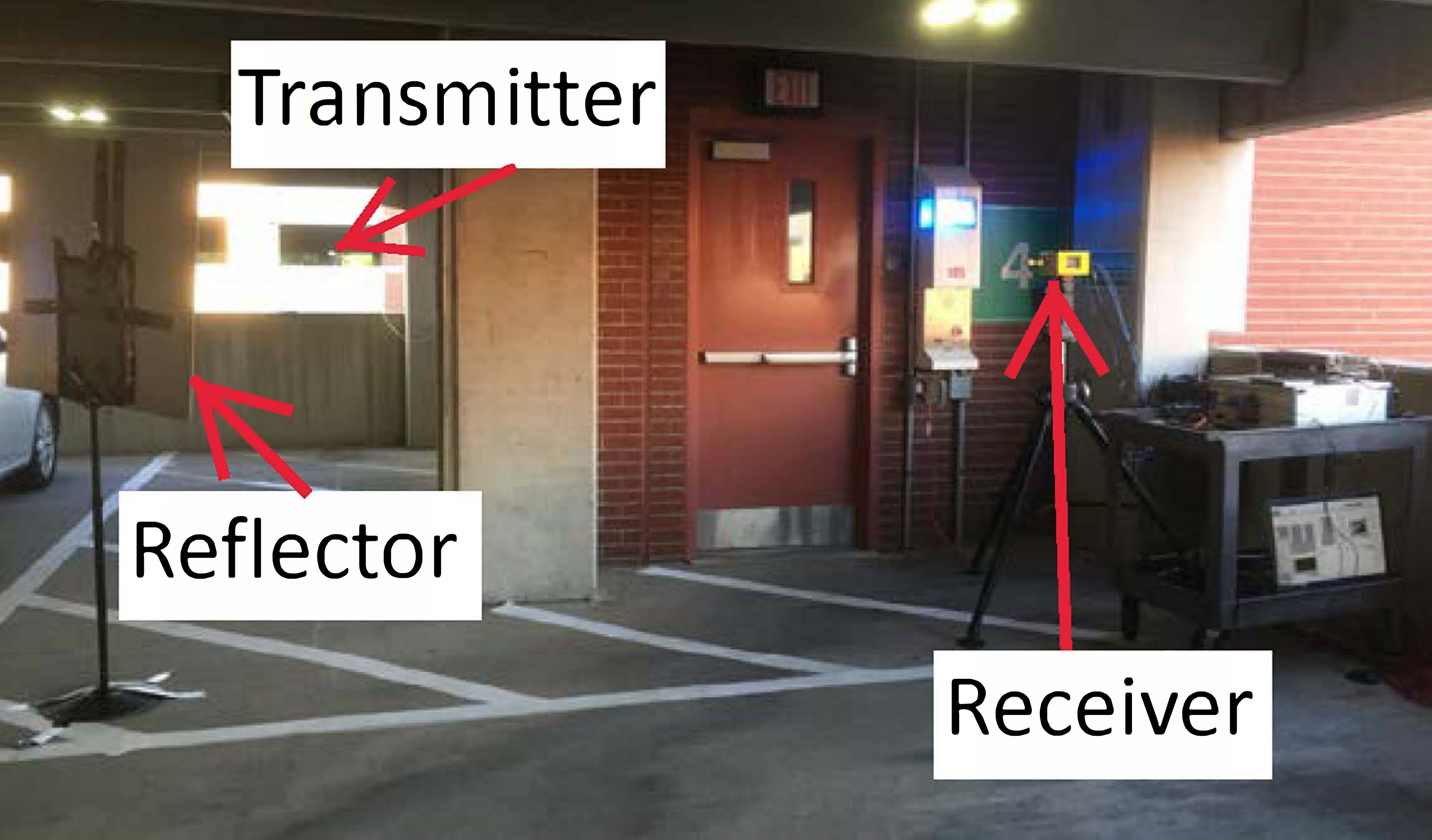}
	 \caption{}
     \end{subfigure}
     \begin{subfigure}{0.43\textwidth}
	\centering
    \centerline{\includegraphics[width=\columnwidth]{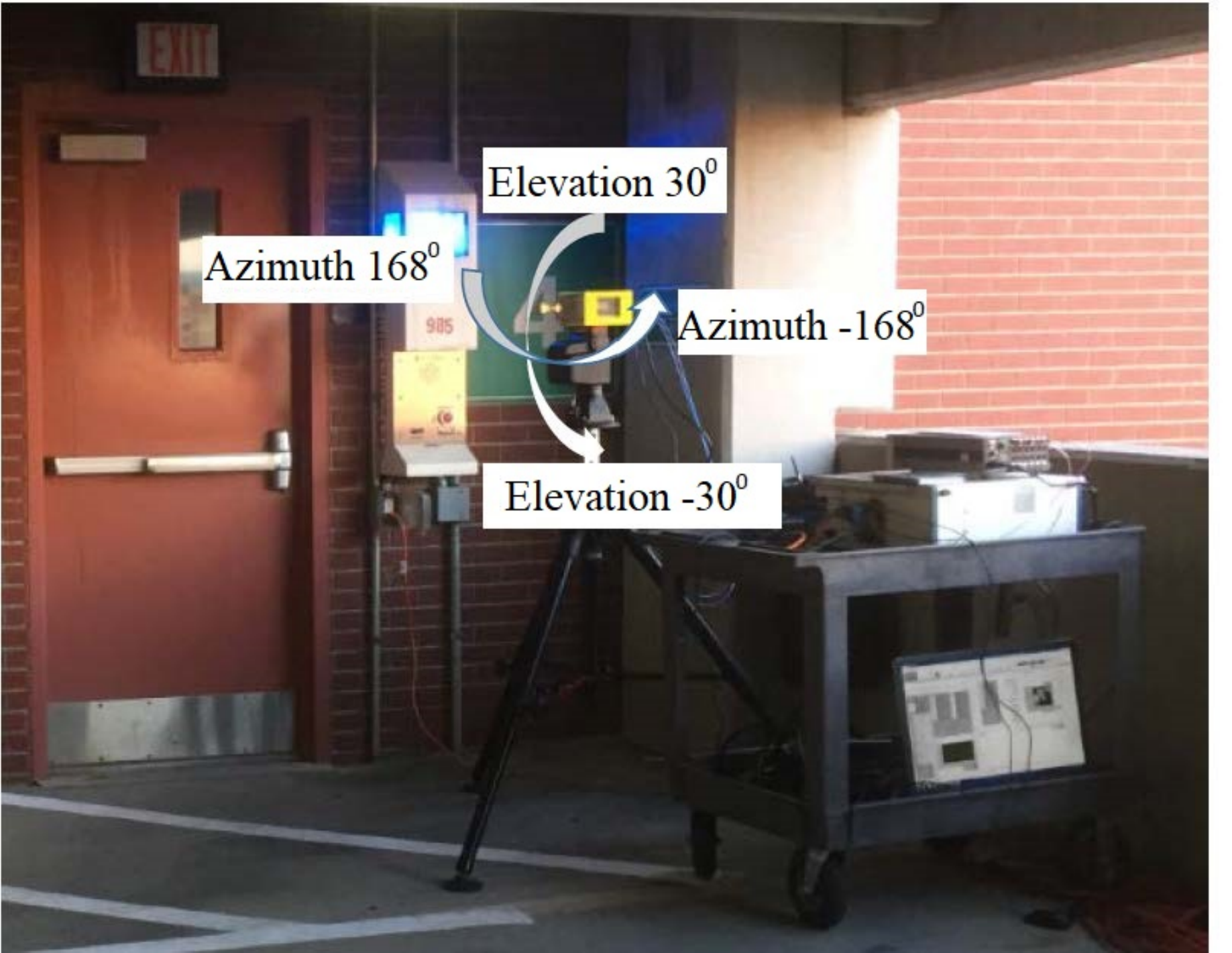}}
	 \caption{}
     \end{subfigure}
    \caption{Measurement setup (a) transmitter shown on the gimbal. During the measurements, the transmitter was not rotated, (b) receiver Location. $0.84\times0.84$~m$^2$ reflector is positioned at approximately $45^{\circ}$ angle to maximize the received power, (c) the orientation of elevation and azimuth angles.}\label{Fig:setup2}\vspace{-4mm}
\end{figure}

\subsection{Outdoor Measurement Setup} 
The outdoor measurements were performed inside two parking buildings next to each other at North Carolina State University (NCSU) campus as shown in~Fig.~\ref{Fig:setup2}. The receiver was located behind a brick wall compartment with a glass window inside. There was no direct LOS between the transmitter and the receiver. However, there was an obstructed line-of-sight~(OLOS) through the window. The receiver antenna was mounted on a rotatable gimbal in order to collect energy from different azimuth and elevation directions. The transmitter antenna was not rotated. Three flat square metallic sheet reflectors of sizes $0.30\times0.30$~m$^2$, $0.61\times0.61$~m$^2$, and $0.84\times0.84$~m$^2$ were used 3.5~m away from the receiver as shown in Fig.~\ref{Fig:setup2}(b). 

For comparison, we considered two scenarios: one without the reflector and one with reflectors. The transmitter antenna was fixed at $0^{\circ}$ elevation angle \emph{facing directly the reflector}. The heights of transmitter and receiver were the same such that the boresight of each antenna point to the center of the reflector. The gimbal at the receiver side scanned the azimuth plane from $-168^{\circ}$ to $168^{\circ}$ with $10^{\circ}$ increments, and the elevation plane from $-30^\circ$ to $30^\circ$ with $10^\circ$ increments shown in Fig.~\ref{Fig:setup2}(c). The transmit power was set to $0$~dBm.
\vspace{-1mm}

\subsection{Ray Tracing Simulation Setup}\label{Section:Ray_tracing_setup}
Simulations for the passive metallic reflectors at mmWave frequencies were performed using Remcom Wireless InSite RT software, replicating the indoor experimental environment as shown in Fig.~\ref{Fig:scenario}(b). The red blocks in the figure represent the individual receiver points in the grid. A sinusoidal sounding signal at $28$~GHz was used, and the transmit power was $0$~dBm. Horn antennas~\cite{Horn_antenna_sage}, similar to used in the measurements, were used at both transmitter and the receiver grid. 

\begin{figure*}[!t]
    \centering
	\begin{subfigure}{0.43\textwidth}
	\centering
	\includegraphics[width=\columnwidth]{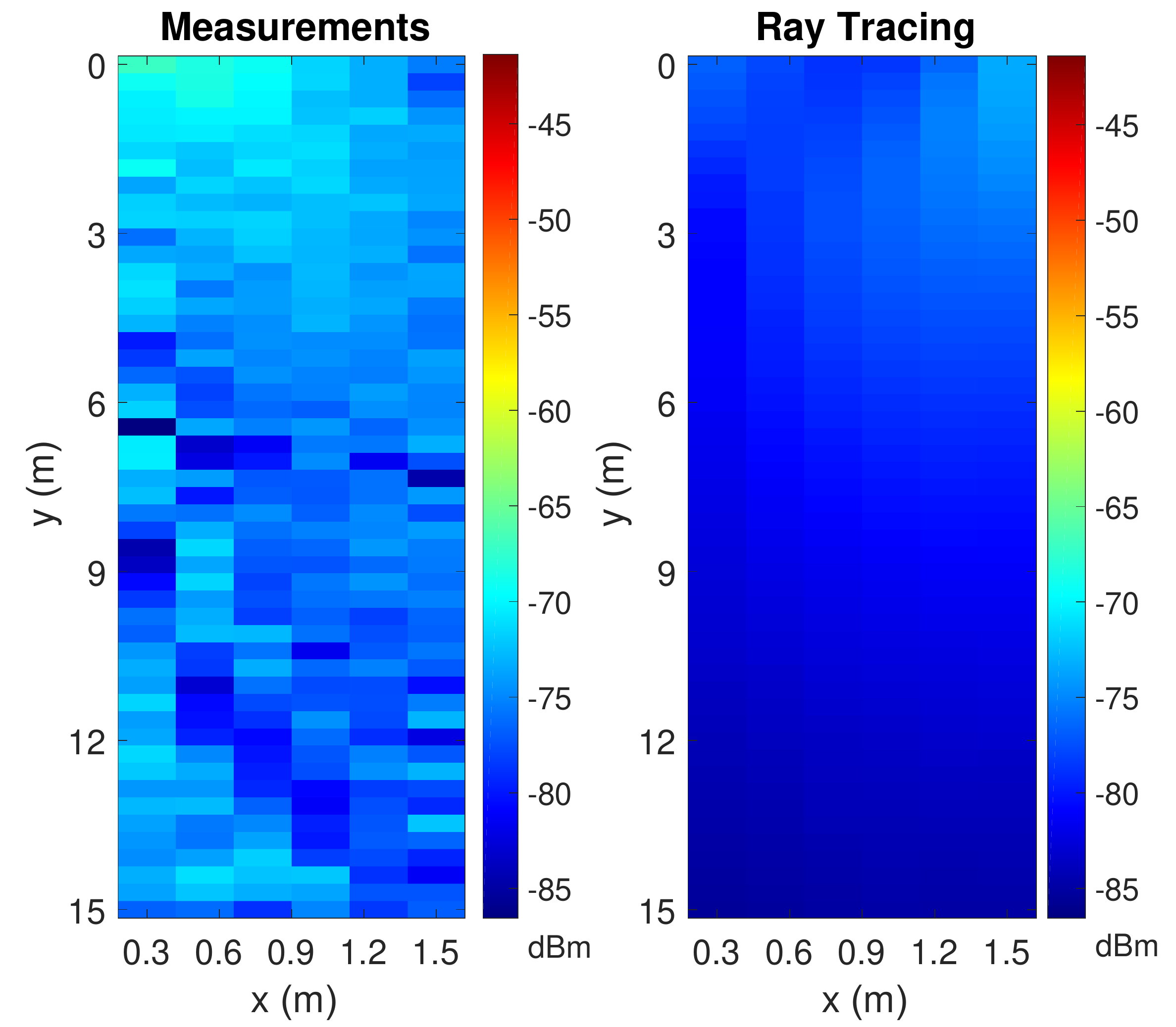}
	\caption{}
    \end{subfigure}			
	\begin{subfigure}{0.43\textwidth}
	\centering
    \includegraphics[width=\columnwidth]{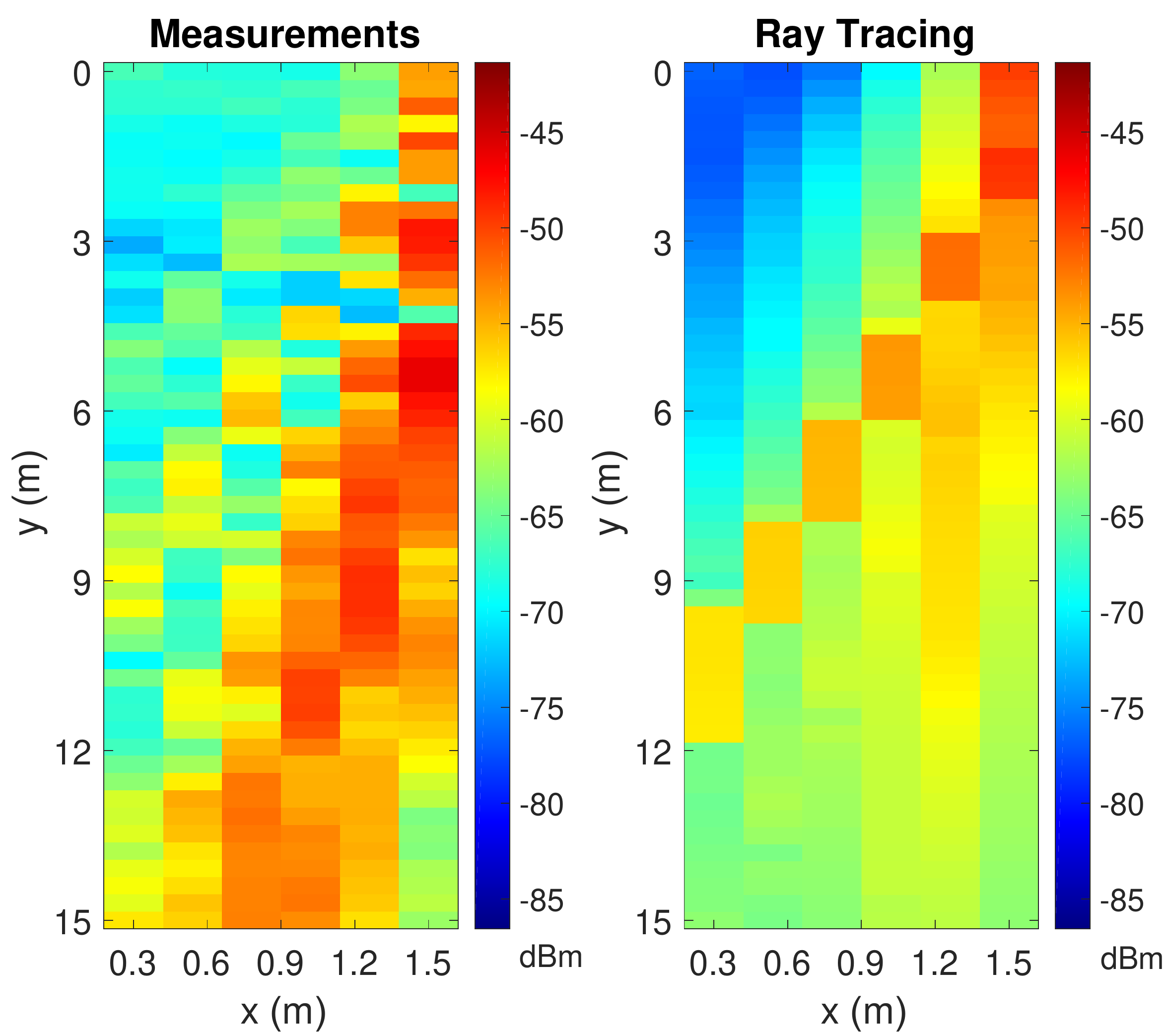}
	 \caption{}
     \end{subfigure}
	\begin{subfigure}{0.43\textwidth}
	\centering
    \includegraphics[width=\columnwidth]{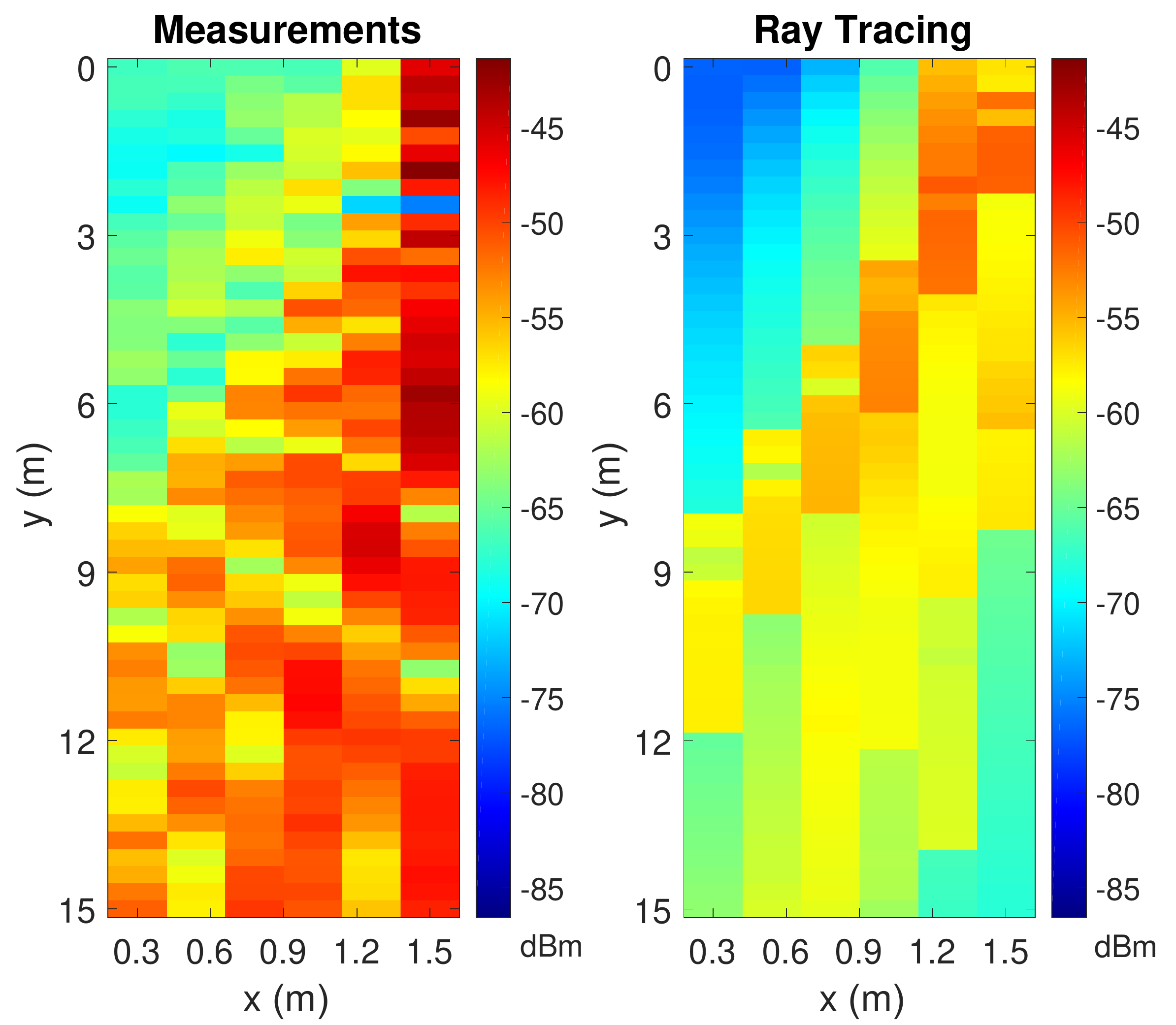}
	 \caption{}
     \end{subfigure}
     \begin{subfigure}{0.43\textwidth}
	\centering
    \includegraphics[width=\columnwidth]{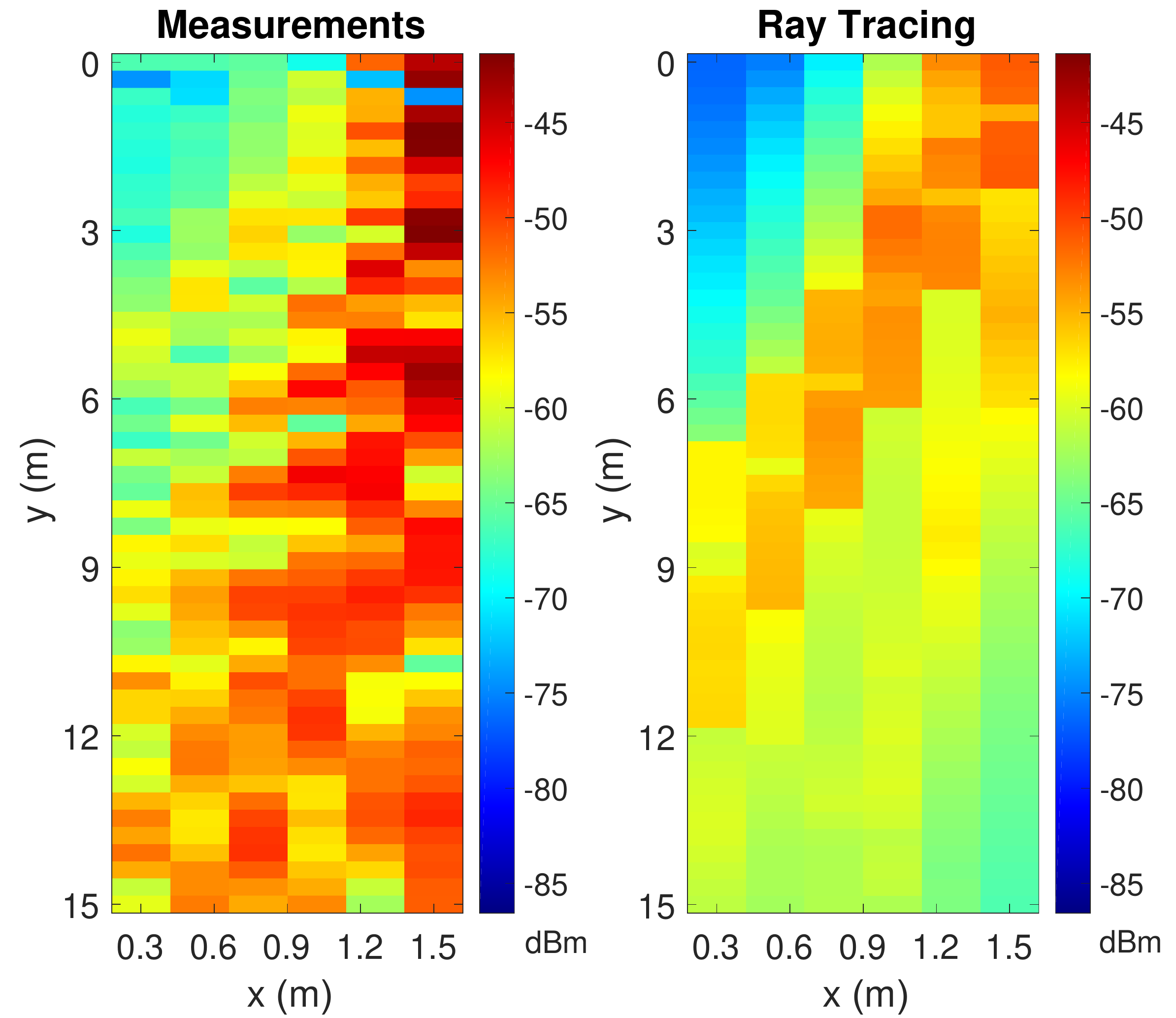}
	 \caption{}
     \end{subfigure}
\caption{Total received power results on the grid shown in Fig.~\ref{Fig:Core_area}(a) for (a) no reflector,  obtained using (left) measurements, and (right) ray tracing simulations, (b) $0.30\times0.30$~m$^2$ flat square aluminum sheet at $\theta = 45^\circ$,  obtained using (left) measurements, and (right) ray tracing simulations, (c) $0.61\times0.61$~m$^2$ flat square aluminum sheet at $\theta = 45^\circ$,  obtained using (left) measurements, and (right) ray tracing simulations, and (d) $0.84\times0.84$~m$^2$ flat square aluminum sheet at $\theta = 45^\circ$,  obtained using (left) measurements, and (right) ray tracing simulations.}\vspace{-2mm} \label{Fig:flat_combine}
\end{figure*}

In addition to specular reflection at mmWave frequencies, diffuse scattering also occurs dominantly due to the comparable size of the wavelength of the transmitted wave and the dimensions of the irregularities of the surfaces that it encounters. In the simulations, the diffuse scattering feature had been used to take into account this factor. The diffuse scattering model used in the simulations was the directive model. Only the diffuse scattering coefficient was changed for different materials, whereas the other model parameters remained the same. Diffuse scattering coefficient of different materials used in the simulations were $0.1$, $0.2$, $0.25$ and $0.3$, for the perfect conductor, concrete, ceiling board, and layered drywall, respectively. The materials with higher roughness were assigned higher diffuse scattering coefficient.

The received power was obtained and summed non-coherently from the received MPCs at a given receiver location. This did not involve the phase of each MPC to be considered in the received power calculation. The similar was done for measurements.  

The walls, floor, ceiling, door and reflector materials were selected such that they were similar to the actual measurement environment setup as much as possible. The \emph{ITU three-layered drywall} was used for walls and ITU \emph{ceiling board} was used for ceilings, \emph{concrete} was used for floor, and a \emph{perfect conductor} was used for the door and the metallic reflector. All the materials were frequency sensitive at $28$~GHz. The dimensions of the simulation setup were the same as in Fig.~\ref{Fig:Core_area}(a).

Ray tracing simulations were also performed for different center frequencies. These frequencies are $1.8$~GHz, $2.4$~GHz, $38$~GHz, and $60$~GHz. No diffuse scattering was used for $1.8$~GHz and $2.4$~GHz center frequencies. This is mainly due to larger wavelengths compared to the size of the surface roughness of the materials in the environment. However, at $38$~GHz, and $60$~GHz, larger diffuse scattering coefficients were used, compared to at $28$~GHz for different materials in the environment. A $0.1$ increase in the diffuse scattering coefficient value for every $10$~GHz increase of the center frequency, compared to at $28$~GHz was used.  

%\vspace{-2mm}
% \begin{figure}
% 	\centering           
% 	\includegraphics[width=\columnwidth]{GSMM_latest_curve_non_coherent_v2.pdf}
% 	\caption{Received power results for metallic curved reflectors, (left) curve angle of $5$~degree, and (right) curve angle of $10$~degree, obtained using ray tracing simulations. } \label{Fig:curved}
% \end{figure} 

 \section{Measurement, Simulation and Analytical Results for Indoor Scenarios with Reflectors} 

In this section we present indoor measurement and ray tracing results at $28$~GHz, and ray tracing results at various frequencies. Analytical results~(from Section~\ref{Section:Power_distribution_modeling}) for the received power distribution over the indoor receiver grid for different reflector sizes/shapes are also provided. 
 \vspace{-2mm}
\subsection{Indoor Measurements and Simulations with Flat Reflectors}\label{Section:Received_power_flat}
In this subsection, empirical and simulation results are presented for the indoor NLOS measurements with and without metallic reflectors for the setup shown in Fig.~\ref{Fig:Core_area}. In measurements shown in Fig.~\ref{Fig:flat_combine}(a) where no reflector is used, we observe slightly higher received power at the top left corner of the receiver grid mostly due to diffraction at the edge of the corridor wall. In the case of simulations, we observe some reflections from the wall opposite to the transmitter; however, the overall received power, in this case, is less than the measurements. %\textcolor{red}{Overall, in case of no reflector scenario, the received power is mainly due to the sidelobes on the right of the transmit antenna towards the receiver antenna~(\ref{Eq:Sum_power}). The received power due to others in (\ref{Eq:Sum_power}) will be negligible when there is no reflector. }  

%\subsection{Coverage with Square Metal Reflectors} \label{Section:Coverage_flat_reflector}
The flat reflectors are oriented at $45^\circ$ in the azimuth plane as shown in Fig.~\ref{Fig:Core_area}(a) for all the measurements. 
For the $0.30\times0.30$~m$^2$ reflector shown in Fig.~\ref{Fig:flat_combine}(b), it can be observed that we have a directional coverage spreading with the distance along the y-grid. The reflections that are perpendicular to the incident plane waves create a strip of dominant coverage area starting from the top right portion of the receiver grid. The width of this dominant coverage area is proportional to the width of the reflector as expected.

The received power decreases exponentially as we move left on the grid. This is because as we move left on the grid, we move away from the optimum reflection angle region. However, as move downward over the grid for a given $x$ value, this decrease becomes smaller. This is due to a small angular difference from the optimum reflection region as we move downward. Moreover, the received power also decreases due to the increase of the distance as we move downward on the grid. However, this decrease due to the increase of distance is small compared to the shift from the optimum reflection region. This is accordingly modeled in Section~\ref{Section:First_order_modeling}. 

The $0.61\times0.61$~m$^2$ and $0.84\times0.84$~m$^2$ reflector measurement results are shown in Fig.~\ref{Fig:flat_combine}(c) and (d), respectively. Similar to $0.30\times0.30$~m$^2$ case, we observe a solid strip of dominant coverage area with a width proportional to the size of the reflector. The received power is distributed similarly across the receiver grid with a better coverage as compared to $0.30\times0.30$~m$^2$ case. Moreover, we observe a similar exponential decrease in the received power as we move away from the optimum reflection angle region. 

\begin{figure*}[!h]
    \centering
	\begin{subfigure}{0.43\textwidth}
	\centering
    \includegraphics[width=\columnwidth]{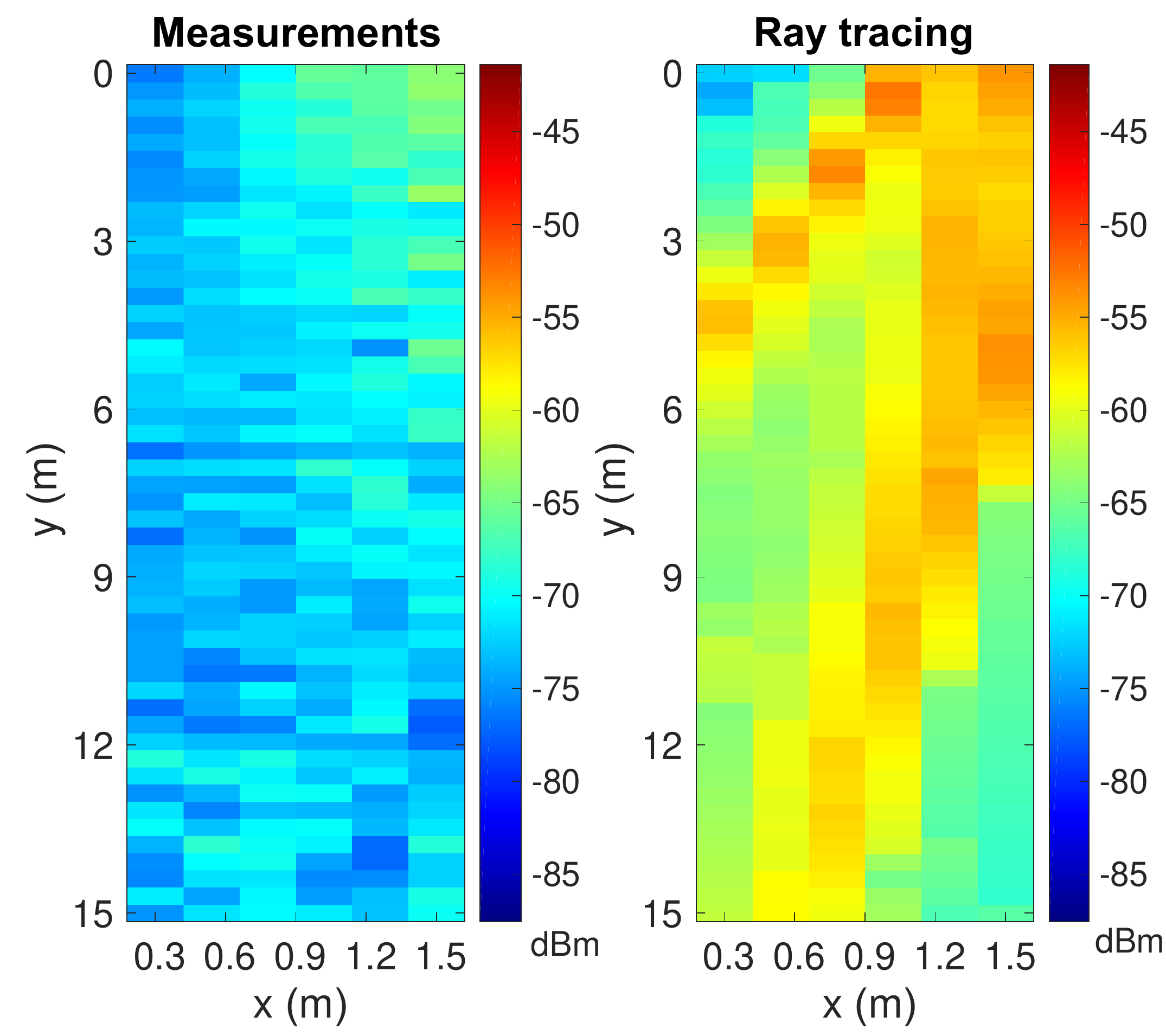}
	\caption{}
    \end{subfigure}       
    \begin{subfigure}{0.43\textwidth}
	\centering           
	\includegraphics[width=\columnwidth]{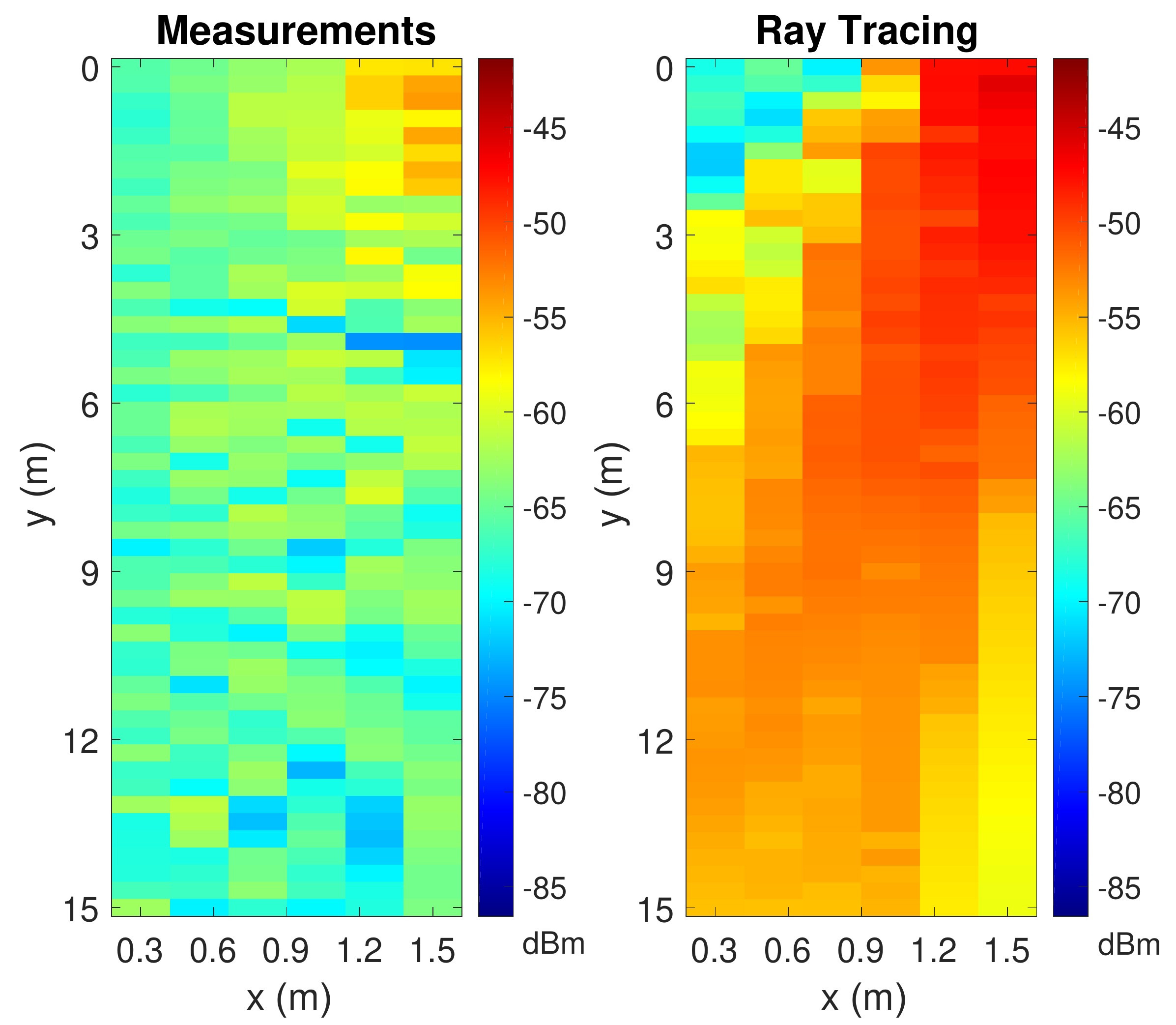}
	\caption{}
    \end{subfigure}    
   \caption{Total received power results for (a) metallic sphere obtained using (left) measurements, and (right) ray tracing simulations; (b) metallic cylinder obtained using (left) measurements, and (right) ray tracing simulations.}\label{Fig:sphere_cylinder_combine} \vspace{-2mm}
\end{figure*}

For all the flat reflector scenarios, we observe power distribution mostly on the right side of the receiver grid, whereas we observe outage at the top left corner of the receiver grid. This is due to the property of directional reflection for the flat reflectors. Three plausible solutions to provide coverage on the top left side of the receiver grid can be; 1) By orienting the reflector at angle less than $45^\circ$ (but will result in reduced power on the right side of the grid); 2) by using secondary reflectors that are oriented accordingly to reuse the reflected energy; 3) using outward curved reflectors e.g. cylinders that can distribute the energy more uniformly on the grid due to divergence phenomenon.

\begin{figure*}[!h]
    \centering
	\begin{subfigure}{0.43\textwidth}
	\centering
    \includegraphics[width=\columnwidth]{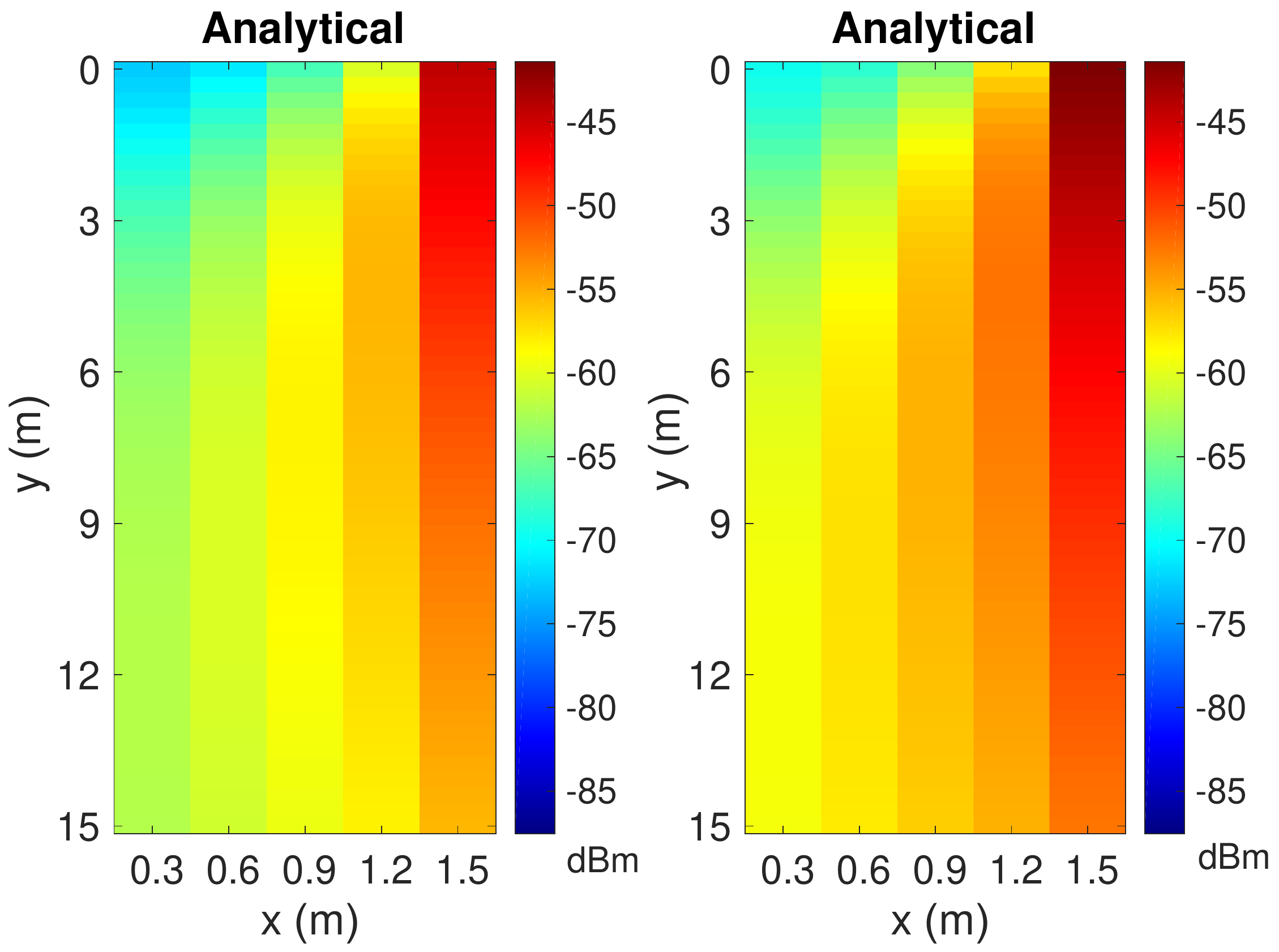}
	\caption{}
    \end{subfigure}       
    \begin{subfigure}{0.43\textwidth}
	\centering           
	\includegraphics[width=\columnwidth]{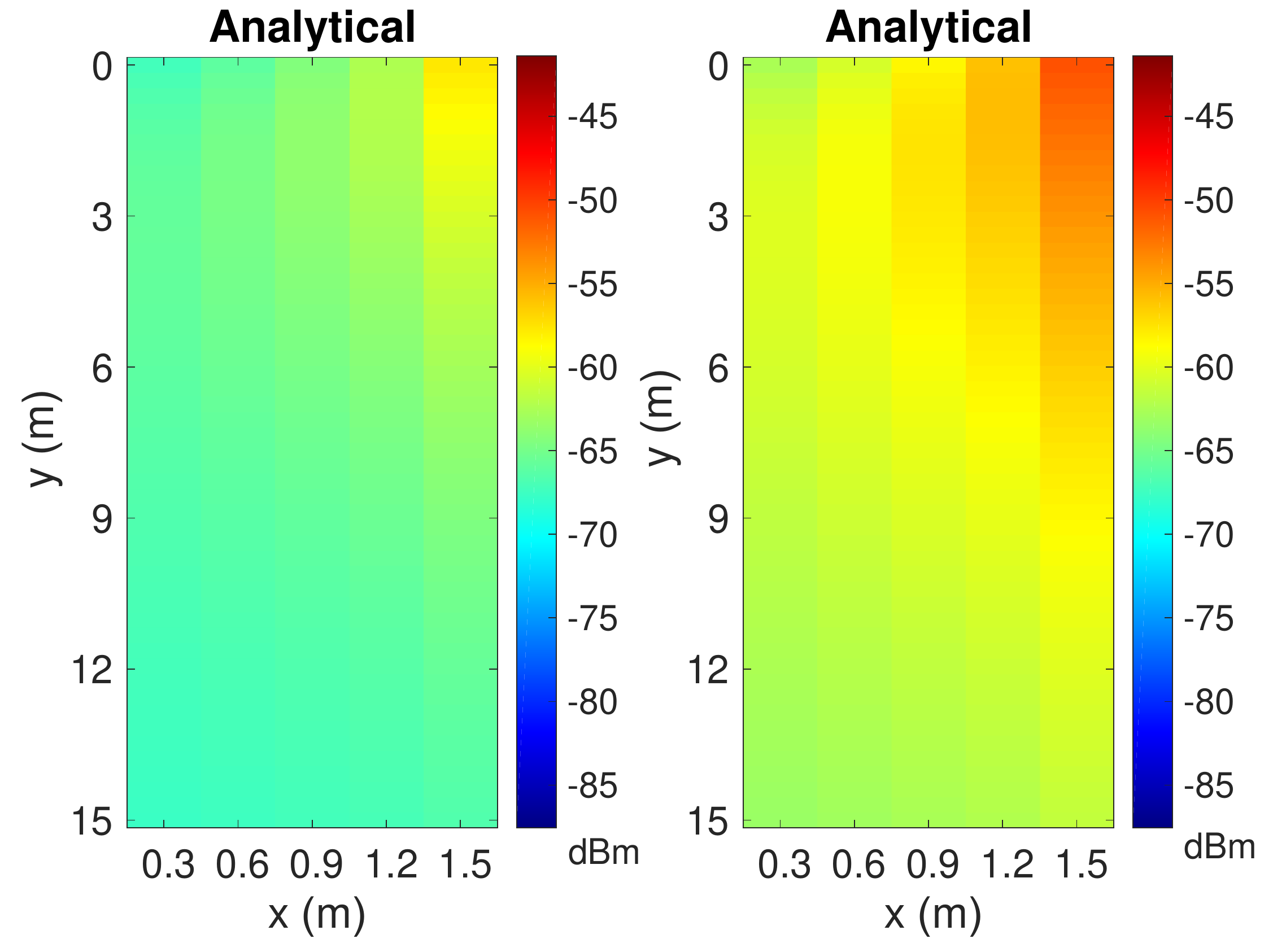}
	\caption{}
    \end{subfigure}    
   \caption{Total received power results obtained analytically from Section~\ref{Section:Power_distribution_modeling} plotted on the grid for (a) flat square metallic sheet reflectors of sizes, $0.30\times0.30$~m$^2$~(left), $0.61\times0.61$~m$^2$~(right), and (b) sphere reflector~(left), cylinder reflector~(right).}\vspace{-4mm} \label{Fig:Analytical} 
\end{figure*}

 \begin{figure*}[!t]
    \centering
	\begin{subfigure}{0.43\textwidth}
	\centering
    \includegraphics[width=\textwidth]{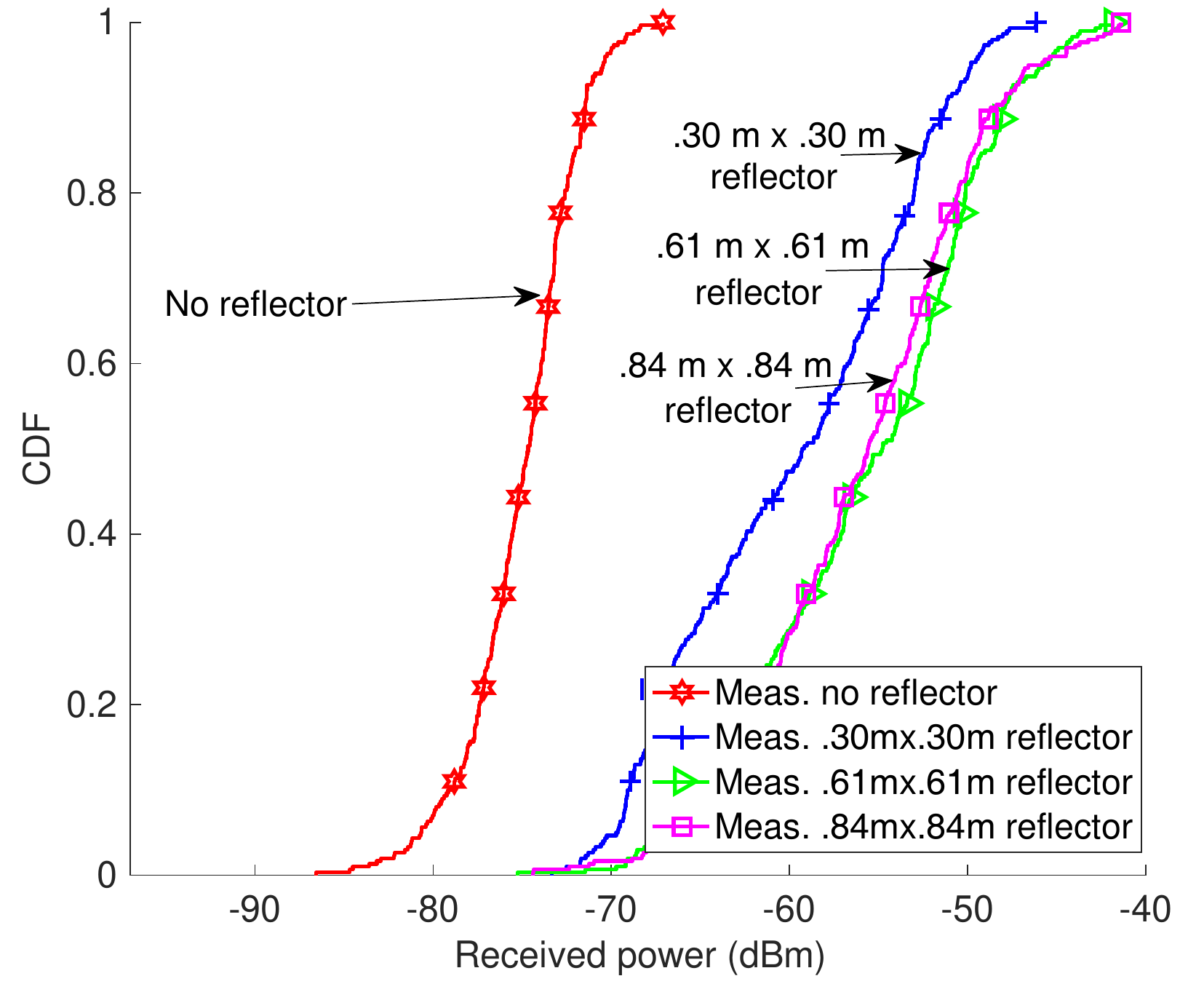}
    \caption{}
    \end{subfigure}	
 	\begin{subfigure}{0.43\textwidth}
    \centering
	\includegraphics[width=\textwidth]{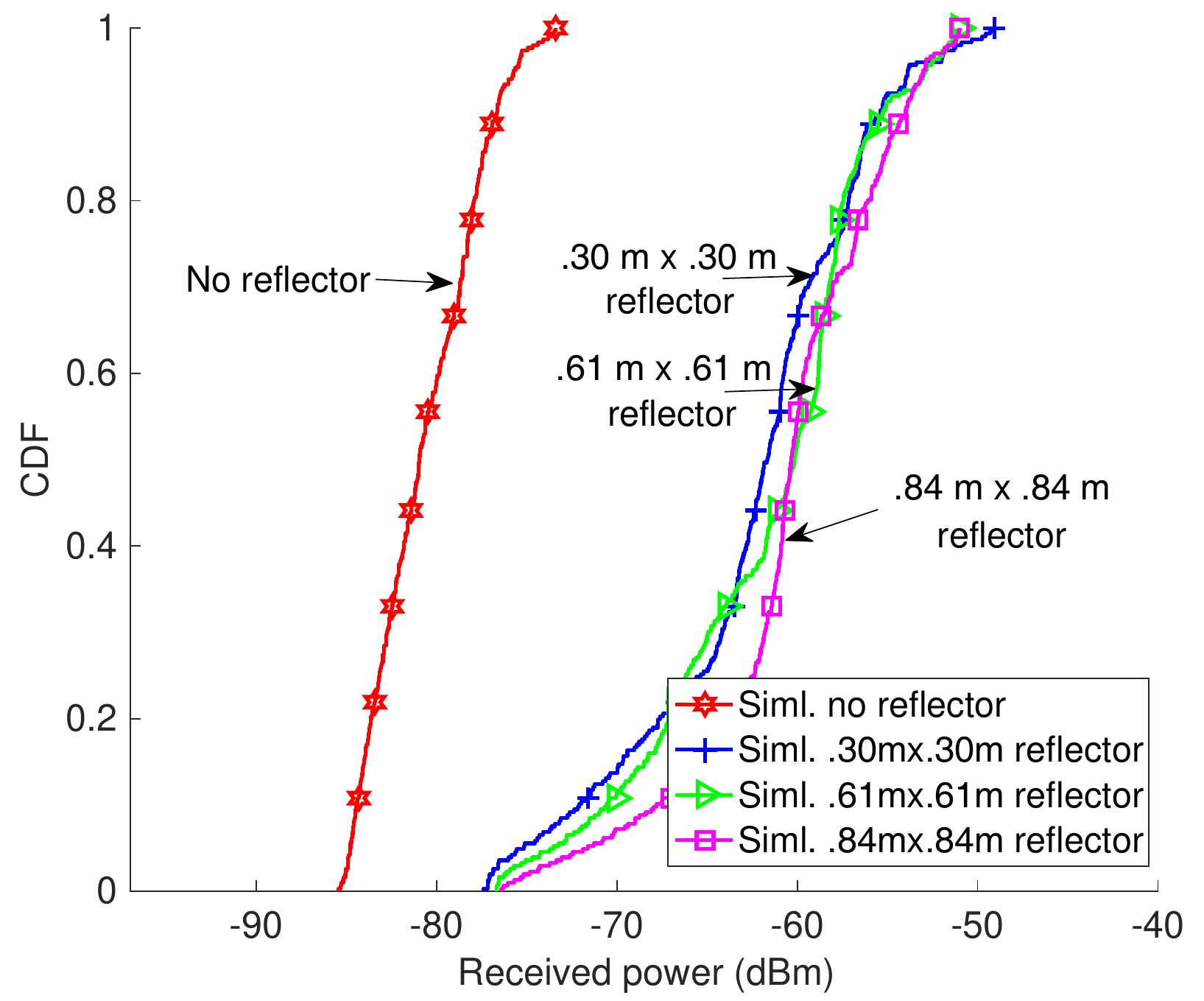}
	  \caption{}  
    \end{subfigure}     
	\begin{subfigure}{0.43\textwidth}
    \centering
	\includegraphics[width=\textwidth]{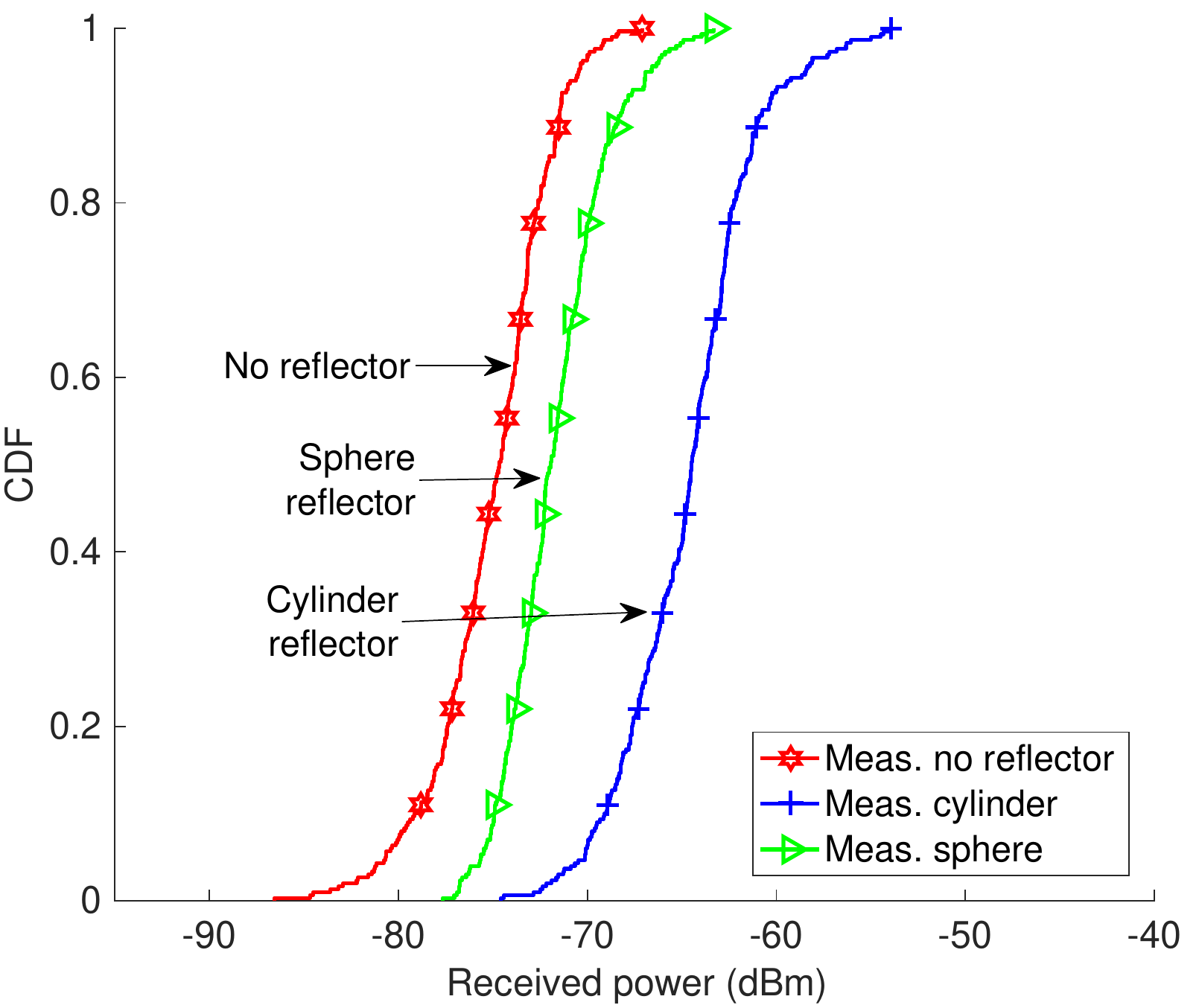}
	 \caption{}  
    \end{subfigure}    
    \begin{subfigure}{0.43\textwidth}
    \centering
	\includegraphics[width=\textwidth]{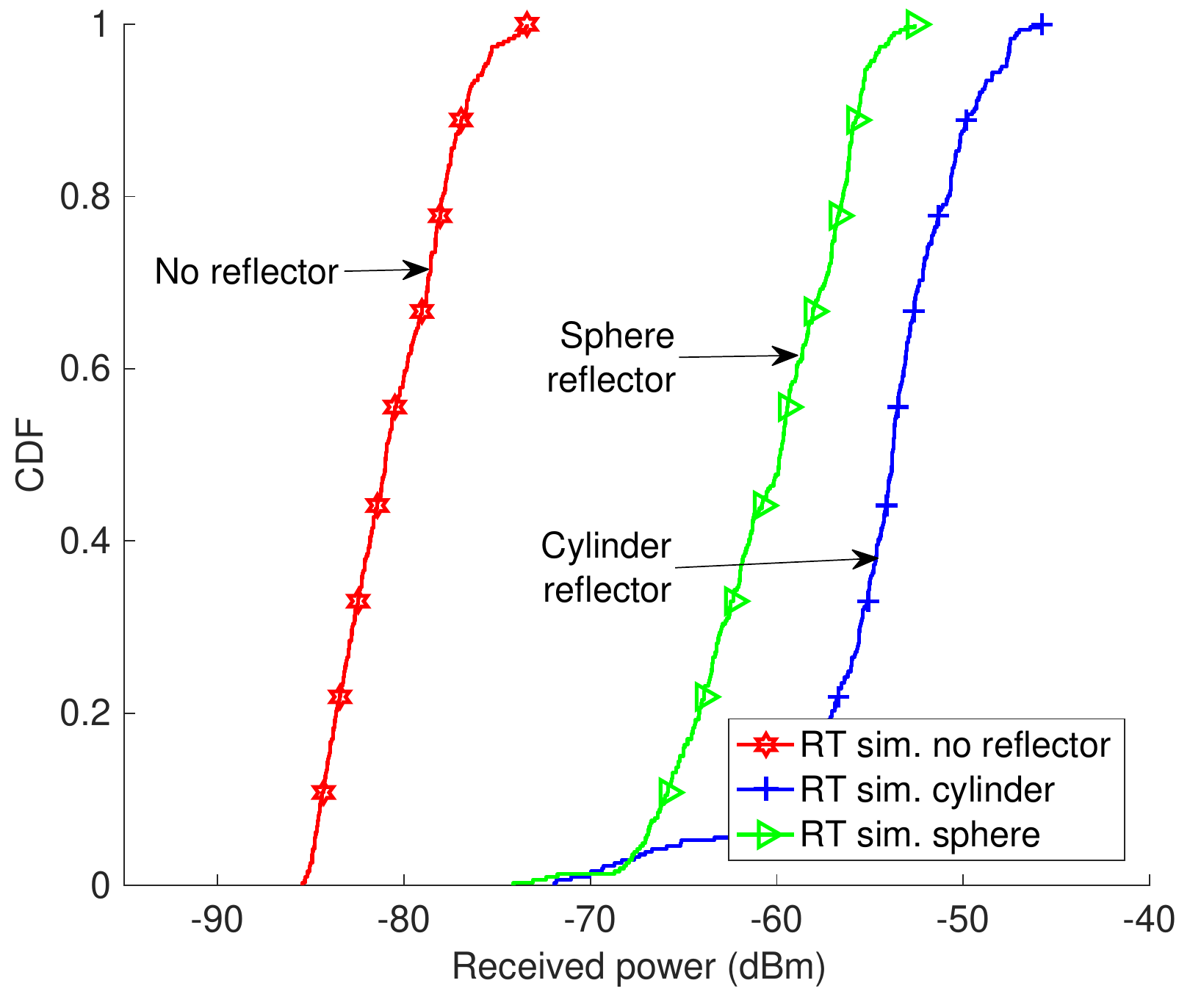}
	 \caption{}  
    \end{subfigure}
        \caption{CDF of total received power for multiple scenarios. (a) no reflector, $0.30\times0.30$~m$^2$, $0.61\times0.61$~m$^2$, and $0.84\times0.84$~m$^2$ flat square sheet reflectors (measurements); (b) no reflector, $0.30\times0.30$~m$^2$, $0.61\times0.61$~m$^2$, and $0.84\times0.84$~m$^2$ flat square sheet reflectors (simulations); (c) no reflector, cylinder reflector, and sphere reflector (measurements); (d) no reflector, cylinder reflector, and sphere reflector (simulations).}\vspace{-4mm} \label{Fig:CDF_combine}
\end{figure*}

\begin{figure*}[!t] 
    \centering
	\begin{subfigure}{0.43\textwidth}
	\centering
    \includegraphics[width=\textwidth]{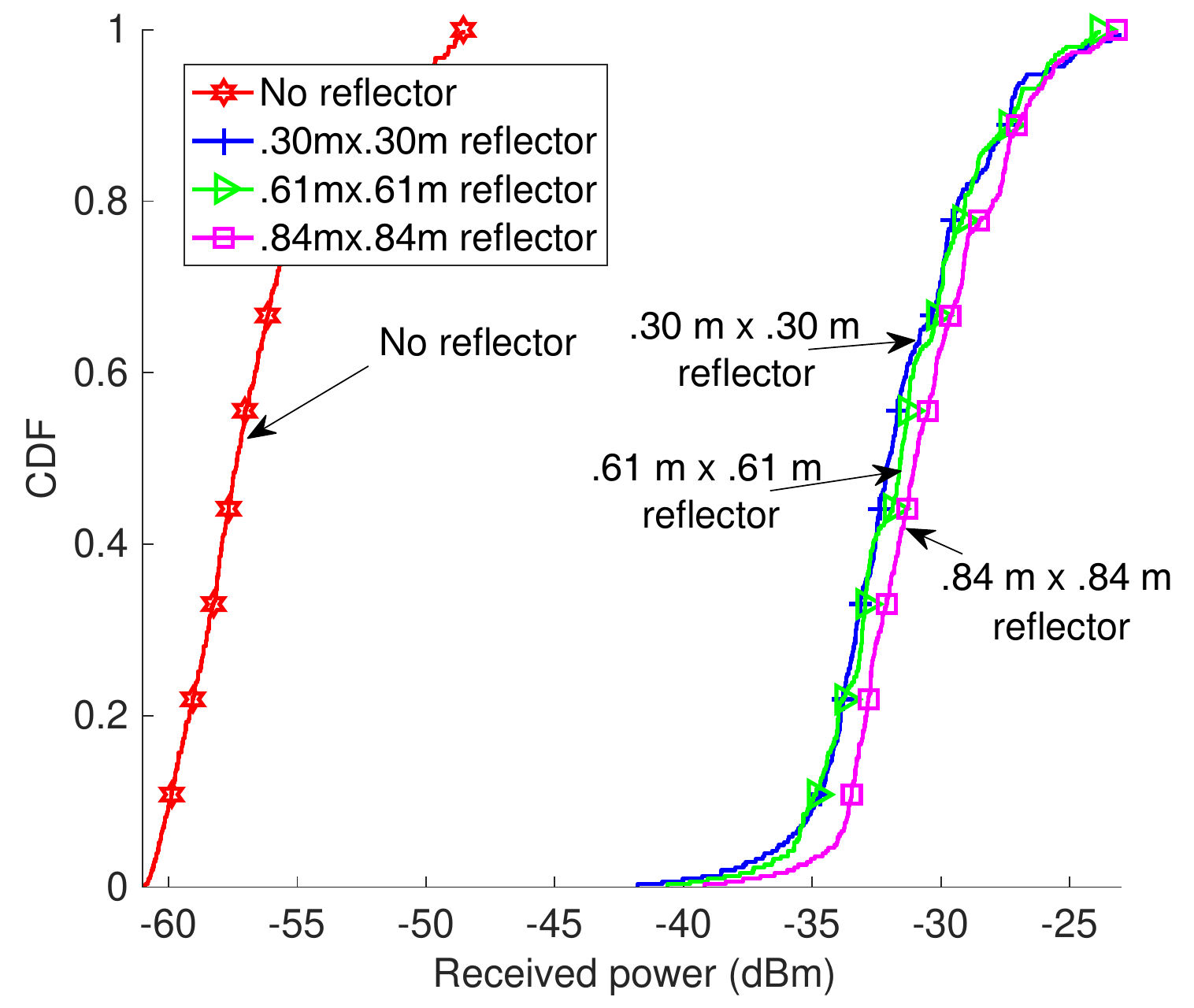}
    \caption{}
    \end{subfigure}	
 	\begin{subfigure}{0.43\textwidth}
    \centering
	\includegraphics[width=\textwidth]{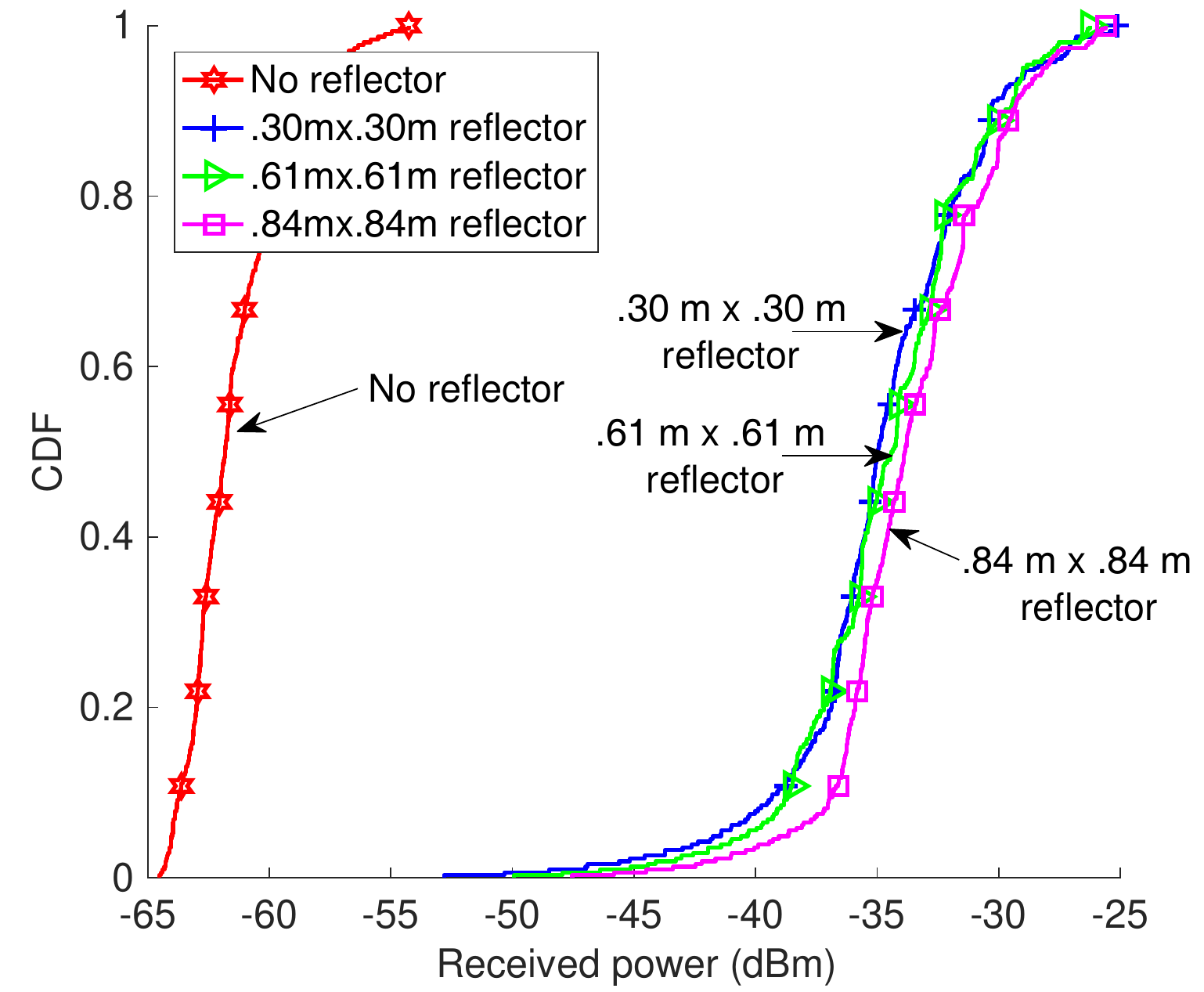}
	  \caption{}  
    \end{subfigure}     
	\begin{subfigure}{0.43\textwidth}
    \centering
	\includegraphics[width=\textwidth]{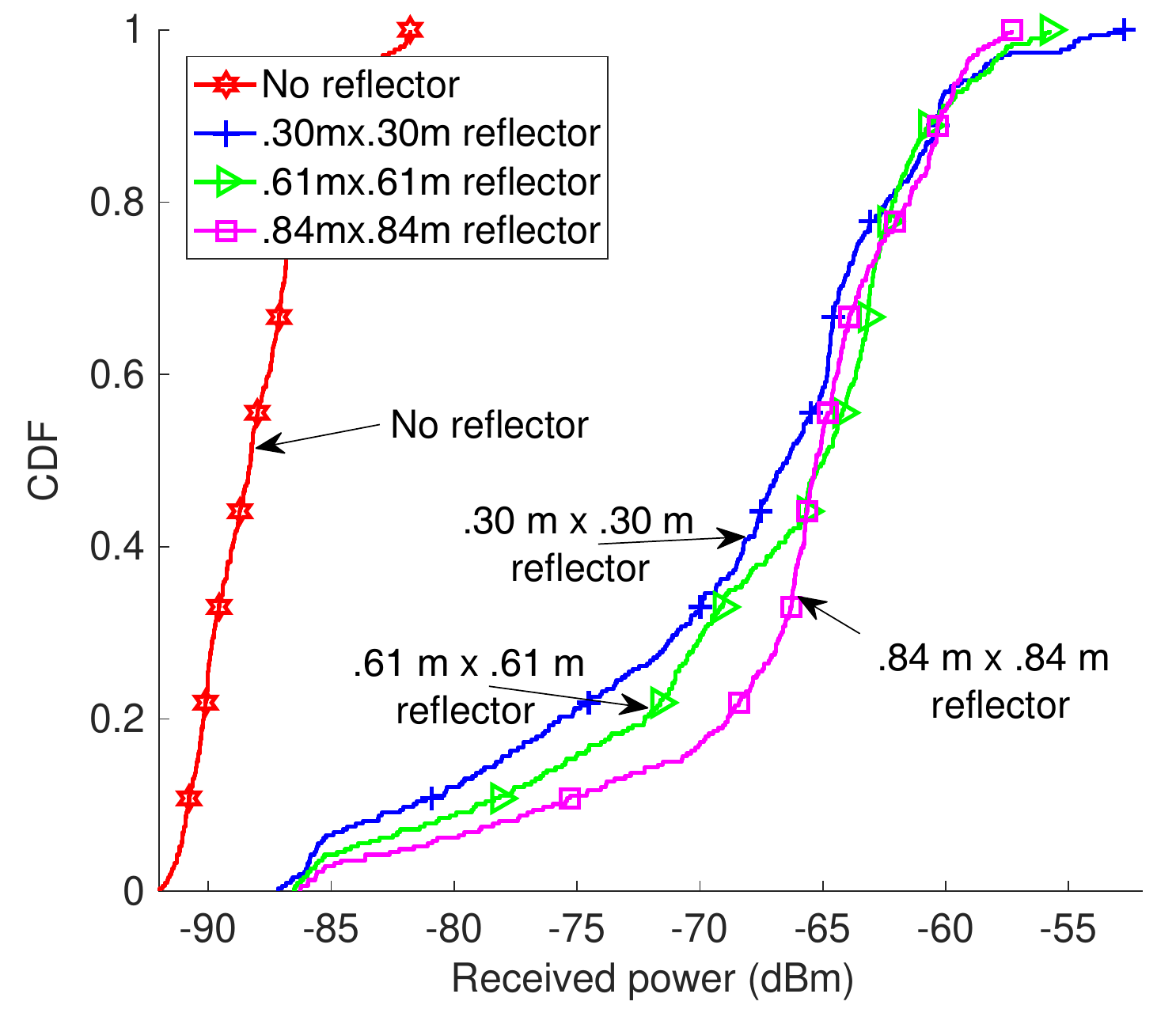}
	  \caption{}  
    \end{subfigure}    
    \begin{subfigure}{0.43\textwidth}
    \centering
	\includegraphics[width=\textwidth]{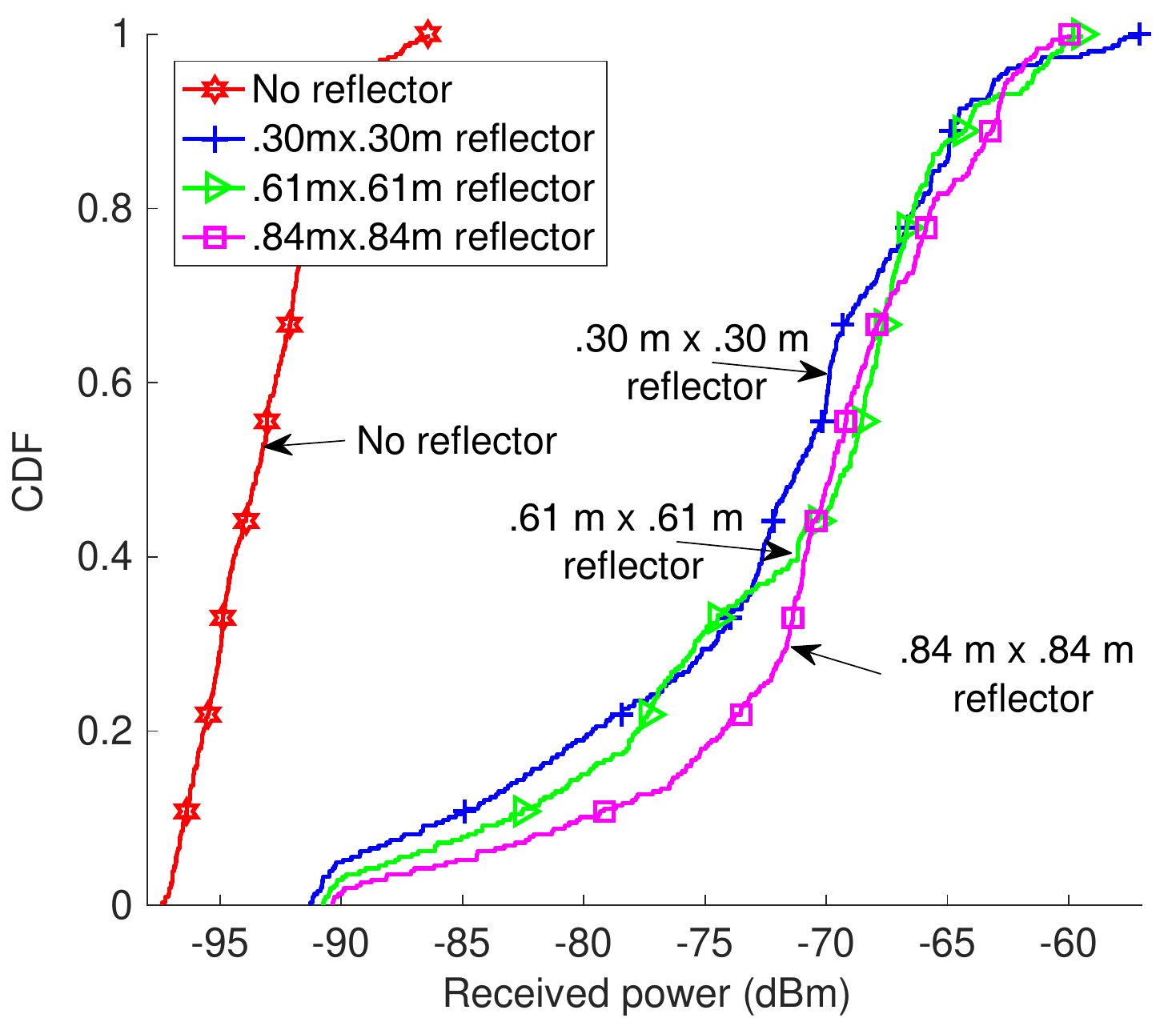}
	  \caption{}  
    \end{subfigure}
\vspace{-1mm}    
    \caption{CDF of total received power at different center frequencies: (a) 1.8~GHz, (b) 2.4~GHz, (c) 38~GHz, and (d) 60~GHz.}\vspace{-3mm} \label{Fig:CDF_diff_freqs}
\end{figure*}

In Fig.~\ref{Fig:flat_combine} side by side comparison of measurements and RT simulations are possible. The distribution of the power on the receiver grid is similar for both the measurements and the simulations for the three different reflector sizes. For all the cases, we observe smaller received power for RT simulations when compared to the measurements. One reason is the presence of additional small scatterers in the environment and approximate diffuse scattering coefficients used in the simulations for real-world materials. Another reason is because of the simple construction of the flat reflectors with a smaller number of reflection points for ray tracing. This is in contrast to complex reflector shapes~(e.g. curved) that have a large number of reflection points. Moreover, a small additional power gain is observed for $0.30\times0.30$~m$^2$ and $0.61\times0.61$~m$^2$ reflectors from the cardboard which was not included in simulations.

\vspace{-1mm}
\subsection{Indoor Measurements Simulations with Non-Flat Reflectors}\label{Section:Received_power_nonflat}
The measurement and simulation results for cylinder and sphere reflectors are shown in Fig.~\ref{Fig:sphere_cylinder_combine}. We observe smaller received power for sphere reflector compared to the cylinder, mainly due to a smaller effective area~(see Section~\ref{Section:Effective_area}). Moreover, the cylinder reflector provides more uniform power distribution on the receiver grid compared to flat reflectors. The reason for this can be explained due to the high divergence of the incoming energy randomly in the surroundings from the top and bottom of the sphere. Similarly, for both the cylinder and sphere, we observe less power as compared to the flat sheet reflector $0.61\times0.61$~m$^2$, though all three have approximately similar cross-section area. The reason for this behavior is due to the small effective area of the cylinder and sphere exposed to the incoming energy as compared to the flat sheet reflector. 

Simulation results in Fig.~\ref{Fig:sphere_cylinder_combine} exhibit larger received power as compared to measurements. This behavior is due to a large number of surface points for curved reflectors~(due to their complex construction) in ray tracing simulations~\cite{ray_tracing}. These additional surface points result in additional reflections for curved reflectors, which yield larger received power in ray tracing simulations. However, for flat reflectors, the number of surface points for ray tracing is smaller compared to curved surfaces. This is due to their simpler construction. Therefore, we observe smaller received power for flat reflectors compared to curved reflectors. 
\vspace{-1mm}
\subsection{Analytical Received Power Results for Indoor}\label{Section:Analytical_results}

The analytical results for end-to-end received power in indoor NLOS area is obtained from Section~\ref{Section:Power_distribution_modeling}. From~(\ref{Eq:Sum_power}), there are three major sources of received power in the indoor setup, $P_{\rm refl}^{(1)}$, $P_{\rm refl}^{(2)}$ and $P_{\rm s}$. However, the major contributor to the received power is $P_{\rm refl}^{(1)}$. Whereas, the received power due to $P_{\rm refl}^{(2)}$ cannot be calculated directly in the corridor due to random nature of the second-order reflections. In addition, the portion of the received power provided by $P_{\rm s}$ is very small. Therefore, we can approximate $P_{\rm s} = -70$~dBm uniformly over the receiver grid for analytical results. This value is obtained from empirical results for no reflector in Fig.~\ref{Fig:flat_combine}(a). The value of $\alpha_{\rm f}$ from (\ref{Eq:PowerEq3}) is set to $0.72$, whereas, $\alpha_{\rm c}= 0.85$ from (\ref{Eq:PowerEq4}). Also, as we used polished metal sheets, therefore, $\Gamma=1$ and $|\rho_{\rm TX}\cdot\rho_{\rm refl}|^2 = 1$. %On other hand, orientation mismatch loss between the reflected plane waves and the receiver antenna's boresight due to the size difference of the reflectors cannot be obtained directly. However, from empirical results $\beta^{1-|\uvec{n}_{\rm rp}\cdot\uvec{n}_{\rm rx}|^2}$ is approximated to $-5$dB. }

The received power calculated analytically on each receiver grid position is shown in Fig.~\ref{Fig:Analytical}. In Fig.~\ref{Fig:Analytical}(a), analytical received power due to $0.30\times0.30$~m$^2$ reflector~(left) and $0.61\times0.61$~m$^2$ reflectors~(right) are shown. The analytical results for $0.84\times0.84$~m$^2$ reflector are similar as $0.61\times0.61$~m$^2$. The received power for $0.30\times0.30$~m$^2$ reflector is smaller than $0.61\times0.61$~m$^2$ reflector. This is mainly due to the factor, $\beta^{1-|\uvec{n}_{\rm rp}\cdot\uvec{n}_{\rm rx}|^2}$, explained in Section~\ref{Section:First_order_modeling}. However, calculating an exact value of $\beta^{1-|\uvec{n}_{\rm rp}\cdot\uvec{n}_{\rm rx}|^2}$ at different grid positions is not possible. Therefore, from empirical results in Fig.~\ref{Fig:flat_combine}(b), (c) and (d), the value of $\beta^{1-|\uvec{n}_{\rm rp}\cdot\uvec{n}_{\rm rx}|^2}$ is approximated to $-3$~dB. This value corresponds to the decrease in received power for $0.30\times0.30$~m$^2$ reflector compared to $0.61\times0.61$~m$^2$ and $0.84\times0.84$~m$^2$ reflectors. On the other hand, from $0.61\times0.61$~m$^2$ to $0.84\times0.84$~m$^2$, this value do not change significantly.

From Fig.~\ref{Fig:Analytical}(a), it can be observed that we have maximum received power around the optimum reflection angle region~(rightmost on the grid). However, as the angular difference from this region increases, the received power decreases accordingly given by $\alpha_{\rm f}^{\Delta \theta}$ from~(\ref{Eq:PowerEq3}). On the other hand, the decrease in the received power due to the distance from the reflector source on the grid as we move left or downward is less significant compared to $\alpha_{\rm f}^{\Delta \theta}$. The value $\alpha_{\rm f}^{\Delta \phi}$ is $1$ as the angle in the elevation plane is set to optimum. Also, there is no change in the elevation plane as we move on the receiver grid. Moreover, there is no decrease in the received power due to the size of the reflector. This is because the sizes of the reflectors are larger than the size of the plane waves. %On the other hand, the smaller received power for $12\times12$~in$^2$ reflector compared to $24\times24$~in$^2$ is due to orientation mismatch of the reflected plane waves with the boresight of the receiver antenna.}

Comparing the analytical results in Fig.~\ref{Fig:Analytical}(a) with the measurement and simulation results in Fig.~\ref{Fig:flat_combine}(b) and (c), we observe a close match. However, the discontinuities in the received power over the grid for measurements are mainly due to higher order reflections in the corridor. In both the measurement and analytical results for $0.61\times0.61$~m$^2$ reflector, the received power approaches to the Friis free space power. 

\begin{figure*}[!t] 
    \centering
	\begin{subfigure}{0.45\textwidth}
	\centering
    \includegraphics[width=\textwidth]{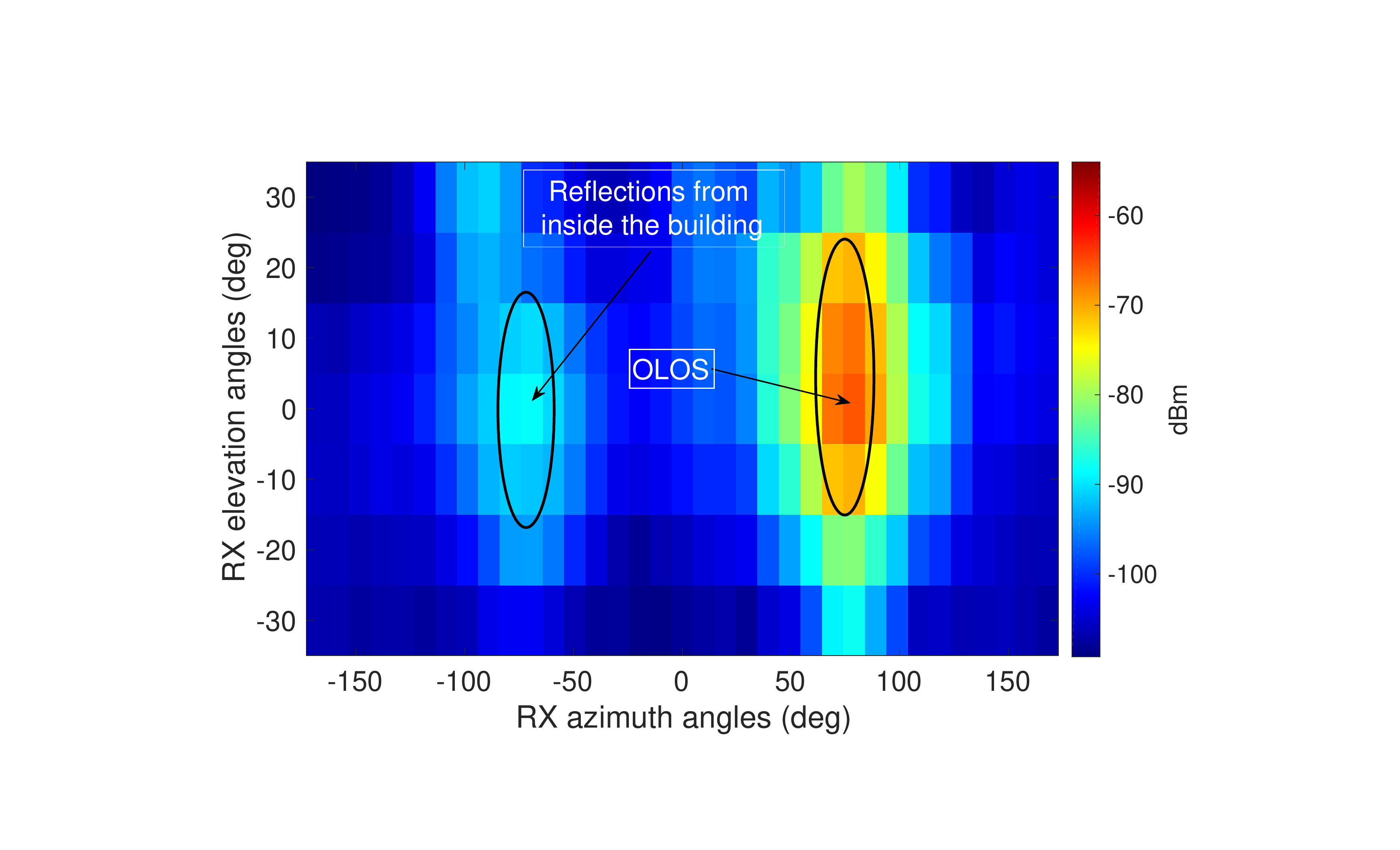}
    \caption{}
    \end{subfigure}	
 	\begin{subfigure}{0.45\textwidth}
    \centering
	\includegraphics[width=\textwidth]{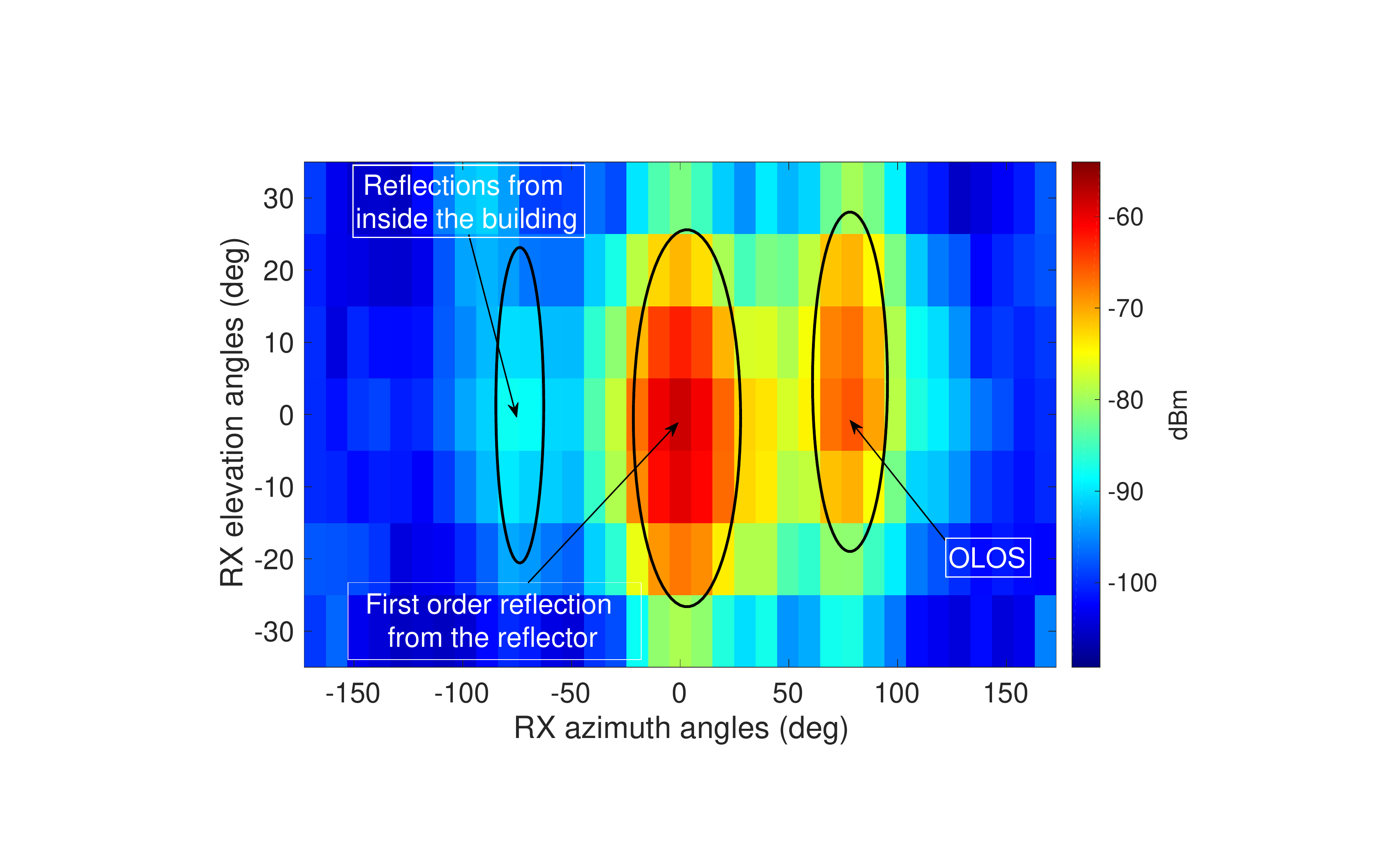}
	  \caption{}  
    \end{subfigure}     
	\begin{subfigure}{0.45\textwidth}
    \centering
	\includegraphics[width=\textwidth]{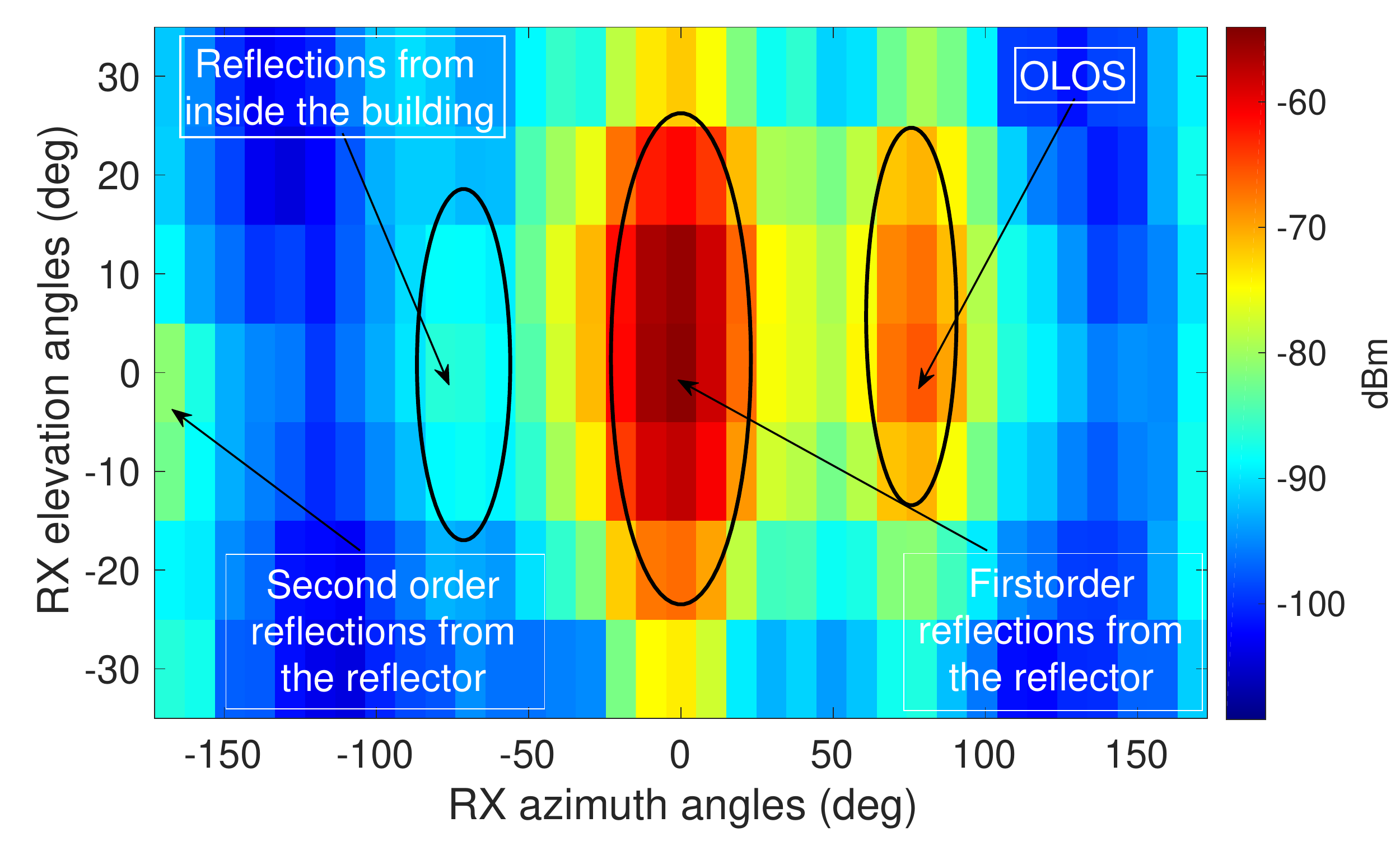}
	  \caption{}  
    \end{subfigure}    
    \begin{subfigure}{0.45\textwidth}
    \centering
	\includegraphics[width=\textwidth]{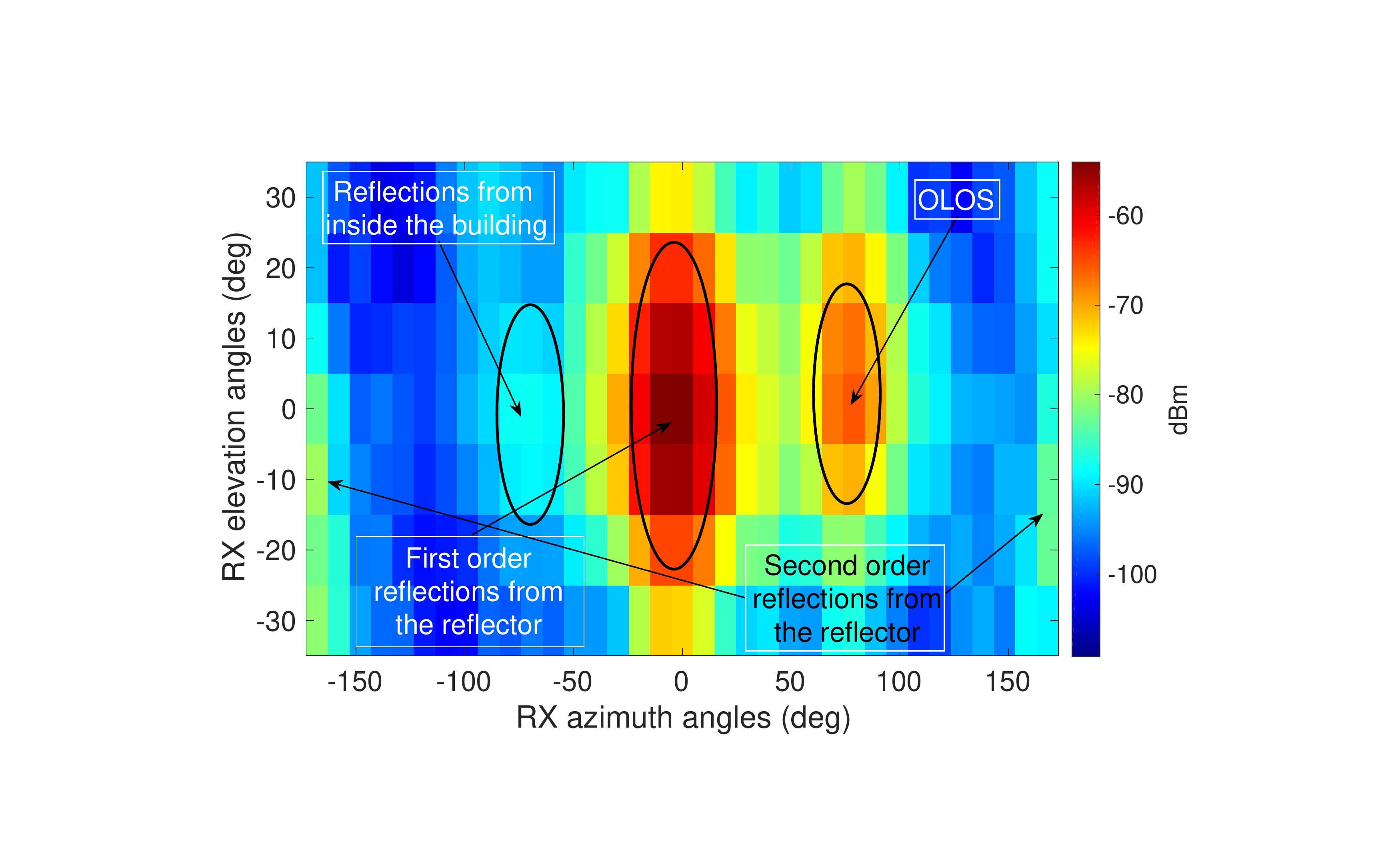}
	  \caption{}  
    \end{subfigure}
    \caption{Total received power at different azimuth and elevation angles at the receiver in the outdoor setting for (a) no reflector, (b) $0.30\times0.30$~m$^2$ reflector, (c) $0.61\times0.61$~m$^2$ reflector, and (d) $0.84\times0.84$~m$^2$ reflector.}\vspace{-4mm} \label{Fig:Outdoor_power}
\end{figure*}

The analytical received power results for sphere and cylinder reflectors are shown in Fig.~\ref{Fig:Analytical}(b). In the analytical modeling, we considered the surfaces of the cylinder and sphere as circularly curved. Therefore, they diverge the incoming power equally in all the directions over the grid. However, the decrease in the received power on the grid is mainly due to decrease in the receiver antenna's boresight gain as we move left on the grid. This decrease is represented by factor $\alpha_{\rm c}^{\Delta \theta}$, whereas, $\alpha_{\rm c}^{\Delta \phi} = 1$. Moreover, the received power on the grid is limited by the effective area of the cylinder and sphere~(see Section~\ref{Section:Effective_area}). Therefore, the cylinder has larger received power compared to sphere due to a larger effective area. Moreover, comparing measurement and analytical results for sphere and cylinder indicate differences, which are mainly due to non-ideal curved surfaces of real cylinder and sphere. 

\vspace{-1mm}
\subsection{CDF of Received Power for Indoor}
The received power across the receiver grid can be captured using a cumulative distribution function (CDF) for each scenario. The CDF plots of received power over the whole receiver grid for flat sheet reflectors and no reflector are shown in Fig.~\ref{Fig:CDF_combine}(a). As observed previously, the received power for the $0.30\times0.30$~m$^2$ reflector is smaller than the $0.61\times0.61$~m$^2$ and $0.84\times0.84$~m$^2$ reflectors, while the low power~(outage) areas are similar. The variance of the received power with different reflectors is higher when compared to no reflector scenario as expected. Another observation from Fig.~\ref{Fig:CDF_combine}(a) is that the received power varies in the range [-$75$,-$40$]~dBm, while for no reflector case it is [-$85$,-$70$]~dBm. This can be related to directional propagation in mmWave bands. In particular, highly directional scattering (reflection) results in power increase in some regions less and in some others more. 

In Fig.~\ref{Fig:CDF_combine}(b), the CDFs of received power from the simulations are shown corresponding to the same reflector scenarios. Results show that simulations match reasonably with measurements. We obtain a median gain of around $20$~dB for the $0.61\times0.61$~m$^2$ and $0.84\times0.84$~m$^2$ reflector scenarios as compared to the no reflector case. 

The CDF of received power for cylinder and sphere reflectors from measurements and simulations are shown in Fig.~\ref{Fig:CDF_combine}(c) and Fig.~\ref{Fig:CDF_combine}(d), respectively. Cylinder reflector exhibits  higher received power in the measurements compared to the sphere. Whereas in simulations, we observe high received power for both cylinder and sphere.

%\subsection{\textcolor{red}{CDF of Received Power at Different Center Frequencies}} 
 A plot of the CDF of the received powers at different center frequencies for different sizes of flat sheet reflectors is provided in Fig.~\ref{Fig:CDF_diff_freqs}. The results are obtained through ray tracing simulations. It can be observed that the mean received power is approximately the same for different sizes of flat reflectors, at a given center frequency. Moreover, the power gain~(comparing the received power in the presence of reflector with no reflector) at $38$~GHz and $60$~GHz is close to obtained at $28$~GHz. However, at $1.8$~GHz and $2.4$~GHz, we observe slightly higher power gain compared to other higher frequencies. This increased gain is attributed to negligible diffuse scattering from different material surfaces in the environment at $1.8$~GHz and $2.4$~GHz~(see Section~\ref{Section:Ray_tracing_setup}). The comparison of the received power results at different center frequencies with the analytical results from Section~\ref{Section:Power_distribution_modeling} provides a close fit.  

 \vspace{-2mm}

\section{Measurement and Analytical Results for Outdoor Scenarios with Reflectors}
In this section, outdoor measurements are presented and analyzed for different sizes of flat sheet reflectors. Moreover, analytical modeling results from Section~\ref{Section:Power_distribution_modeling} and their comparison with the measurement results  are also provided.
\vspace{-1mm}

\subsection{Outdoor Measurement Results}
The measurement results for received power in outdoor are obtained at a single receiver position. The receiver antenna is rotated from $-168^\circ$ to $168^\circ$ in the azimuth plane and $-30^\circ$ to $30^\circ$ in the elevation plane. The received power is obtained at each receiver antenna rotation. On the other hand, the transmit antenna is not rotated. Fig.~\ref{Fig:Outdoor_power}(a) shows the received power when there is no reflector. In this case, the maximum power of -$65.6$~dBm is observed at the azimuth angle of $81^\circ$ and an elevation angle of $0^{\circ}$. This received power corresponds to the OLOS path. Fig.~\ref{Fig:Core_area}(e) shows the geometry of the OLOS path. This OLOS path is from the transmitter through the glass window of the brick compartment and see-through glass of the door. The obstruction loss due to the glasses is given by $\eta = -10.86$~dB from (\ref{Eq:OLOS}). From Fig.~\ref{Fig:Outdoor_power}(a), there is also another strong reflection from inside the building at around $-70^\circ$ azimuth and $0^\circ$ elevation.

Fig.~\ref{Fig:Outdoor_power}(b) shows the received power with the $0.30\times0.30$~m$^2$ sheet reflector. We observe a strong reflection from the reflector at $0^\circ$ azimuth and $0^\circ$ elevation angles. This is because the center of the reflector is aligned to the boresight of the receiver antenna in azimuth and elevation planes. In addition, we observe the received power from other sources as observed previously for no reflector in Fig.~\ref{Fig:Outdoor_power}(a). Similarly, we observe reflections from the $0.61\times0.61$~$^2$ and $0.84\times0.84$~m$^2$ sheet reflectors at the same position as shown in Fig.~\ref{Fig:Outdoor_power}(c) and (d), respectively. The received power due to first order reflections from these reflectors are large compared to the $0.30\times0.30$~m$^2$ reflector. This is mainly because of the orientation mismatch for the reflected plane waves towards the receiver from the $0.30\times0.30$~m$^2$ reflector~(see Section~\ref{Section:First_order_modeling}). 

\begin{figure}[!t]
\centering\vspace{-2mm}
\centerline{\includegraphics[width=0.42\textwidth]{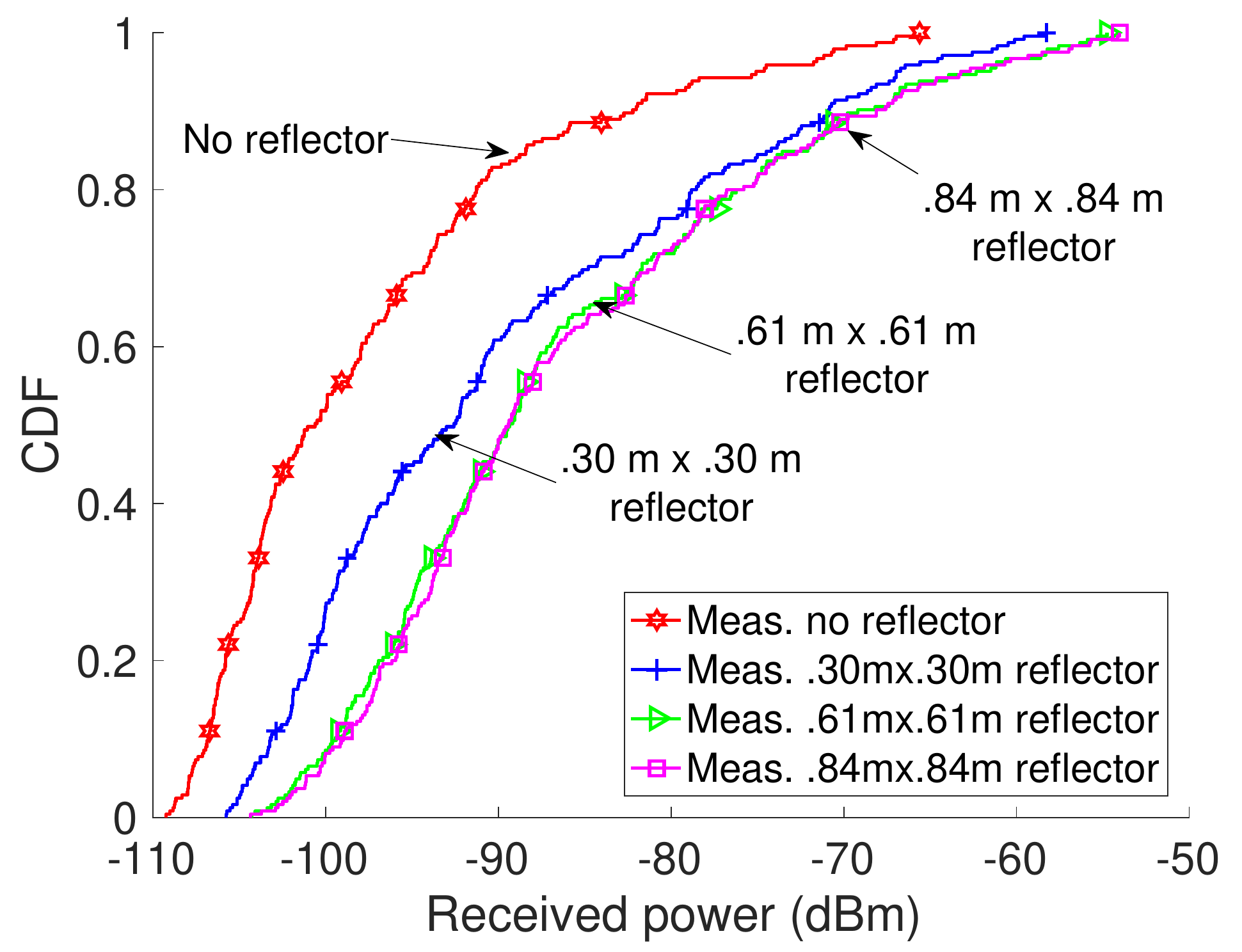}}
\caption{CDF of total received power for $0.30\times0.30$~m$^2$, $0.61\times0.61$~m$^2$ and $0.84\times0.84$~m$^2$ metallic sheet reflectors.} \vspace{-4mm}\label{Fig:outdoor_CDF}
\end{figure}

\begin{figure}[!t]
\centering
\centerline{\includegraphics[width=0.42\textwidth]{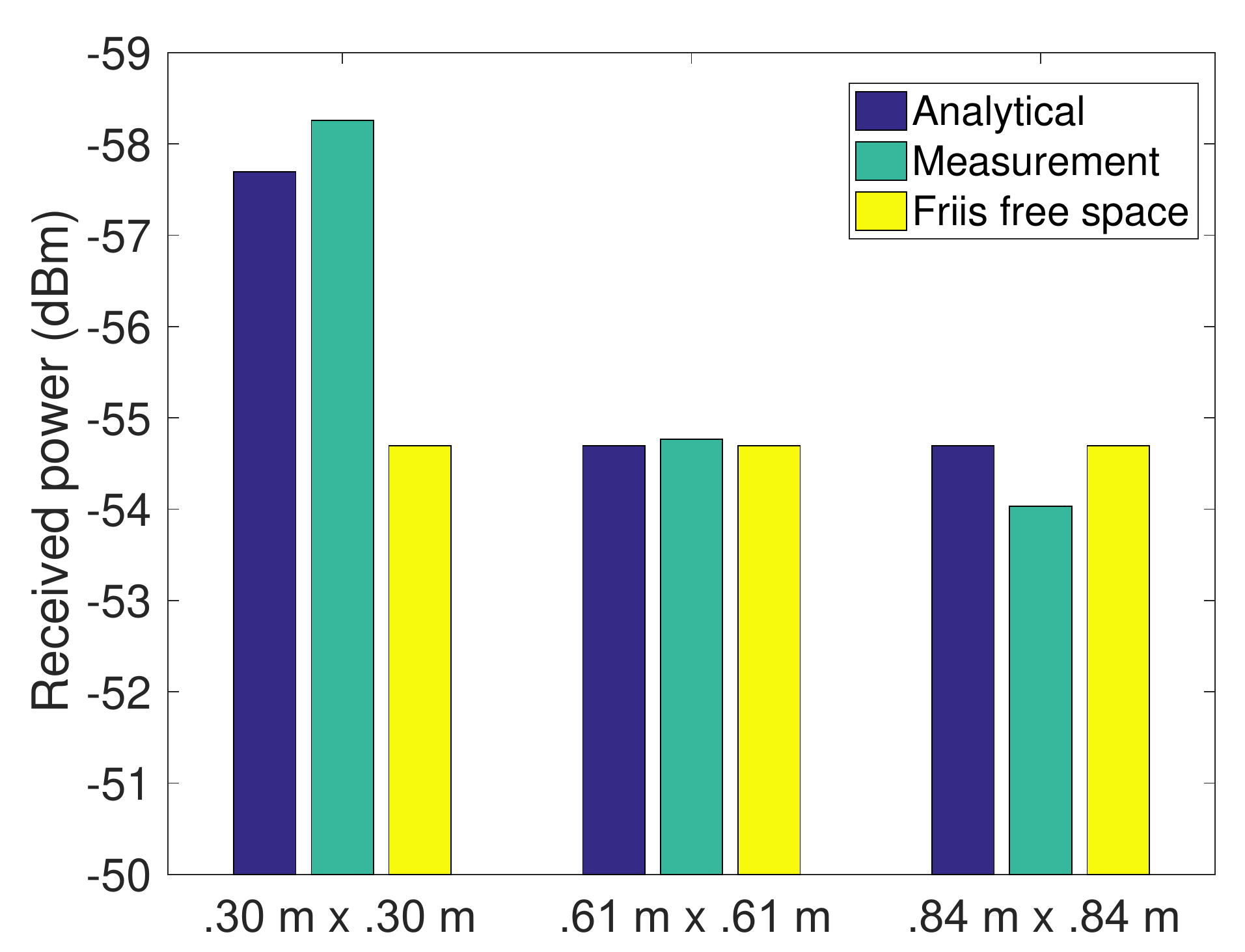}}
\caption{Total outdoor received power at $0^\circ$ azimuth and $0^\circ$ elevation receiver angles. The results are obtained for three sizes of flat square metallic reflectors. The analytical results are obtained from Section~\ref{Section:First_order_modeling}.} \vspace{-4mm}\label{Fig:outdoor_bar}
\end{figure}

Moreover, we observe secondary reflections at $\pm 168^\circ$ in the azimuth plane for $0.61\times0.61$~m$^2$ and $0.84\times0.84$~m$^2$ reflectors shown in Fig.~\ref{Fig:Outdoor_power}(c) and (d), respectively. This is because of spreading of the incoming plane waves over a large surface area of these reflectors. Therefore, the reflected plane waves are also larger. These large reflected plane waves create secondary reflections from surrounding objects. The secondary reflection at $-168^\circ$ is due to a car parked parallel to the receiver on the left side. On the other hand, the reflection at $167^\circ$ is from the wall on the right side of the receiver. These secondary reflections are not observed for $0.30\times0.30$~m$^2$ reflector due to the smaller size of the reflected plane waves. The CDFs of total received power captured at different angular positions for the outdoor measurements in Fig.~\ref{Fig:Outdoor_power} are shown in Fig.~\ref{Fig:outdoor_CDF}. It can be observed that we have a maximum gain of around $11$~dB for $0.61\times0.61$~m$^2$ and $0.84\times0.84$~m$^2$ reflectors compared to no reflector. %This gain is smaller compared to the indoor measurements. The reason for this small gain is due to larger received power for the no reflector outdoor compared to indoor. The large received received power for no reflector in outdoor is due to contribution from the OLOS, shown in Fig.~\ref{Fig:Outdoor_power}(a). }  

\vspace{-1mm}
\subsection{Analytical Received Power Results for Outdoor}
The empirical and analytical received power results for outdoor due to flat reflectors of different sizes are shown in Fig.~\ref{Fig:outdoor_bar}. These results are obtained for $0^\circ$ azimuth and elevation angles of the receiver antenna. The gain of the transmit and receiver antennas are maximum at this angular position. Moreover, we consider only first order reflections from reflectors at this angular position from Section~\ref{Section:First_order_modeling}. Therefore, from (\ref{Eq:PowerEq3}), there is no decrease in the received power due to the area of the reflectors as $A_{\rm refl}>A_{\rm pw}$. In addition, the orientation mismatch losses due to the transmitter, reflector and receiver positions given by $\alpha_{\rm f}^{\Delta \theta}$ and $\alpha_{\rm f}^{\Delta \phi}$ are negligible. 

The received power results from measurement and analytical model for the $0.30\times0.30$~m$^2$ reflector are shown in Fig.~\ref{Fig:outdoor_bar}. It is observed that we have lower received power compared to the Friis free space received power. This is due to orientation mismatch of the reflected plane waves towards the receiver from the small-sized reflector, given by $\beta^{1-|\uvec{n}_{\rm rp}.\uvec{n}_{\rm rx}|^2}$ in (\ref{Eq:PowerEq3}). This orientation mismatch loss is approximately $3$~dB. %IG: This was negative, loss can not be negative
However, for the $0.61\times0.61$~m$^2$ and $0.84\times0.84$~m$^2$ reflectors, we observe the received power from measurement and analytical results approach the Friis free space power.    
  
\section{Conclusions}\label{Section:Concluding Remarks}
In this work, channel measurements at $28$~GHz were carried out in NLOS indoor and outdoor scenarios. Metallic sheet reflectors of different shapes and sizes were used to enhance the received power, yielding a better signal coverage in the NLOS region. The reflected power from a metallic reflector was taken as a secondary source of transmission that can be further extended to other reflectors. This helps to provide coverage in any kind of NLOS scenario. Moreover, there is a minimum size of flat reflector required comparable to the size of the plane waves. This ensures the maximum received power approach to the Friis free space power. It was observed that the received power at a given point for flat reflectors is more sensitive to the orientation of the reflector compared to the size of the reflector. For the cylinder and sphere reflectors, we observed divergence of the received power according to their sizes and shapes. The measurement results were compared with RT simulations and analytical results, which are observed to  provide a close agreement. Ray tracing simulations at different center frequencies indicated a similar trend of reflected received power for flat sheet reflectors.

%\section*{Acknowledgement}
% This work has been supported in part by NASA under the Federal Award ID number
% NNX17AJ94A and by DOCOMO Innovations, Inc.

\bibliographystyle{IEEEtran}
\balance
%\bibliography{Main}


\end{document}